\PassOptionsToPackage{unicode}{hyperref}
\PassOptionsToPackage{hyphens}{url}
\PassOptionsToPackage{dvipsnames,svgnames,x11names}{xcolor}
\documentclass[
  12pt]{article}

\usepackage{amssymb,amsmath,amsfonts,amsthm}
\usepackage{diagbox}
\usepackage[linesnumbered,ruled,vlined]{algorithm2e}
\usepackage{algpseudocode}
\usepackage{ragged2e}
\usepackage{setspace} 
\usepackage{color}
\allowdisplaybreaks

\DeclareMathOperator*{\argmin}{arg\,min}
\usepackage{iftex}
\usepackage{bm}
\ifPDFTeX
  \usepackage[T1]{fontenc}
  \usepackage[utf8]{inputenc}
  \usepackage{textcomp} 
\else 
  \usepackage{unicode-math}
  \defaultfontfeatures{Scale=MatchLowercase}
  \defaultfontfeatures[\rmfamily]{Ligatures=TeX,Scale=1}
\fi
\usepackage{lmodern}
\ifPDFTeX\else  
\fi
\IfFileExists{upquote.sty}{\usepackage{upquote}}{}
\IfFileExists{microtype.sty}{
  \usepackage[]{microtype}
  \UseMicrotypeSet[protrusion]{basicmath} 
}{}
\makeatletter
\@ifundefined{KOMAClassName}{
  \IfFileExists{parskip.sty}{%
    \usepackage{parskip}
  }{
    \setlength{\parindent}{0pt}
    \setlength{\parskip}{6pt plus 2pt minus 1pt}}
}{
  \KOMAoptions{parskip=half}}
\makeatother
\usepackage{xcolor}
\makeatletter
\ifx\paragraph\undefined\else
  \let\oldparagraph\paragraph
  \renewcommand{\paragraph}{
    \@ifstar
      \xxxParagraphStar
      \xxxParagraphNoStar
  }
  \newcommand{\xxxParagraphStar}[1]{\oldparagraph*{#1}\mbox{}}
  \newcommand{\xxxParagraphNoStar}[1]{\oldparagraph{#1}\mbox{}}
\fi
\ifx\subparagraph\undefined\else
  \let\oldsubparagraph\subparagraph
  \renewcommand{\subparagraph}{
    \@ifstar
      \xxxSubParagraphStar
      \xxxSubParagraphNoStar
  }
  \newcommand{\xxxSubParagraphStar}[1]{\oldsubparagraph*{#1}\mbox{}}
  \newcommand{\xxxSubParagraphNoStar}[1]{\oldsubparagraph{#1}\mbox{}}
\fi
\makeatother

\usepackage{longtable,booktabs,array}
\usepackage{calc} 
\usepackage{etoolbox}
\makeatletter
\patchcmd\longtable{\par}{\if@noskipsec\mbox{}\fi\par}{}{}
\makeatother
\IfFileExists{footnotehyper.sty}{\usepackage{footnotehyper}}{\usepackage{footnote}}
\makesavenoteenv{longtable}
\usepackage{graphicx}
\makeatletter
\def\maxwidth{\ifdim\Gin@nat@width>\linewidth\linewidth\else\Gin@nat@width\fi}
\def\maxheight{\ifdim\Gin@nat@height>\textheight\textheight\else\Gin@nat@height\fi}
\makeatother
\setkeys{Gin}{width=\maxwidth,height=\maxheight,keepaspectratio}
\makeatletter
\def\fps@figure{htbp}
\makeatother

\makeatletter
\@ifpackageloaded{caption}{}{\usepackage{caption}}
\AtBeginDocument{%
\ifdefined\contentsname
  \renewcommand*\contentsname{Table of contents}
\else
  \newcommand\contentsname{Table of contents}
\fi
\ifdefined\listfigurename
  \renewcommand*\listfigurename{List of Figures}
\else
  \newcommand\listfigurename{List of Figures}
\fi
\ifdefined\listtablename
  \renewcommand*\listtablename{List of Tables}
\else
  \newcommand\listtablename{List of Tables}
\fi
\ifdefined\figurename
  \renewcommand*\figurename{Figure}
\else
  \newcommand\figurename{Figure}
\fi
\ifdefined\tablename
  \renewcommand*\tablename{Table}
\else
  \newcommand\tablename{Table}
\fi
}
\@ifpackageloaded{float}{}{\usepackage{float}}
\floatstyle{ruled}
\@ifundefined{c@chapter}{\newfloat{codelisting}{h}{lop}}{\newfloat{codelisting}{h}{lop}[chapter]}
\floatname{codelisting}{Listing}

\makeatother
\makeatletter
\@ifpackageloaded{caption}{}{\usepackage{caption}}
\@ifpackageloaded{subcaption}{}{\usepackage{subcaption}}
\makeatother

\ifLuaTeX
  \usepackage{selnolig}  
\fi
\usepackage[]{natbib}
\usepackage{bookmark}

\IfFileExists{xurl.sty}{\usepackage{xurl}}{} 
\hypersetup{
  pdftitle={A Physics-Informed Convolutional Long Short Term Memory Statistical Model for Fluid Thermodynamics Simulations},
  pdfauthor={Luca Menicali},
  pdfkeywords={3 to 6 keywords, that do not appear in the title},
  colorlinks=true,
  linkcolor={blue},
  filecolor={Maroon},
  citecolor={Blue},
  urlcolor={Blue},
  pdfcreator={LaTeX via pandoc}}

\newcommand{\anon}{1}

\begin{document}

\def\spacingset#1{\renewcommand{\baselinestretch}%
{#1}\small\normalsize} \spacingset{1}


\if1\anon
{
  \title{\bf A Physics-Informed Spatiotemporal Deep Learning Framework for Turbulent Systems}
  \author{Luca Menicali\hspace{.2cm}\\
    Department of Applied and Computational Mathematics and Statistics, \\ University of University of Notre Dame\\
    and \\
    Andrew Grace \\
    Department of Environmental and Civil Engineering, \\ University of Notre Dame \\ 
    and \\
    David H. Richter \\
    Department of Environmental and Civil Engineering, \\ University of Notre Dame \\ 
    and \\
    Stefano Catruccio \\
    Department of Applied and Computational Mathematics and Statistics, \\ University of University of Notre Dame}
  \maketitle
} \fi

\if0\anon
{
  \bigskip
  \bigskip
  \bigskip
  \begin{center}
    {\LARGE\bf A Physics-Informed Spatiotemporal Deep Learning Framework for Turbulent Systems}
\end{center}
  \medskip
} \fi

\bigskip
\begin{abstract}
Fluid thermodynamics underpins atmospheric dynamics, climate science, industrial applications, and energy systems. However, direct numerical simulations (DNS) of such systems can be computationally prohibitive. To address this, we present a novel physics-informed spatiotemporal surrogate model for Rayleigh-B\'enard convection (RBC), a canonical example of convective fluid flow. Our approach combines convolutional neural networks, for spatial dimension reduction, with an innovative recurrent architecture, inspired by large language models, to model long-range temporal dynamics. Inference is penalized with respect to the governing partial differential equations to ensure physical interpretability. Since RBC exhibits turbulent behavior, we quantify uncertainty using a conformal prediction framework. This model replicates key \textit{physical} features of RBC dynamics while significantly reducing computational cost, offering a scalable alternative to DNS for long-term simulations.
\end{abstract}

\noindent%
\vfill

\newpage
\spacingset{1.8} 

\section{Introduction \& Background} \label{sec:intro}

Modeling fluid thermodynamics is critical for advancing our understanding of complex physical systems, including atmospheric dynamics \citep{trenberth2003seamless}, climate science \citep{held1994proposal}, and industrial processes \citep{singh2014heat}. Accurate thermodynamic models enable the prediction of phase transitions, heat transfer, and fluid motion---key to optimizing engineering systems and mitigating environmental impacts \citep{gyftopoulos2012thermodynamics}. However, Direct Numerical Simulation (DNS) is computationally expensive due to the wide range of spatial and temporal scales and strong nonlinearities involved \citep{xu2015direct}. As a result, statistical surrogate models have emerged as promising alternatives, capable of learning from limited data and enabling longer simulations at significantly lower cost.

Surrogates are effective for modeling turbulence only if they can capture the nonlinear, multiscale structure of the governing dynamics. To this end, reduced-order models such as Proper Orthogonal Decomposition and Koopman operator theory have provided principled, interpretable, and computationally efficient frameworks for simulating dominant flow structures \citep{rowley2005model, brunton2021modern}. Proper
orthogonal decomposition-based surrogate modeling combined with Gaussian process emulation has further enabled efficient reduced-order representations with built-in uncertainty quantification by treating reduced states as time-independent \citep{Mak02102018}. More recently, operator learning approaches, such as the Fourier Neural Operator, have shown promise in approximating Partial Differential Equations (PDE) solution operators, scaling to high-resolution turbulent flows \citep{li2020fourier}. These methods aim to learn solution operators rather than simple input-output maps. However, most are deterministic and lack mechanisms for uncertainty quantification, limiting their reliability in sensitive or chaotic regimes.

An alternative paradigm uses supervised learning to emulate turbulent flows directly from data. Convolutional Neural Networks (CNNs) have been used to predict flow fields from boundary conditions or coarse inputs, while Recurrent Neural Networks (RNNs) model temporal evolution \citep{guo2016convolutional, maulik2021rom}. Convolutional Recurrent Neural Networks (CRNNs) — which combine CNN-based spatial encoding with RNN-based temporal modeling — have shown success in spatiotemporal forecasting of turbulent flows. These models typically employ an encoder to compress spatial structure, an RNN to model dynamics in reduced space, and a decoder to reconstruct high-dimensional output. Applications include two-dimensional cylinder flows \citep{BEIKI2023}, turbulence reconstruction from sparse data \citep{fukami2021global}, and short-term forecasting in low-turbulence settings \citep{mohan2019compressed, data_driven_RBC, straat2025}. These models are however currently limited in that their architectures are suboptimal in long-range forecasting and are not formulated as \textit{statistical} models, so they don't allow formal uncertainty quantification.

In parallel, a growing body of work in \textit{scientific machine learning} seeks to improve model interpretability by embedding physical laws directly into learning algorithms. Physics-Informed Neural Networks (PINNs) incorporate PDE constraints into inference via penalized optimization, leading to predictions that better align with physical theory \citep{cuomo2022scientific, RAISSI2019}. While PINNs have been applied to systems like the 1D Burgers' equation, laminar flows, and Reynolds-averaged Navier-Stokes, their extension to fully turbulent, high-dimensional systems remains an open challenge \citep{rao2020physics, rans}. Key limitations include a focus on weakly turbulent regimes, limited long-range accuracy, and lack of the quantification of predictive uncertainty.

In this work, we propose a novel Physics-Informed Convolutional Recurrent Neural Network (PI-CRNN) for efficient, long-range forecasting of fluid thermodynamic systems. Our model integrates four key components: 
\begin{itemize}
    \item spatial feature extraction via a convolutional autoencoder
    \item temporal modeling through a sequence-to-sequence ConvLSTM architecture
    \item  physics-informed inference by enforcing mass, momentum, and energy conservation
    \item uncertainty quantification using conformal prediction \citep{convlstm, bonas_quantile}
\end{itemize}

While previous work has explored CRNNs and PINNs separately, to our knowledge, this is the first model that embeds a ConvLSTM within the reduced space of a convolutional autoencoder while enforcing PDE constraints and providing distribution-free prediction intervals. PI-CRNN model is intended to be used as an offline-trained statistical surrogate, which is crucial for a variety of scientific and engineering applications where long time series of a turbulent physical process are needed but DNS is considerably more expensive. 

Our testbed is Rayleigh-B\'enard convection (RBC), a canonical example in fluid thermodynamics featuring nonlinear buoyancy-driven flow between differentially heated surfaces \citep{rbc}. RBC exhibits high-dimensional turbulence and is highly sensitive to initial conditions, making it ideal for evaluating long-term forecast accuracy and uncertainty quantification. While the governing physical equations, i.e., the data-generating process, are \textit{deterministic}, this sensitivity results in \textit{de facto} stochastic behavior \citep{lorenz1963}. We incorporate physical constraints by enforcing the Navier-Stokes equations via penalized optimization \citep{navier1822memoire, stokes1845friction}, and quantify uncertainty through conformal prediction, which guarantees calibrated prediction intervals under minimal assumptions. The turbulent nature of RBC, coupled with the computational cost of DNS, motivates the development of a statistical surrogate capable of capturing the spatiotemporal dynamics and associated uncertainties at a fraction of the computational expense. PI-CRNN offers a unified, principled framework for surrogate modeling of complex thermodynamic systems, balancing computational efficiency, physical plausibility, and statistical rigor.

The remainder of the manuscript is structured as follows: Section \ref{sec:data} summarizes the turbulent convection data and scientific background; Section \ref{sec:methods} introduces the PI-CRNN model; Section \ref{sec:inference} presents the physics-informed learning framework; Section \ref{sec:results} reports the results; and Section \ref{sec:conclusion} concludes with a discussion of limitations and future directions.

\section{Data and Scientific Background} \label{sec:data}

The data employed in this work is numerically simulated two-dimensional Rayleigh-B\'enard  convection (RBC). RBC is a canonical fluid flow and is a well-studied model of turbulent thermal convection \citep{loh24}. Turbulent thermal convection is a fundamental natural process that governs heat transport in oceanographic, climatic, and astrophysical settings \citep{GROSSMANN_LOHSE_2000, RevModPhys.81.503, annurev.fluid.010908.165152}, and is crucial for HVAC systems and natural draft cooling towers for large-scale nuclear or thermal power stations \citep{stanford2003hvac}. Due to fluid turbulence (the chaotic and seemingly random evolution of the state of the fluid flow), the physics underlying RBC defies simple analytical treatment. Therefore, understanding and modeling the transport processes in these contexts requires numerical or statistical modeling. 
 
A particularly important application across many natural science and engineering disciplines is that of particle dispersion in turbulent natural convection. Microscopic particles, such as cloud droplets or drizzle, dynamically respond to turbulent thermal convection.  
Modern approaches to studying particle dispersion often take one of several flavors. For instance, one may study particle dispersion based on artificially prescribed flow statistics (e.g., with Gaussian white noise). The fundamental drawback to this method is that the underlying flow statistics miss the spatial and temporal correlations characteristic of turbulent convection. The result is that the particle response may be fundamentally incorrect. An alternative approach is to numerically solve both the highly non-linear and multi-scale Navier-Stokes equations and the equations of motion for tens of millions of individual particles simultaneously. This method properly represents the statistics of the turbulent convective flow and thus the particle-fluid interactions, but can be computationally expensive and time consuming, thereby rendering ensemble calculations infeasible at large enough scales. Thus, finding an alternative way to efficiently generate ensembles of the turbulent convective flow state, ensuring that underlying statistics are properly represented, is of immense value. 

Importantly, RBC is the mathematical tool by which we study the underlying turbulent processes governing heat and particle transport in the aforementioned contexts. 
RBC is driven by a gravitational instability due to its density. The density of a fluid is often controlled by its temperature, and for many natural settings, increasing (decreasing) the temperature leads to a decrease (increase) in density. If cool dense fluid is initially found below warmer, less dense, fluid, vertical motion is induced as the system tries to attain a stable state. In the RBC flow considered here, this flow state is driven by a difference in temperature between the warm lower plate and the cold upper plate, which produces convection cells as warm (thus less dense) fluid rises and cold (thus more dense) fluid falls, as seen in Figure S1 of the supplementary material. 
The dataset, generated via DNS, includes horizontal and vertical velocity components, pressure, and temperature with high spatial and temporal resolution. RBC is governed by the Navier-Stokes equations under the Boussinesq approximation, which encode mass conservation, momentum, and energy conservation for a thermally stratified fluid \citep{kun24}. The Rayleigh number for this data (a quantity identifying the amount of turbulence in the system) is $\text{Ra} = 2.54 \times 10^{8}$. A low Ra (less than $10^{6}$) corresponds to laminar (non-turbulent) flow, while in our case we have strongly turbulent convection, making the dataset both scientifically realistic and statistically challenging. This data offers an ideal testbed: the underlying physics are well established, yet the resulting flow is high-dimensional, nonlinear, and turbulent, posing a significant challenge for long-range forecasting. This motivates the development of a surrogate model that can capture both uncertainty and dynamics efficiently. Further details on the data and governing equations are in Section S.2 of the supplementary material.

In light of these observations, we address the following research questions:
\begin{itemize}
\item[\textbf{Q1.}] Can a computationally efficient surrogate model reproduce the essential spatial and temporal statistics of strongly turbulent convection?
\item[\textbf{Q2.}] Can such a surrogate enable accurate long-range forecast of turbulent thermal convection?
\end{itemize}

The data was generated on a $n_{x} \times n_{z}$ grid, $n_{x}=n_{z}=256$, with a temporal resolution of $\Delta_{t} = 0.1$ seconds for a total of $T=2{,}500$ time steps. We denote the RBC spatiotemporal process with $d=4$ covariates observed on a $n_{x} \times n_{z}$ grid by $\bm{Y}_{t}(x,z) = \left( u_{t}(x,z), w_{t}(x,z), p_{t}(x,z), \theta_{t}(x,z) \right)$, where $u_{t}(x,z)$ and $w_{t}(x,z)$ are the horizontal and vertical velocity components, respectively, $p_{t}(x,z)$ is pressure, $\theta_{t}(x,z)$ is temperature, $(x,z)$ are spatial coordinates, and $t=1,\ldots,T$. For simplicity, we hereto omit the spatial coordinates, so we denote one observation \textit{over the entire spatial domain} by $\bm{Y}_{t} = \left( u_{t}, w_{t}, p_{t}, \theta_{t} \right) \in \mathbb{R}^{n_{x} \times n_{z} \times d}$. We also denote the sequence of $b$ observations \textit{before} $t$ by $\mathcal{Y}^{-}_{t;b} = \left(\bm{Y}_{t-b}, \ldots, \bm{Y}_{t-2}, \bm{Y}_{t-1}\right)$ and the sequence of $a$ observations \textit{starting at} $t$ by $\mathcal{Y}^{+}_{t;a} = \left(\bm{Y}_{t}, \bm{Y}_{t+1}, \ldots, \bm{Y}_{t+a-1}\right)$. 

\section{Methods} \label{sec:methods}

As detailed in Section \ref{sec:data}, $\bm{Y}_{t}$ is a high-dimensional process, where one observation consists of $D = n_{x} \times n_{z} \times d$ points. At $\text{Ra} = 2.54 \times 10^{8}$, RBC exhibits turbulent dynamics, making direct modeling of $\bm{Y}_{t}$ both computationally expensive and statistically challenging. To address this, we model the temporal evolution of $\bm{Y}_{t}$ through a reduced space representation, which we denote by $\bm{\tilde{Y}}_{t}$, of dimensions $\tilde{D} \ll D$. Indeed, $\bm{\tilde{Y}}_{t}$ is intended to capture the essential spatial structure of $\bm{Y}_{t}$ while enabling efficient forecasting. This reduced representation naturally leads to a state-space formulation, where spatial compression and temporal evolution can be modeled separately. 

The state-space model is, formally,
\begin{subequations} \label{mod:state_space}
\begin{align}
    \text{Observation Equation:} &\quad \bm{Y}_{t} = \mathcal{SD}\left( \underbrace{\mathcal{SE}\left( \bm{Y}_{t} \mid \bm{W}_{\text{en}} \right)}_{\bm{\tilde{Y}}_{t}} \mid \bm{W}_{\text{de}} \right) + \bm{\eta}_{t}, \label{eq:ss_obs} \\
    \text{State Equation:} &\quad \bm{\tilde{Y}}_{t} = \mathcal{T}\left( \bm{\tilde{Y}}_{t-1}, \ldots, \bm{\tilde{Y}}_{t-b} \mid \bm{W}^{\text{time}} \right) + \tilde{\bm{\epsilon}}_{t}, \label{eq:ss_state}
\end{align}
\end{subequations}
where $\bm{\tilde{Y}}_{t} \in \mathbb{R}^{\Tilde{n}_{x} \times \Tilde{n}_{z} \times \tilde{d}}$, $\tilde{D} = \Tilde{n}_{x} \times \Tilde{n}_{z} \times \tilde{d}$, is the reduced space representation of $\bm{Y}_{t}$. In \eqref{eq:ss_obs}, $\mathcal{SE}(\cdot)$ denotes the spatial encoder and $\mathcal{SD}(\cdot)$ denotes the spatial decoder, so the composition of the two is an \textit{autoencoder} as a (possibly nonlinear) function of some parameters $\bm{W}^{\text{space}}$. Lastly, $\bm{\eta}_{t}$ is the error term associated with the data reduction which we assume to be Gaussian with mean zero.  In \eqref{eq:ss_state}, $\mathcal{T}(\cdot)$ models the temporal evolution of $\bm{\tilde{Y}}_{t}$ as a (possibly nonlinear) function of some parameters $\bm{W}^{\text{time}}$. The error term $\tilde{\bm{\epsilon}}_{t}$ captures the prediction uncertainty in reduced space and we assume it to be Gaussian with mean zero. Diagnostics on the Gaussian assumptions are available in Section S.6 of the supplementary material. This modeling approach allows for 1) the complexity of the temporal structure of $\bm{Y}_{t}$ to be greatly simplified in reduced space, since $\bm{\tilde{Y}}_{t}$ only retains key features and 2) the computational burden to be greatly reduced, since the dimensions of $\tilde{\bm{Y}}_{t}$ are $\frac{\tilde{D}}{D}$\% of those of $\bm{Y}_{t}$.

In the observation equation \eqref{eq:ss_obs}, we implement a convolutional autoencoder (CAE), a spatially-aware neural network model where a spatial encoder compresses the input $\bm{Y}_{t}$ into $\bm{\tilde{Y}}_{t}$ (reduced space), while a spatial decoder inputs $\bm{\tilde{Y}}_{t}$ and returns it to the original dimensions $D$ \citep{cae}. As such, if $\bm{Y}_{t} \approx \mathcal{SD}(\bm{\tilde{Y}}_{t} \mid \bm{W}_{\text{de}})$ then the spatial encoder $\mathcal{SE}(\cdot)$ successfully compressed the features of $\bm{Y}_{t}$ into $\bm{\tilde{Y}}_{t}$. In the state equation \eqref{eq:ss_state}, the temporal structure of $\bm{\tilde{Y}}_{t}$ is modeled in time with a RNN $\mathcal{T}(\cdot)$ using a sequence-to-sequence framework, i.e., a model that inputs a sequence and outputs a sequence (not necessarily of the same length). We denote the parameters in the temporal model by $\bm{W}^{\text{time}}$ and all the parameters in our spatiotemporal model \eqref{mod:state_space} by $\bm{W}=(\bm{W}^{\text{space}}, \bm{W}^{\text{time}})$.

In the following Sections \ref{sec:cae} and \ref{sec:rnn} we detail the CAE and RNN, respectively. Section \ref{sec:crnn} summarizes the spatiotemporal model.

\subsection{Spatial Model: Convolutional Autoencoder} \label{sec:cae}

\begin{figure}[!ht]
    \centering
    \includegraphics[width=\linewidth]{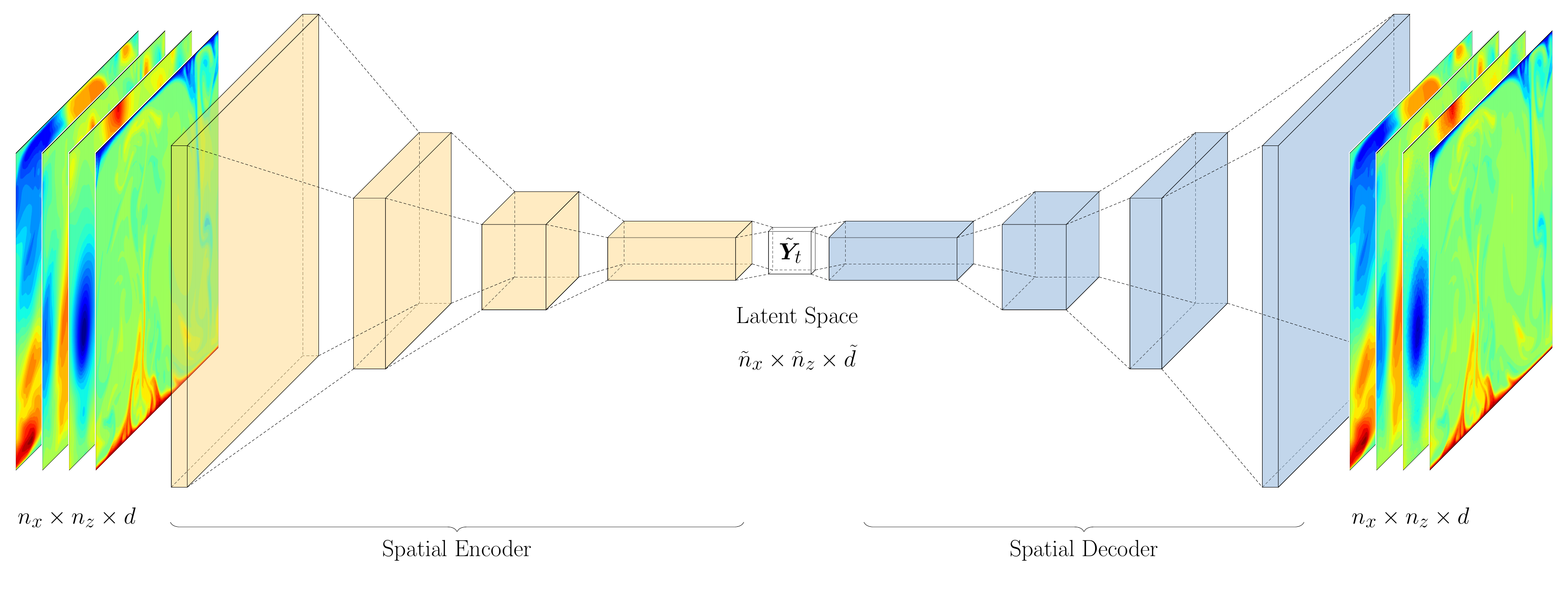}
    \caption{Architecture of the CAE model. The input $\bm{Y}$ is projected onto reduced space through a series of convolutions, \eqref{eq:convolution} in yellow, to $\bm{\tilde{Y}}$ and subsequently projected back onto the original dimensions, $n_{x} \times n_{z} \times d$, through a series of deconvolutions, in blue.}
    \label{fig:cae_architecture}
\end{figure}

Throughout this section we omit the $t$ subscript from $\bm{Y}_{t}$ for simplicity. We model the spatial structure of $\bm{Y}$ with a CAE, whose key principle is to first project the input onto a reduced space, $\tilde{\bm{Y}} \in \mathbb{R}^{\Tilde{n}_{x} \times \Tilde{n}_{z} \times \tilde{d}}$, referred to as the spatial encoder, and subsequently projected back onto $\mathbb{R}^{n_{x} \times n_{z} \times d}$, referred to as the spatial decoder. Figure \ref{fig:cae_architecture} provides an illustration for the CAE model's architecture, where yellow blocks represent the encoder and blue block represent the decoder.  Both the spatial encoder and spatial decoder are \textit{nonlinear} functions and the dimension-reconstruction in the decoder is \textit{not} generally the inverse of the dimension-reduction performed in the encoder. In CAEs the input undergoes convolutions, followed by a nonlinear activation function $g$ such as the rectified linear unit, sigmoid, or hyperbolic tangent function \citep{goodfellow2016}. 

The spatial encoder consists of $\ell=1, \ldots, L$ convolutional layers with $K^{\ell}$ kernels in the $\ell$-th layer, where each kernel is a two-dimensional matrix $r^{\ell}_{x} \times r^{\ell}_{z}$ (since the data is observed in two dimensions). The kernels each perform a local convolution of the input every $s=2$ elements (\textit{strides}) and are summed across the kernels from the $(\ell-1)$-th layer. Strided convolutions combine downsampling and feature extraction in a single operation, reducing computational cost compared to separate convolution and pooling layers \citep{springenberg2019strided}. As such, the $\ell$-th convolutional layer with $K^{\ell}$ kernels and stride $s$ maps yields $K^{\ell}$ \textit{feature maps}, each one of dimensions $n_{x}^{\ell} \times n_{z}^{\ell}$, i.e., a three-dimensional matrix $n_{x}^{\ell} \times n_{z}^{\ell} \times K^{\ell}$, where $n_{x}^{\ell} = \lfloor n_{x}^{\ell-1} / s \rfloor$ and $n_{z}^{\ell} = \lfloor n_{z}^{\ell-1} / s \rfloor$. In the $\ell$-th layer, denote by $Z^{\ell}_{i,j,f}$ the $(i,j)$ entry of the $f$-th feature map, by $w^{\ell}_{a,b,\tilde{f},f}$ the $(a,b)$ entry of the $f$-th kernel for the $\tilde{f}$-th channel of the previous layer, $\tilde{f} \in \{1,\ldots,K^{\ell-1} \}$, and by $w^{\ell}_{0,f}$ the bias term. The spatial encoder performs the following:
\begin{gather}
    Z^{\ell}_{i,j,f} = g \left( \sum_{\tilde{f}=1}^{K^{\ell-1}} \sum_{a=1}^{r_{x}^{\ell}} \sum_{b=1}^{r_{z}^{\ell}} Z^{\ell-1}_{si+a,sj+b;\tilde{f}} \cdot w^{\ell}_{a,b;\tilde{f},f} + w^{\ell}_{0,f} \right). \label{eq:convolution}
\end{gather}
If we denote the weights in the $\ell$-th layer of the spatial encoder by $\bm{W}^{\ell} = \left\{ \left(w^{\ell}_{a,b,\tilde{f},f}, w^{\ell}_{0,f} \right) : a=1, \ldots, r_{x}^{\ell}, ~b=1, \ldots, r_{z}^{\ell},~\tilde{f}=1,\ldots, K^{\ell-1},~f=1,\ldots,K^{\ell} \right\}$ and $\bm{Z}^{\ell} = \{Z^{\ell}_{i,j,f} : i = 1, \ldots, n^{\ell}_{x},~ j = 1, \ldots, n^{\ell}_{z},~f=1,\ldots,K^{\ell} \}$, then, the spatial encoder $\mathcal{SE}(\cdot \mid \bm{W}_{\text{en}})$ can be written in matrix form
\begin{gather}
    \bm{Z}^{\ell} = g\left( \bm{Z}^{\ell-1} \ast \bm{W}^{\ell} \right),~ \ell=1,\ldots,L, \label{mod:spatial_encoder}
\end{gather}
where $\ast$ denotes the strided ($s$) convolution operation \eqref{eq:convolution}. Additionally, note that $\bm{Z}^{0}=\bm{Y}$, $\bm{Z}^{L}=\bm{\tilde{Y}}$, with $\left(n_{x}^{L}, n_{z}^{L}, K^{L} \right) = \left(\Tilde{n}_{x}, \Tilde{n}_{z}, \tilde{d} \right)$. So for $\ell=1$ we have $K^{0} = d$, the four RBC variables, and for $\ell=L$ we have $K^{L} = \tilde{d}$, i.e., the number of feature maps in $\bm{\tilde{Y}}$ (the reduced space representation of $\bm{Y}$). We denote the parameters in the spatial encoder by $\bm{W}_{\text{en}} = \left\{\bm{W}^{\ell} : \ell=1,\ldots,L\right\}$ and the number of parameters to estimate  is
\begin{gather}
    \left| \bm{W}_{\text{en}} \right| = \sum_{\ell=1}^{L} \left(r_{x}^{\ell} \cdot r_{z}^{\ell} \cdot K^{\ell-1} + 1 \right) \cdot K^{\ell}. \label{eq:num_pars_conv}
\end{gather}

The spatial decoder consists of $L+1$ deconvolutional layers, which are similar to convolutional layers but \textit{increase} the number of dimensions of the feature maps. In the $\ell$-th deconvolutional layer, denote the number of kernels by  $\dot{K}^{\ell}$ and the size of each kernel by $\dot{r}_{x}^{\ell} \times \dot{r}_{z}^{\ell}$. In the first $L$ layers of the decoder, the number of kernels in the $\ell$-th layer is the same as the number of kernels in the $(L-\ell-1)$-th layer of the encoder, so the spatial encoder and spatial decoder ``mirror'' each other, where $( \dot{n}_{x}^{\ell}, \dot{n}_{z}^{\ell}, \dot{K}^{\ell} ) = \left(n_{x}^{L-\ell-1}, n_{z}^{L-\ell-1}, K^{L-\ell-1} \right)$. Since the $L$-th layer of the decoder yields $K^{1}$ feature maps (i.e., the number of feature maps in the first layer of the encoder), the spatial decoder includes one final (output) layer with $\dot{K}^{L+1}=d$ kernels, so that the output matches the dimensions of the data, $n_{x} \times n_{z} \times d$.  The deconvolution operation is similar to \eqref{eq:convolution} except it increases the dimensions of the feature maps. Indeed, the $\ell$-th deconvolutional layer with $\dot{K}^{\ell}$ kernels and stride equal to $s$ yields a feature map of dimensions  $\dot{n}_{x}^{\ell} \times \dot{n}_{z}^{\ell} \times \dot{K}^{\ell}$, where $\dot{n}_{x}^{\ell} = s\dot{n}_{x}^{\ell-1}$ and $\dot{n}_{z}^{\ell} = s\dot{n}_{z}^{\ell-1}$. The input to the spatial decoder $\mathcal{SD}(\cdot \mid \bm{W}_{\text{de}})$ is the $L$-th layer's output in the spatial encoder, $\bm{Z}^{L} = \bm{\tilde{Y}}$. In the $\ell$-th layer, if denote by $\dot{Z}^{\ell}_{i,j,f}$ the $(i,j)$ entry of the $f$-th feature map, by $\dot{w}^{\ell}_{a,b,\tilde{f},f}$ the $(a,b)$ entry of the $f$-th kernel for the $\tilde{f}$-th channel of the previous layer, $\tilde{f} \in \{1,\ldots,\dot{K}^{\ell-1} \}$, and by $\dot{w}^{\ell}_{0,f}$ the bias term. The spatial decoder performs the following:
\begin{gather*}
    \dot{Z}^{\ell}_{i,j,f} = g \left( \sum_{\tilde{f}=1}^{\dot{K}^{\ell-1}} \sum_{a=1}^{\dot{r}_{x}^{\ell}} \sum_{b=1}^{\dot{r}_{z}^{\ell}} \dot{Z}^{\ell-1}_{\lfloor \frac{i}{s} \rfloor +a,\lfloor \frac{j}{s} \rfloor+b;\tilde{f}} \cdot \dot{w}^{\ell}_{a,b,\tilde{f},f} + \dot{w}^{\ell}_{0,f} \right).
\end{gather*}
If we denote the weights in the $\ell$-th layer of the spatial decoder by $\dot{\bm{W}}^{\ell} = \left\{ \left(\dot{w}^{\ell}_{a,b,\tilde{f},f}, \dot{w}^{\ell}_{0,f} \right) : a=1, \ldots, \dot{r}_{x}^{\ell},~b=1, \ldots, \dot{r}_{z}^{\ell}, \tilde{f}=1,\ldots, K^{\ell-1},~f=1,\ldots,\dot{K}^{\ell} \right\}$ and $\dot{\bm{Z}}^{\ell} = \{\dot{Z}^{\ell}_{i,j,f} : i = 1, \ldots, \dot{n}^{\ell}_{x},~ j = 1, \ldots, \dot{n}^{\ell}_{z},~f=1,\ldots,\dot{K}^{\ell} \}$, then, the spatial decoder $\mathcal{SD}\left(\cdot \mid \bm{W}_{\text{de}} \right)$ can be written in matrix form
\begin{subequations} \label{mod:spatial_decoder}
\begin{gather}
    \dot{\bm{Z}}^{\ell} = g\left( \dot{\bm{Z}}^{\ell-1} \dot{\ast} \dot{\bm{W}}^{\ell} \right),~ \ell=1,\ldots,L, \\
    \widehat{\bm{Y}} = g \left( \dot{\bm{Z}}^{L} \dot{\ast} \dot{\bm{W}}^{L+1} \right), \label{eq:dec_output_layer}
\end{gather}
\end{subequations}
where $\dot{\ast}$ denotes the strided ($s$) deconvolution operation. Additionally, $\dot{\bm{Z}}^{0}=\bm{\tilde{Y}}$, $\dot{\bm{Z}}^{L}=\bm{Z}^{1}$, with $(\dot{n}_{x}^{L}, \dot{n}_{z}^{L}, \dot{K}^{L}) = \left(n_{x}, n_{z}, K^{1}\right)$. We denote the parameters of the spatial decoder by $\bm{W}_{\text{de}} = \left\{\dot{\bm{W}}^{\ell}: \ell=1,\ldots,L,L+1 \right\}$ and the number of parameters in $\bm{W}_{\text{de}}$ can be derived along the same lines of \eqref{eq:num_pars_conv}. 

\subsection{Temporal Model: Recurrent Neural Network} \label{sec:rnn}

The temporal structure of $\bm{Y}_{t}$ is modeled in reduced space, i.e., we model the sequence $\tilde{\mathcal{Y}}^{+}_{t;a} = \left( \tilde{\bm{Y}}_{t}, \tilde{\bm{Y}}_{t+1}, \ldots, \tilde{\bm{Y}}_{t+a-1} \right)$, as a function of the previous $b$ observations $\tilde{\mathcal{Y}}^{-}_{t;b} = \left( \tilde{\bm{Y}}_{t-b}, \tilde{\bm{Y}}_{t-b+1}, \ldots, \tilde{\bm{Y}}_{t-1} \right)$. The temporal model comprises of two components: a context builder, which learns the temporal structure of the (reduced space) input sequence into a context representation, and a sequence generator, which learns how to generate the future sequence given the context representation. This \textit{sequence-to-sequence} approach fundamentally differs from autoregressive strategies that predict only a single future timestep at each iteration and then feed that prediction back into the model to forecast subsequent steps. The problem is well-documented in highly nonlinear systems and multistep forecasting, where autoregressive methods tend to diverge from physically realistic trajectories \citep{pascanu2013difficulty, seq2seq, bias_multistep2016, guen2019shapetimedistortionloss}. In Figure S12 of the supplementary material we provide evidence with our data. 

The sequence-to-sequence formulation implemented in this work generates the entire future sequence $\tilde{\mathcal{Y}}^{+}_{t;a}$ in a single forward pass, conditioned on the context of $\tilde{\mathcal{Y}}^{-}_{t;b}$. As such, this approach minimizes temporal inconsistencies due to long-range iterative forecasting. We denote the context builder by $\mathcal{T}_{\text{con}}(\cdot \mid \bm{W}_{\text{con}})$ and the sequence generator by  $\mathcal{T}_{\text{gen}}(\cdot \mid \bm{W}_{\text{gen}})$. The modeling approach is similar to that of a language translation model, wherein $\mathcal{T}_{\text{con}}$ learns the ``meaning'' of $\tilde{\mathcal{Y}}^{-}_{t;b}$ and $\mathcal{T}_{\text{gen}}$ ``translates'' it into $\tilde{\mathcal{Y}}^{+}_{t;a}$. The temporal model is therefore the composition of the sequence generator and the context builder, 
\begin{gather*}
    \mathcal{T}\left(\cdot \mid \bm{W}^{\text{time}} \right) =  \mathcal{T}_{\text{gen}}\left( \mathcal{T}_{\text{con}}\left( \cdot \mid \bm{W}_{\text{con}} \right) \mid \bm{W}_{\text{gen}} \right).
\end{gather*}
Both the context builder and sequence generator are a ConvLSTM model, an RNN that is able to capture short- and long-term temporal dependencies \textit{while also} accounting for the spatial structure of the data which, in this case, is a sequence of $b$ three-dimensional matrices, each of dimensions $\tilde{n}_{x} \times \tilde{n}_{z} \times \tilde{d}$ \citep{lstm, convlstm}. At time step $t$, a ConvLSTM model takes as input $\tilde{\bm{Y}}_{t}$ along with the hidden and cell state matrices $\left(\bm{H}_{t-1}, \bm{C}_{t-1} \right)$, which carry temporal and memory information from previous time steps, and outputs their respective value for the next time step, formally,
\begin{subequations} \label{eq:convlstm_con}
\begin{align}
    \bm{I}_{t} &= \sigma\left( \bm{W}^{k}_{y,i} \ast \tilde{\bm{Y}}_{t} + \bm{W}^{k}_{h,i} \ast \bm{H}_{t-1} + \bm{W}^{k}_{c,i} \circ \bm{C}_{t-1} + \bm{W}^{k}_{0,i} \right), \\
    \bm{F}_{t} &= \sigma\left( \bm{W}^{k}_{y,f} \ast \tilde{\bm{Y}}_{t} + \bm{W}^{k}_{h,f} \ast \bm{H}_{t-1} + \bm{W}^{k}_{c,f} \circ \bm{C}_{t-1} + \bm{W}^{k}_{0,f} \right), \\
    \bm{C}_{t} &= \bm{F}_{t} \circ \bm{C}_{t-1} + \bm{I}_{t} \circ f\left( \bm{W}^{k}_{y,c} \ast \tilde{\bm{Y}}_{t} + \bm{W}^{k}_{h,c} \ast \bm{H}_{t-1} + \bm{W}^{k}_{0,c} \right), \\
    \bm{O}_{t} &= \sigma\left( \bm{W}^{k}_{y,o} \ast \tilde{\bm{Y}}_{t} + \bm{W}^{k}_{h,o} \ast \bm{H}_{t-1} + \bm{W}^{k}_{c,o} \circ \bm{C}_{t} + \bm{W}^{k}_{0,o} \right), \\
    \tilde{\bm{H}}_{t} &= \bm{O}_{t} \circ f(\bm{C}_{t}).
\end{align}
\end{subequations}
In \eqref{eq:convlstm_con}, $\sigma(\cdot)$ denotes the sigmoid function, $\circ$ denotes the Hadamard product, and $f(\cdot)$ denotes the hyperbolic tangent function. Additionally, $\bm{I}_{t}$ is the input gate activation matrix at time $t$, $\bm{O}_{t}$ is the output gate activation matrix, and ``$\ast$'' denotes the convolution operation \eqref{eq:convolution}, in this case with stride $s=1$, kernel size $\tilde{r}_{x} \times \tilde{r}_{z}$, and number of kernels $\tilde{K}$. The parameters to learn are $\bm{W}_{k} = \left\{\bm{W}^{k}_{x,m}: x \in \{y,h,c,0 \},~ m \in \{i,f,c,o \} \right\},~ k \in \{\text{con}, \text{gen} \}$.

The temporal structure of the input sequence $\tilde{\mathcal{Y}}^{-}_{t;b} = \left(\tilde{\bm{Y}}_{t-b}, \tilde{\bm{Y}}_{t-b+1}, \ldots, \tilde{\bm{Y}}_{t-1} \right)$ is learned and stored in \textit{context matrices} $\left(\bm{H}_{t-1}, \bm{C}_{t-1} \right)$, which capture the short- and long-term dynamics of $\tilde{\mathcal{Y}}^{-}_{t;b}$, respectively. As such, the context matrices are initialized $\bm{H}_{t-b-1} = \bm{C}_{t-b-1} = \bm{0}$ and $\mathcal{T}_{\text{con}}$ is defined as in \eqref{eq:convlstm_con} and operates recursively over the input sequence, as described by the following recurrent relationship,
\begin{gather}
    \left(\tilde{\bm{Y}}_{t-j+1}, \bm{H}_{t-j}, \bm{C}_{t-j} \right) = \mathcal{T}_{\text{con}} \left(\tilde{\bm{Y}}_{t-j}, \bm{H}_{t-j-1}, \bm{C}_{t-j-1} \mid \bm{W}_{\text{con}} \right),~j = b, b-1, \ldots, 2, 1 \label{eq:tau_con}.
\end{gather}
Importantly, only the final hidden and cell states, $\left(\bm{H}_{t-1}, \bm{C}_{t-1} \right)$, are retained and passed to the sequence generator and the output of each step in the context builder, $\tilde{\bm{Y}}_{t-j+1}$, is discarded. This reflects the context builder's role to summarize the past into context. A visual representation of \eqref{eq:tau_con} is shown in Figure S2 of the supplementary material.

The sequence generator $\mathcal{T}_{\text{gen}}$ learns how to ``translate'' the context matrices $\left(\bm{H}_{t-1}, \bm{C}_{t-1} \right)$ into the (reduced space) output sequence $\tilde{\mathcal{Y}}^{+}_{t;a} = \left(\tilde{\bm{Y}}_{t}, \tilde{\bm{Y}}_{t+1}, \ldots, \tilde{\bm{Y}}_{t+a-1} \right)$. Therefore, the inputs to the sequence generator are $\left( \tilde{\bm{Y}}_{t-1}, \bm{H}_{t-1}, \bm{C}_{t-1} \right)$, the last observation of the input sequence and the context matrices. The sequence generator is also implemented as a ConvLSTM and operates recursively as defined in \eqref{eq:convlstm_con}, formally,
\begin{gather}
    \left(\tilde{\bm{Y}}_{t+j}, \bm{H}_{t+j}, \bm{C}_{t+j} \right) = \mathcal{T}_{\text{gen}} \left(\tilde{\bm{Y}}_{t+j-1}, \bm{H}_{t+j-1}, \bm{C}_{t+j-1} \mid \bm{W}_{\text{gen}} \right),~j = 0, 1, \ldots, a-1 \label{eq:tau_gen}.
\end{gather}
Importantly, unlike in \eqref{eq:tau_con}, the output at each step in \eqref{eq:tau_gen} is fed back into the sequence generator and used as input for the next prediction. This recursive mechanism allows the model to generate multi-step forecasts in a single forward pass while maintaining temporal consistency. A key advantage of this approach is that it avoids the error accumulation and discontinuities inherent in traditional iterative forecasting methods, where entire predicted sequences are repeatedly fed back into the model. Instead, the ConvLSTM's hidden and cell states propagate sequential dependencies internally, enabling stable long-horizon predictions. A visual representation of \eqref{eq:tau_gen} is shown in Figure S2 the supplementary material.

\subsection{Spatiotemporal Model: Convolutional Recurrent Neural Network} \label{sec:crnn}

The final output sequence is obtained by reconstructing the output of the temporal model, $\tilde{\mathcal{Y}}^{+}_{t;a}$, back onto the original dimensions using the spatial decoder \eqref{mod:spatial_decoder}. Given input and output sequences, $\mathcal{Y}^{-}_{t;b} = \left(\bm{Y}_{t-b}, \bm{Y}_{t-b+1}, \ldots, \bm{Y}_{t-1} \right)$ and $\mathcal{Y}^{+}_{t;a} = \left(\bm{Y}_{t}, \bm{Y}_{t+1}, \ldots, \bm{Y}_{t+a-1} \right)$, respectively, our state-space model \eqref{mod:state_space} is then 
\begin{gather*}
    \tilde{\bm{Y}}_{t-i} = \mathcal{SE} \left(\bm{Y}_{t-i} \mid \bm{W}_{\text{en}} \right),\quad i=b,b-1, \ldots, 2, 1, \\
    \left(\cdot, \bm{H}_{t-1}, \bm{C}_{t-1}\right) =  \mathcal{T}_{\text{con}}\left( \tilde{\mathcal{Y}}^{-}_{t;b}, \bm{H}_{t-b-1}, \bm{C}_{t-b-1} \ \middle\vert \ \bm{W}_{\text{con}} \right), \\
    \left( \bm{\tilde{Y}}_{t+j}, \bm{H}_{t+j}, \bm{C}_{t+j}\right) = \mathcal{T}_{\text{gen}} \left( \bm{\tilde{Y}}_{t+j-1}, \bm{H}_{t+j-1}, \bm{C}_{t+j-1}  \ \middle\vert \ \bm{W}_{\text{gen}} \right), \quad j=0,1, \ldots, a-1 \\
    \widehat{\mathcal{Y}}^{+}_{t;a} = \mathcal{SD} \left(\tilde{\mathcal{Y}}^{+}_{t;a} \mid \bm{W}_{\text{de}} \right).
\end{gather*}
We hereto refer to the spatiotemporal model as $\mathcal{ST}(\cdot \mid \bm{W})$, $\bm{W}=(\bm{W}^{\text{space}}, \bm{W}^{\text{time}})$.  As a neural network with a convolutional (spatial) and a recurrent (temporal) component, this model is a CRNN, which may be summarized as
\begin{gather*}
    \overbrace{\mathcal{Y}^{-}_{t;b} \xrightarrow{\shortstack{\text{Spatial} \\ \text{Encoder}}} \underbrace{\tilde{\mathcal{Y}}^{-}_{t;b}  \xrightarrow{\shortstack{\text{Context} \\ \text{Builder}}} \shortstack{\text{Context} \\ $\left(\bm{H}_{t-1}, \bm{C}_{t-1}\right)$} \xrightarrow{\shortstack{\text{Sequence} \\ \text{Generator}}} \tilde{\mathcal{Y}}^{+}_{t;a}}_{\text{Temporal Model $\mathcal{T}(\cdot)$}}  \xrightarrow{\shortstack{\text{Spatial} \\ \text{Decoder}}} \mathcal{Y}^{+}_{t;a}}^{ \text{Spatiotemporal Model $\mathcal{ST(\cdot)}$}}.
\end{gather*}

\section{Inference} \label{sec:inference}

Inference is performed in two stages, wherein we first estimate the spatial parameters $\bm{W}^{\text{space}}$ and then the temporal parameters $\bm{W}^{\text{time}}$ conditional on $\widehat{\bm{W}}^{\text{space}}$. We adopt this strategy since the CAE and RNN serve two distinct (and purposely decoupled) functions. Indeed, the CAE can focus on reconstructing spatial structure, while the RNN can focus on modeling the temporal evolution in the reduced space. While joint training may offer the benefit of task-specific representations optimized end-to-end for forecasting performance, we found in practice that decoupled training yields accurate reconstructions and forecasts, while being computationally tractable. Section \ref{sec:train_test} describes the training and testing set, Section \ref{sec:cae_training} discusses the inferential process for the CAE, and Section \ref{sec:pi_training} presents the training strategy for the temporal portion of the CRNN conditional on the estimated CAE. Inference is performed in \texttt{Python 3.12.0} and \texttt{Tensorflow 2.13}. The model was trained using 8 CPU cores for general processing and a dedicated NVIDIA A10 GPU for accelerated computation.

\subsection{Training and Testing set} \label{sec:train_test}
The first $n=2{,}000$ (of ouf $T=2{,}500$) observations were used for model training, while the remaining $T-n=500$ were used for testing. For the CAE, the training and testing set are $\mathcal{D}_{\text{train}}^{\text{CAE}} = \left\{ \bm{Y}_{1}, \bm{Y}_{2}, \ldots, \bm{Y}_{n}  \right\}$ and $\mathcal{D}_{\text{test}}^{\text{CAE}} = \left\{ \bm{Y}_{n+1}, \bm{Y}_{n+2}, \ldots, \bm{Y}_{T}  \right\}$, respectively. For the RNN, since it is a sequence-to-sequence model, its input and output is a sequence of observations, so its training set is constructed by dividing $\mathcal{D}_{\text{train}}^{\text{CAE}}$ into $\tilde{n}$ input and output sequences, 
$\mathcal{D}_{\text{train}}^{\text{RNN}} = \left\{\left( \mathcal{Y}^{-}_{t;b}, \mathcal{Y}^{+}_{t;a} \right) : t = b, b+o, b+2o, \ldots, n-a \right\} \subset \mathcal{D}_{\text{train}}^{\text{CAE}}$ and $    \mathcal{D}_{\text{test}}^{\text{RNN}} = \left\{\left( \mathcal{Y}^{-}_{t+n;b}, \mathcal{Y}^{+}_{t+n;a} \right) : t = b, b+o, b+2o, \ldots, n-a \right\} \subset \mathcal{D}_{\text{test}}^{\text{CAE}}$, where $o$ is the offset between sequences. The number of training sequences is therefore $\tilde{n} = \frac{n - a - b}{o} + 1$, and we set $o=8$, the smallest $o$ that could fit in memory. Note that all the sequences in $\mathcal{D}_{\text{train}}^{\text{RNN}}$ are subsets of $\mathcal{D}_{\text{train}}^{\text{CAE}}$. Having first trained the CAE, we can represent each input sequence in $\mathcal{D}_{\text{train}}^{\text{RNN}}$ in the reduced space and obtain $\tilde{\mathcal{Y}}^{-}_{t;b} = \left( \tilde{\bm{Y}}_{t-b}, \tilde{\bm{Y}}_{t-b+1},\ldots, \tilde{\bm{Y}}_{t-1} \right)$, where $t = b, b+o, b+2o, \ldots, n-a$, so that the training and testing sets for the temporal model are $\tilde{\mathcal{D}}_{\text{train}}^{\text{RNN}} = \left\{\left( \tilde{\mathcal{Y}}^{-}_{t;b}, \tilde{\mathcal{Y}}^{+}_{t;a} \right) : t = b, b+o, b+2o, \ldots, n-a \right\}$ and $   \tilde{\mathcal{D}}_{\text{test}}^{\text{RNN}} = \left\{\left( \tilde{\mathcal{Y}}^{-}_{t+n;b}, \tilde{\mathcal{Y}}^{+}_{t+n;a} \right) : t = b, b+o, b+2o, \ldots, n-a \right\}$.

\subsection{Convolutional Autoencoder Inference} \label{sec:cae_training}

In CAE training, each observation is treated independently, since the objective is to compress any one observation into a space of lower dimensions. We find $\widehat{\bm{W}}^{\text{space}}$ by minimizing the mean squared error with respect to the data,
\begin{gather}
    \widehat{\bm{W}}^{\text{space}} = \argmin_{\bm{W}^{\text{space}}} \text{MSE}_{\text{space}}\left(\bm{W}^{\text{space}}\right) = \argmin_{\bm{W}^{\text{space}}} \left\{ \sum_{i} \lambda^{\text{space}}_{i}\mathcal{L}^{\text{space}}_{i} \left(\bm{W}^{\text{space}}\right) \right\}, \label{eq:loss_function_cae}
\end{gather}
where $\bm{\lambda}^{\text{space}}  = \left\{ \lambda^{\text{space}}_{i} : i=u,w,p,\theta \right\}$ are weights balancing the four terms' respective contributions. The terms  $\mathcal{L}^{\text{space}}_{i}$ are the mean square error with respect to each of the four variables. Denote by $\bm{Y}_{t}\left(\bm{W}^{\text{space}}\right) = \left(u_{t}\left(\bm{W}^{\text{space}}\right), w_{t}\left(\bm{W}^{\text{space}}\right), p_{t}\left(\bm{W}^{\text{space}}\right), \theta_{t}\left(\bm{W}^{\text{space}}\right) \right)$ the reconstruction at $t$ as a function of $\bm{W}^{\text{space}}=\left(\bm{W}_{\text{en}}, \bm{W}_{\text{de}} \right)$, where $\bm{Y}_{t}\left(\bm{W}^{\text{space}}\right) = \mathcal{SD}\left( \mathcal{SE} \left( \bm{Y}_{t} \mid \bm{W}_{\text{en}} \right) \mid \bm{W}_{\text{de}} \right)$. The loss term for $\mathcal{L}^{\text{space}}_{u}$ is then
\begin{gather*}
     \mathcal{L}^{\text{space}}_{u} \left(\bm{W}^{\text{space}}\right) = \frac{1}{n}\sum_{t=1}^{n} \left[u_{t} - u_{t}\left(\bm{W}^{\text{space}}\right) \right]^{2},
\end{gather*}
and $\mathcal{L}^{\text{space}}_{w}$, $\mathcal{L}^{\text{space}}_{p}$, and $\mathcal{L}^{\text{space}}_{\theta}$ are similar. 

The $\bm{\lambda}^{\text{space}}$ hyperparameters  balance the model training and are not chosen via cross-validation. They are instead implemented via a dynamic method that automatically balances the four terms in \eqref{eq:loss_function_cae} and lessens the risk that the optimization ends up at a local minimum \citep{wang_experts_2023}. This ensures that the model does not disproportionately focus on optimizing one output at the expense of others, thereby promoting more balanced training. The approach follows established practices in multi-task learning, where adaptive loss weighting has been shown to improve convergence and final model performance \citep{wang_understanding_2020}. Technical details are available in Section S.4 of the supplementary material.

\subsection{Recurrent Neural Network Physics-Informed Inference} \label{sec:pi_training}

We estimate the temporal parameters $\bm{W}^{\text{time}}$ \textit{conditional on} $\widehat{\bm{W}}^{\text{space}}$, the estimated CAE parameters. We also incorporate the RBC governing equations in the inferential process, making it a \textit{physics-informed} model, so henceforth we refer to it as ``PI-CRNN''.  Formally, we find $\widehat{\bm{W}}^{\text{time}}$ by minimizing a penalized target function:
\begin{align}
    \widehat{\bm{W}}^{\text{time}} &= \argmin_{\bm{W}^{\text{time}}} \mathcal{L}\left(\bm{W}^{\text{time}} \right) \nonumber \\
    &= \argmin_{\bm{W}^{\text{time}}} \left\{ \sum_{i} \lambda^{\text{Data}}_{i}\mathcal{L}^{\text{Data}}_{i}\left( \bm{W}^{\text{time}} \right) + \sum_{j} \lambda_{j}^{\text{PDE}} \mathcal{L}^{\text{PDE}}_{j}(\bm{W}^{\text{time}}) \right\}, \label{eq:loss_function}
\end{align}
where $i \in \{u, w, p, \theta\}$ and $j \in \{\text{Mass},~ \text{Mom-}u,~ \text{Mom-}w,~ \text{Energy}\}$. The term $\mathcal{L}^{\text{Data}}_{i}$ represents the loss term with respect to the observed data, e.g.,
\begin{gather*}
    \mathcal{L}^{\text{Data}}_{u} \left(\bm{W}^{\text{time}}\right) = \frac{1}{a \cdot \tilde{n}}\sum_{t=1}^{\tilde{n}} \sum_{r=1}^{a} \left(u_{t;r} - u_{t;r}\left(\bm{W}^{\text{time}} \right) \right)^{2},
\end{gather*}
where $u_{t;r}$ is the $r$-th element of the output sequence (out of $\tilde{n}$ training sequences) starting at $t$ (in this case, for the horizontal component of velocity) and $u_{t;r}\left(\bm{W}^{\text{time}} \right)$ its prediction as a function of $\bm{W}^{\text{time}}$. The terms $\mathcal{L}^{\text{PDE}}_{j}$ instead represent the loss terms with respect to the governing PDEs, e.g.,
\begin{align*}
    \mathcal{L}^{\text{PDE}}_{\text{Mass}} \left(\bm{W}\right) &= \frac{1}{a \cdot \tilde{n}}\sum_{t=1}^{\tilde{n}} \sum_{r=1}^{a} \left(\frac{\partial u_{t;r}\left(\bm{W}^{\text{time}} \right)}{\partial x} + \frac{\partial w_{t;r}\left(\bm{W} ^{\text{time}} \right)}{\partial z} \right)^{2},
\end{align*}
and the vector $\bm{\lambda}^{\text{time}} = \left( \lambda^{\text{Data}}_{i}, \lambda^{\text{PDE}}_{j}\right),~i \in \{u, w, p, \theta\},~j \in \{\text{Mass},~ \text{Mom-}u,~ \text{Mom-}w,~ \text{Energy}\}$, contains weights balancing the eight terms' respective contributions to the objective function $\mathcal{L}(\bm{W}^{\text{time}})$. The derivatives in the PDE penalty are computed using a central difference method, detailed in the supplementary material. The dynamic loss balancing for $\bm{\lambda}^{\text{time}}$ in \eqref{eq:loss_function} is similar to \eqref{eq:loss_function_cae} and detailed in Section S.4 of the supplementary material.

We hereto denote the forecast at time $t$ by $\widehat{\bm{Y}}_{t}=\left(\widehat{u}_{t}, \widehat{w}_{t}, \widehat{p}_{t}, \widehat{\theta}_{t}\right)$ to simplify the notation. After training the PI-CRNN model, we predict new realizations of RBC using input sequences in the testing set $\mathcal{D}^{\text{RNN}}_{\text{test}}$ as an input and forecasting over a time horizon of $R$ time steps. The procedure to generate forecasts is outlined in Algorithm S3 of the supplementary material.

\section{Results} \label{sec:results}

We apply the PI-CRNN model described in Section \ref{sec:methods} to the dataset in Section \ref{sec:data} and address questions Q1 and Q2 in the same Section. We fist focus on Q1 and compare our model's spatial performance against competing statistical approaches and then the temporal one. In regard to the dimension-reduction (and subsequent reconstruction) of the data, we compare the CAE with principal component analysis (PCA; \citet{jolliffe2002principal}), independent component analysis (ICA; \citet{comon1994independent}), fixed rank Kriging (FRK; \citet{Cressie2008}), and proper orthogonal decomposition (POD; \citet{POD}). Details on PCA, ICA, FRK and POD are available in Section S.6.1 of the supplementary material. Then, we address Q2 by comparing the forecast obtained using our PI-CRNN with a non-physics-informed CRNN, an echo state network (ESN), and an autoregressive integrated moving average model (ARIMA). We also assess the ability of PI-CRNN to forecast well beyond the training forecast horizon.

We assess our model's performance according to its two stated objectives: efficient reduced space representation and long-range forecasting of RBC. For spatial reconstruction, we analyze the CAE's ability to represent high-dimensional spatial data into a reduced space and subsequently reconstruct it back onto the original dimensions. The metrics we employ are the mean squared error and the structural similarity index measure (SSIM, \cite{ssim}) between the reconstruction and ground truth over the testing set. The SSIM was originally conceived to quantify the similarity between images by comparing visual fidelity with respect to the original data. For long-range forecasting, we assess adherence to the governing PDE as well as physical properties of RBC that should be preserved. We focus on 1) the Nusselt number: a measure of how efficiently heat is transferred through a fluid by convection as compared to conduction; 2) the probability density functions (PDFs) of the velocity component and temperature; 3) dissipation rate of thermal fluctuations: loss of temperature variance due to heat diffusion and conduction in the fluid, respectively \citep{emran2012conditional}. The forecast we generate is of 7 turnovers $\tau_{c}$, or 37.8 seconds, i.e., $R=378$, which is considerably longer (hence more challenging) than previous attempts to forecast similar turbulent flows \citep{mohan2019compressed, mo2019deep, tang2020deep}. A thorough explanation on the reliability of the forecast as a function of time is available in Section S.6 of the supplementary material. Technical details on each metric of assessment are also available in Section S.6 of the supplementary material. Due to RBC's extreme sensitivity to initial conditions, even minute discrepancies, e.g., numerical perturbations, can cause forecasts to diverge rapidly over time. This inherent unpredictability makes uncertainty quantification essential. Rather than relying on point predictions alone, it is crucial to capture the range of plausible future states the system may evolve into. Therefore, in our work we also aim at assessing the quality and reliability of the predictive uncertainty provided by our PI-CRNN. 

The following sub-sections are organized as follows. In Section \ref{sec:results_spatial} we analyze the spatial portion of our model, while in Section \ref{sec:results_forecast} we assess the PI-CRNN forecast and its adherence to the underlying physics. Finally, in Section \ref{sec:results_twin} we discuss the long-range forecasting ability of PI-CRNN against existing numerical methods and in Section \ref{sec:limitations} we discuss limitations.

\subsection{Spatial Analysis (Dimension Reduction)} \label{sec:results_spatial}

The reconstruction performance of our CAE presents an alternative to established reduced order modeling techniques in fluid mechanics, which seek to ease the computational burden in numerical simulations without sacrificing key physical properties of the system of interest. The complexity of a dynamical system such as RBC is often reduced with basic decomposition methods. Such approaches predicate the decomposition of a random vector field $\bm{Y}(x, t)$ into a finite sum, e.g., $\bm{Y}(x, t) = \sum_{i=1}^{M}  a_{i}(t) \phi_{i}(x)$, where $\phi_{i}(x)$ are the spatial basis functions and $a_{i}(t)$ are temporal coefficients, such that the the approximation improves as $M \rightarrow \infty$. Common choices of $\phi_{i}$ include Fourier series, Legendre polynomials, and Chebyshev polynomials \citep{chatterjee2000introduction}. Our CAE model accomplishes the same objective in nonlinear fashion. This approach makes the physical interpretation of $\bm{\tilde{Y}}_{t}$ more challenging, but provides increased flexibility granted by the nonlinear activation function $g$ in the convolution step \eqref{mod:spatial_encoder} and deconvolution step \eqref{mod:spatial_decoder}. 

In Table \ref{tab:spatial_results} we report the $\text{MSE}_{\text{space}}$ and SSIM, along with the interquartile range (IQR) for the CAE, PCA, ICA, FRK and POD. For PCA and ICA we chose the highest possible number of principal components $n$ and used them to reconstruct the $T-n$ validation observations. For a fair comparison with CAE, in the case of FRK we interpolated in space (independently for each realization of RBC) using every fourth grid point, since the CAE reduces the original dimensions by a factor of 16. For POD, we chose the maximum possible number of modes. Details on PCA, ICA, FRK and POD are available in Section S.6.1 of the supplementary material. In Figure \ref{fig:cae_results} we show one validation observation $\bm{Y}_{t}$ and our reconstructed estimate $\widehat{\bm{Y}}_{t}$.

The results clearly highlight that other approaches are suboptimal in capturing the nonlinear nature of this data, as the CAE yields a validation MSE of $\mathcal{O}(10^{-6})$, FRK of $\mathcal{O}(10^{-4})$, and PCA and ICA of $\mathcal{O}(10^{-3})$. FRK delivers the second best reconstruction MSE, which is over 10 times worse than that of the CAE, most likely driven by the spatial nonstationarity of RBC in the vertical direction. The suboptimal performance of PCA and ICA, both linear methods, can be attributed to the highly nonlinear nature of the data, though PCA has been shown to be effective in RBC simulations with lower Rayleigh numbers, i.e., less turbulence \citep{data_driven_RBC}. The results in terms of SSIM follow the same pattern of $\text{MSE}_{\text{space}}$. The relatively high SSIM for FRK, PCA, and ICA can be attributed to a fundamental difference in how it evaluates reconstruction quality compared to $\text{MSE}_{\text{space}}$. SSIM assesses perceptual similarity, making it more sensitive to the preservation of global patterns and spatial coherence rather than small-scale errors. Our result is significant as it reveals how effective the CAE is in extracting the spatial features of RBC \textit{while at the same time} reducing the dimensions of the data by a factor of 16. As such, reconstructions whose MSE is $\mathcal{O}(10^{-6})$ imply that the CAE retains the key features in the reduced space representation, as is apparent from Figure \ref{fig:cae_results}. 

\renewcommand{\arraystretch}{1.25}
\begin{table}[!ht]
    \centering
    \begin{tabular}{|l|c c|}
        \hline
        Dimension-Reduction &  & \\ 
        Method & $\text{MSE}_{\text{space}}$ (IQR) $\times 10^{-3}$ & SSIM (IQR) \\ \hline \hline
        \textbf{CAE} & $\bm{6.62 \times 10^{-3}~\left(8.70 \times 10^{-2}\right)}$ & $\bm{0.99~ \left( 2.36 \times 10^{-4} \right)}$ \\
        FRK & $1.65 \times 10^{-1} ~\left(1.86 \times 10^{-2}\right)$ & 0.98~$\left(2.86 \times 10^{-3} \right)$ \\
        POD & $1.85 ~\left(1.32 \right)$ & 0.97~$\left( 2.81 \times 10^{-2} \right)$\\
        PCA & $2.34 ~\left(1.52 \right)$ & 0.96~$\left( 3.39 \times 10^{-2} \right)$\\
        ICA & $4.17 ~\left(3.51 \right)$ & 0.96~$\left( 4.81 \times 10^{-2} \right)$\\  \hline
    \end{tabular}
    \caption{Comparison of dimension-reduction methods in terms of MSE and SSIM on the validation set. In parenthesis we report the IQR across the spatial domain.}
    \label{tab:spatial_results}
\end{table}

Figure \ref{fig:cae_results} illustrates the performance of the CAE on a representative example from the validation set at $t=2{,}125$. The top row displays the original data fields for the four variables ($u$, $w$, $p$, $\theta$, from left to right), while the bottom row shows the corresponding reconstructions obtained from the CAE after reducing the data dimensionality by 93.75\%. Despite this significant compression, the CAE successfully preserves the main spatial structures and fine-scale patterns. In particular, the reconstruction captures the large-scale flow structures, sharp gradients, and coherent features present in the original data. This provides strong visual evidence for the model’s ability to faithfully represent the essential dynamics of RBC. The high fidelity of the reconstructions further supports the quantitative results, as reflected by the low $\text{MSE}_{\text{space}}$ and the high SSIM from Table \ref{tab:spatial_results}.

\begin{figure}[!ht]
    \centering
    \textbf{Data}
    \includegraphics[width=\linewidth]{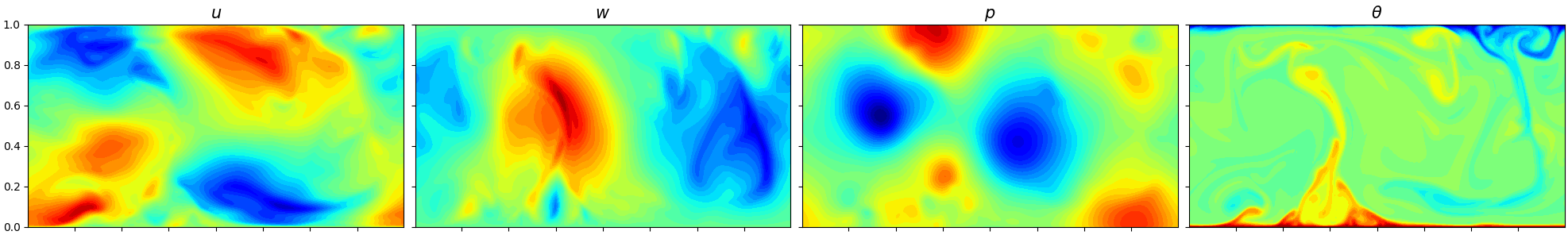} \\
    \textbf{Reconstruction}
    \includegraphics[width=\linewidth]{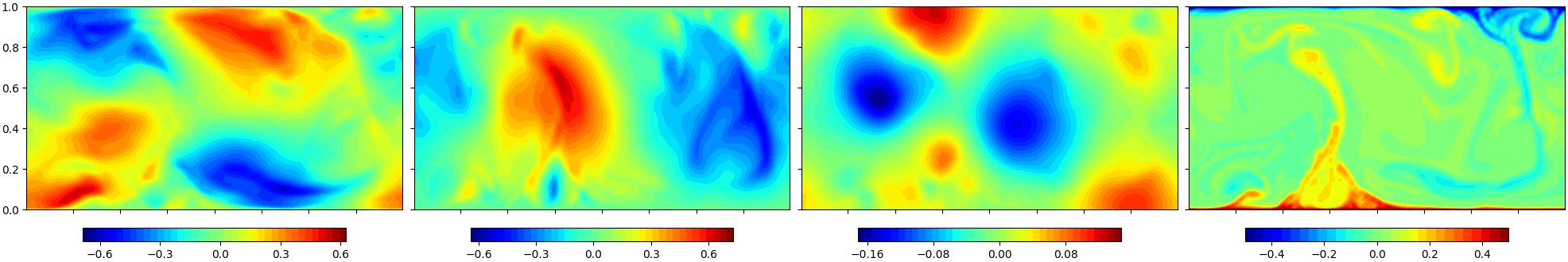} 
    \caption{CAE results for one observation in the validation set, $t=2{,}125$, showing the data (top row) and the reconstruction (bottom row) after reducing the dimensions of the data by 93.75\%. The $\text{MSE}_{\text{space}}$ and SSIM for this observation are $6.12 \times 10^{-6}$ and $0.99$, respectively.}
    \label{fig:cae_results}
\end{figure}

\subsection{Temporal Analysis} \label{sec:results_forecast}

We compare our forecast of $\tau_c=7$ turnover times with the ESN, CRNN, and ARIMA. The CRNN model is identical to PI-CRNN, but is not informed with the governing PDE, so the loss function we minimize is similar to \eqref{eq:loss_function} but without the ``PDE'' terms. The ESN is a type of recurrent neural network with a similar structure as a LSTM network, but with considerably less computational demand dictated by the sparseness of its parameter matrices \citep{desn}. The main advantage of ESNs is the absence of backpropagation through time (which is the computational bottleneck of LSTM and ConvLSTM), since only the output weights are trained. Since we perform physics-informed inference, we hereby denote it by ``PI-ESN''. The forecast is generated by recursively applying the ESN $R$ times. Lastly, the ARIMA model is one of the most popular approaches in time series analysis thanks to its simplicity and ease of implementation. Similar to the ESN, we generate a forecast of length $R$ by recursively applying ARIMA. Technical details on PI-ESN, CRNN, and ARIMA are available in Section S.6.2 of the supplementary material.

The results in Table \ref{tab:all_results} report the comparison on forecast performance against the aforementioned methods, showing (I) $\text{MSE}_{\text{RBC}}$, the mean squared error with respect to the governing PDE, (II) the mean squared error with respect to the estimate of the Nusselt number (Nu), (III) the Kullback-Leibler (KL) divergence between the true and estimated PDFs of $\widehat{u}_{t}$, $\widehat{w}_{t}$, and $\widehat{\theta}_{t}$, and (IV) the mean squared error with respect to the dissipation rate of thermal fluctuations $\varepsilon^{\theta}$. In each case, the IQR (computed across forecasts with $E=50$ different starting sequences) is reported in parenthesis.

Our PI-CRNN yields the forecast that most closely resembles a time series of RBC that is physically-consistent, with a $\text{MSE}_{\text{RBC}}$ of $0.01 ~\left(3.12 \times 10^{-3}\right)$, which is approximately 97\% smaller than the second best competing approach, CRNN. While PI-ESN attains the second-lowest $\text{MSE}_{\text{RBC}}$, its PDFs deviate considerably from ground truth, indicating systematic biases in its representation of coherent structures. CRNN ranks third in terms of $\text{MSE}_{\text{RBC}}$, yet it better preserves physical realism, producing more accurate distributions and rare-event statistics, making it a stronger overall model than PI-ESN. Similarly, the key physical attributes of the forecast, represented here by the Nusselt number, probability density functions of the velocity and temperature, and the dissipation rate of thermal fluctuations, are best preserved in PI-CRNN. The error with respect to a perfectly physical forecast, captured by $\text{MSE}_{\text{RBC}}$, increases over time, as small-scale discrepancies propagate through time. Nonetheless, overall statistical adherence is mostly preserved, as indicated by a small mean squared error with respect to the true Nusselt number. The dissipation is a much stricter benchmark, since it depends on accurately capturing small-scale gradients, and it presents a more difficult challenge for PI-CRNN to capture. The CRNN model yields the second best results, with a $\text{MSE}_{\text{RBC}}$ of 0.74 $\left(9.39 \times 10^{-2}\right)$, primarily thanks to the same sequence-to-sequence approach implemented in PI-CRNN, though performance degrades quickly due to the absence of direct embedding of the physics in the inferential process. Indeed, Nu and the PDFs, both of which require a long (physically-consistent) forecast in order to obtain reasonable estimates, are markedly worse. The PI-ESN model yields results that are comparable to CRNN, but it suboptimally captures the distributional characteristics of RBC. This is likely due to fact that ESNs' reservoir weights are in fact \textit{random} and untrained, limiting their ability to control information flow over long horizons. The worst performing model is (expectedly) ARIMA, which is unable to capture the highly chaotic, nonlinear nature of RBC. In Section S.6 of the supplementary material we report a comprehensive comparison of these three physical metrics. Figures S5-S7 show the Nusselt number, PDFs, and dissipation, respectively. Figure S11 of the supplementary material show additional visual evidence for the superior performance of PI-CRNN. 

\renewcommand{\arraystretch}{1.25}
\begin{table}[!ht]
    \centering
    \resizebox{\textwidth}{!}{
    \begin{tabular}{|l|c c c c|}
        \hline 
        & (I) & (II) & (III) & (IV) \\ 
        & $\text{MSE}_{\text{RBC}}$ & $\Delta$Nu & KL & Dissipation \\ \hline \hline
        \textbf{PI-CRNN} & $\bm{0.01~\left(3.12 \times 10^{-3} \right)}$ & $\bm{22.11~(40.45)}$ & $\bm{5.05~(2.91)}$ & $\bm{3.55 \times 10^{-3}~\left(1.08 \times 10^{-3} \right)}$ \\
        CRNN & 0.74 $\left(9.39 \times 10^{-2}\right)$ & 2,544.28 (1,277.46) & 10.86 (8.92) &  $4.89 \times 10^{-3}~\left(1.99 \times 10^{-3} \right)$ \\
        PI-ESN & 0.08 $\left(5.50 \times 10^{-2}\right)$& 31.14 (406.02) & 2613.94 (375.12) & $1.02 \times 10^{-2}~\left(5.81 \times 10^{-3} \right)$ \\
        ARIMA & 7.34 $\left(1.39 \times 10^{-1}\right)$& 710.75 (7,214.42) & 3755.95 (823.93) & $2.94 \times 10^{-1}~\left(7.92 \times 10^{-3} \right)$ \\ \hline
    \end{tabular}}
    \caption{Comparison of temporal methods in terms of physical metrics: (I) is the mean squared error with respect to the governing equations, (II) is the mean squared error with respect to the true Nusselt number, (III) is the Kullback-Leibler divergence between the forecast's PDFs and those of the data, and (IV) is the mean squared error with respect to the dissipation rate of thermal fluctuations.}
    \label{tab:all_results}
\end{table}

\begin{figure}[!ht]
    \centering
    \includegraphics[width=1\linewidth]{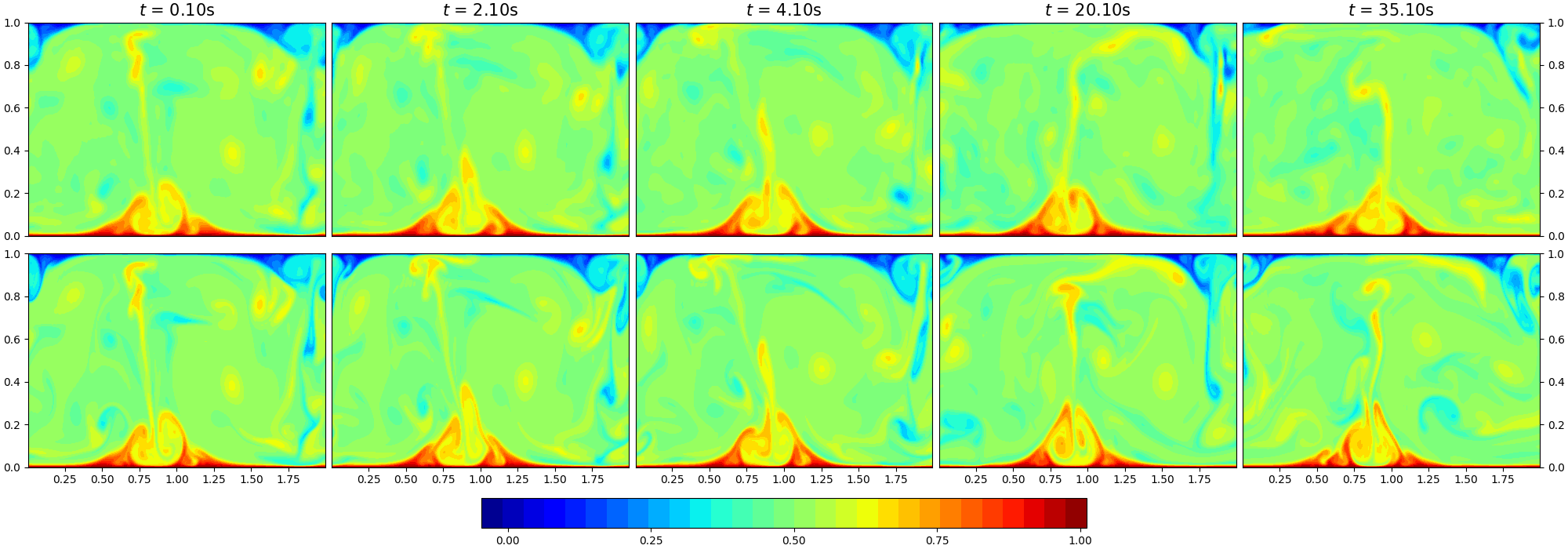}
    \caption{Snapshots of the temperature forecast at selected times. The first three panels correspond to predictions within the first turnover time, highlighting the model’s short-term accuracy, while the last two panels show long-term predictions, where errors have accumulated and deviations from the DNS trajectory become more apparent. An animation is available in the online code associated with the manuscript.}
    \label{fig:forecast}
\end{figure}

Figure \ref{fig:forecast} presents a sequence of snapshots from the temperature forecast $\widehat{\theta}_{t}$ at selected times. The first three snapshots correspond to predictions within the first turnover time, highlighting the model’s short-term accuracy, while the last two panels show long-term predictions, where errors with respect to the DNS data have accumulated. PI-CRNN is able to capture the key dynamical features of plume formation, growth, and lateral motion over time. Although the mean squared error with respect to the true data increases gradually with forecast lead time, indicating some accumulation of small deviations, the residuals of the governing equations remain relatively constant, especially over the first two turnover times (Figures S8-S9 of the supplementary material). This suggests that while the forecast slowly diverges from the realized RBC for $t>n$, the model consistently preserves the underlying physical constraints. An animation showing the full temporal evolution is available in the online code associated with the manuscript.

\subsection{PI-CRNN as a Surrogate} \label{sec:results_twin}

We now assess the PI-CRNN efficacy as a statistical surrogate to produce physically consistent, long-range forecasts. As summarized in Table \ref{tab:stoch_simulator}, the model demonstrates strong generative capabilities, as we are able to produce $\tau_c=7$ turnover times worth of physical realizations of future RBC states in under 20 seconds. By comparison, the runtime required for a direct numerical simulation (DNS, the numerical approach used to generate the training data) is \textit{more than five times larger}. In DNS, partial differential equations are integrated in time by evaluating spatial and temporal derivatives at each time step, meaning that each new realization must be computed sequentially from previous ones. By contrast, our PI-CRNN decouples the learning of flow dynamics from explicit numerical integration by embedding the structure of the governing RBC equations into the parameter estimates through the physics-informed training process outlined in Section \ref{sec:pi_training}. PI-CRNN does not require expensive physical solvers, enabling its use as an ``off-the-shelf'' surrogate capable of producing statistically faithful forecasts of RBC. 

\renewcommand{\arraystretch}{1.25}
\begin{table}[!ht]
    \centering
    \begin{tabular}{|l|c c|}
        \hline
        \diagbox{Method}{Cost} & Per Step & Entire Simulation \\  \hline \hline
        Direct simulation & 0.20 s & 75.6 s  \\
        PI-CRNN & \textbf{0.04} s & \textbf{14.5 s} \\
        \hline
    \end{tabular}
    \caption{Comparison of turbulence generation methods in terms of computational demand. We show the per-step computational time and the total for $R=378$ time steps.}
    \label{tab:stoch_simulator}
\end{table}

It must be noted that PI-CRNN does not yield results with \textit{exact} adherence to the governing PDE at every time step. Small deviations of the surrogate forecast from the original simulation accumulate over time, but they are mitigated by penalizing against non-physical solutions, leading to the retention of key statistical properties of RBC, e.g., the Nusselt number, as discussed in Section \ref{sec:results_forecast}. In fact, even with seemingly identical initial conditions, ensemble members of a DNS are expected to diverge from one another in an instantaneous sense (though they will still adhere to the governing equations). This divergence, which arises from the chaotic nature of fluid flows, helps to motivate our understanding of the key statistical features of RBC, rather than instantaneous features associated with individual ensemble members. By contrast, a purely data-driven approach such as CRNN (defined in Section \ref{sec:results_forecast}), does not preserve any physical properties of RBC and deviations from the governing PDE grow significantly over time. Additional details and results for CRNN are provided in the supplementary material, where we show visual evidence for the superior physical performance of PI-CRNN over CRNN.  

\subsection{Limitations} \label{sec:limitations}

PI-CRNN presents a few structural shortcomings that could be mitigated in future research. First and foremost, maintaining physical consistency over longer horizons remains a key limitation: small violations of the governing equations in the learned context dynamics can accumulate, leading to drift from physically plausible states in the strongly chaotic regime. Extending PI-CRNN's reliability horizon would be beneficial for applications in which the relevant physics unfolds over longer timescales, e.g., particle dispersion. The approach is also computationally expensive to train, the result of a large parameter space and the physics-based loss terms that require evaluating spatial derivatives on fine grids, resulting in substantial memory and time costs. Finally, the model has been trained and validated for only a single set of physical parameters (one combination of Rayleigh and Prandtl numbers), which restricts its ability to generalize across different convection regimes. Incorporating multiple regimes would enable systematic analysis of particle transport, dispersion, and accumulation under varying flow conditions, allowing robust comparison of transport statistics that are relevant to real-world geophysical and industrial convection systems.

\section{Conclusion} \label{sec:conclusion}

In this work we present a spatiotemporal, physics-informed convolutional recurrent neural network model, PI-CRNN, which acts as a surrogate to generate long-range forecasts of RBC convection---a buoyancy-driven flow that manifests between surfaces exhibiting a temperature gradient. Simulating RBC convection by solving the underlying PDEs is time consuming and presents scalability challenges. Our proposed model is considerably faster and offers a scalable and efficient alternative for long-range forecasting of RBC.

Our approach compresses the spatial representation of RBC using a convolutional autoencoder, and models its temporal evolution using a recurrent neural network within a sequence-to-sequence framework inspired by natural language processing. This approach fundamentally differs from classical strategies that predict a single future timestep and iteratively feed the extended sequence back into the temporal model. In contrast, the sequence-to-sequence formulation generates the entire future sequence in a single forward pass, conditioned on the context of the input sequence, thereby minimizing the risk of compounding errors. The approach is also physics-informed, as model parameters are estimated via penalized optimization with respect to the governing equations of RBC.

PI-CRNN’s features, namely its convolutional structure, recurrent memory, and physics-based loss, are generalizable to other nonlinear systems in thermodynamics, climate, and geophysical flows. While the current implementation focuses on fixed physical parameters, our framework is readily extensible to different parameter regimes. This is particularly important for capturing the wide range of turbulent flow behaviors that arise under varying levels of convection. Mesoscale processes emerging from nonlinear interactions between small-scale dynamics and larger structures such as thermal plumes could also benefit from such an approach.

Future work will generalize PI-CRNN by incorporating additional spatiotemporally-aware neural network architectures suitable for high-dimensional data, such as vision transformers \citep{dosovitskiy2021an}. Unlike convolutional neural networks that rely on local spatial relationships, vision transformers use self-attention mechanisms to capture global spatial dependencies across the entire domain. This capability allows them to efficiently capture long-range interactions between distant regions, a key feature in thermodynamic systems, where large-scale structures such as plumes and vortices exert system-wide influence.

\section*{Code and Data Availability}

The code for this work can be found in the following GitHub repository: \url{github.com/Env-an-Stat-group/26.Menicali.JASA}. The numerical solver for the RBC used in this work can be freely downloaded from \url{git.uwaterloo.ca/SPINS/SPINS_main}.

\if1\anon
{
\section*{Acknowledgments}
Stefano Castruccio was partially funded by NSF grant OAC-2347239 and the Notre Dame Provost Strategic Framework Grants. David Richter was partially funded by NSF grant AGS-2227012. Andrew Grace was partially funded by the Engineering Research Council of Canada  fund number PDF - 587364 - 2024.
}\fi

\bibliography{refs.bib}

\clearpage

\renewcommand\thefigure{S\arabic{figure}}
\renewcommand{\thepage}{S\arabic{page}}
\renewcommand{\thesection}{S.\arabic{section}}
\renewcommand{\thetable}{S\arabic{table}}
\renewcommand{\theequation}{S\arabic{equation}}

\setcounter{page}{1}
\setcounter{equation}{0}
\setcounter{figure}{0}
\setcounter{section}{0}
\setcounter{table}{0}

\begin{center}
{\Large Supplementary material for `A Physics-Informed Spatiotemporal Deep Learning Framework for Turbulent Systems'}    
\end{center}

\setcounter{section}{1}
\section{Data \& Governing PDE} \label{sec:governing_pde}

The data employed in this work is a simulated two-dimensional RBC flow for dry air, a type of gravity-driven turbulent flow that occurs between two parallel horizontal plates \citep{rbc}.  Turbulent convection is a ubiquitous process which finds applications in both industrial and environmental applications of fluid mechanics \citep{GROSSMANN_LOHSE_2000, RevModPhys.81.503, annurev.fluid.010908.165152}. The difference in temperature between the warm lower plate and the cold upper plate produces convection cells as warm (thus less dense) fluid rises and cold (thus more dense) fluid falls. The experimental dataset for this work was simulated by DNS in a 2 \textit{m} $\times$ 1 \textit{m} rectangular domain and is comprised of two velocity components (horizontal and vertical), pressure, and temperature, so that the data comprises of $d=4$ variables varying in space and time. The spatial resolution is $n_{x} \times n_{z}$, $n_{x}=n_{z}=256$, and the temporal resolution is $\Delta_{t} = 0.1$ seconds for a total of $T=2{,}500$ time steps. We denote the grid points in the horizontal and vertical direction by $G_{x}$ and $G_{z}$, respectively. The grid in the horizontal direction is evenly spaced, i.e., observations are separated by $R_{x} \approx 0.78~cm$, so $G_{x} = \{(k-1)R_{x}: k=1, \ldots, n_{x}\}$. The grid in the vertical direction is non-uniform, with finer resolution near the upper and lower boundaries (where most small-scale variations in the data occur). The grid points in the vertical direction are the Chebyshev-Gauss-Lobatto points, $G_{z} = \left\{\frac{1}{2}\left[1-\cos\left(\pi \cdot \frac{k-1}{n_{z}-1} \right)\right] : k=1,\ldots,n_{z} \right\}$ \citep{Gottlieb1977, Boyd2001}. One simulation of RBC was generated in under one hour using 8 CPU cores and requiring about 8 MB of memory per time step. Technical details on the numerical method used to generate the data can be found in \cite{spins_dns}.

\begin{figure}[!ht]
    \centering
    \includegraphics[width=\linewidth]{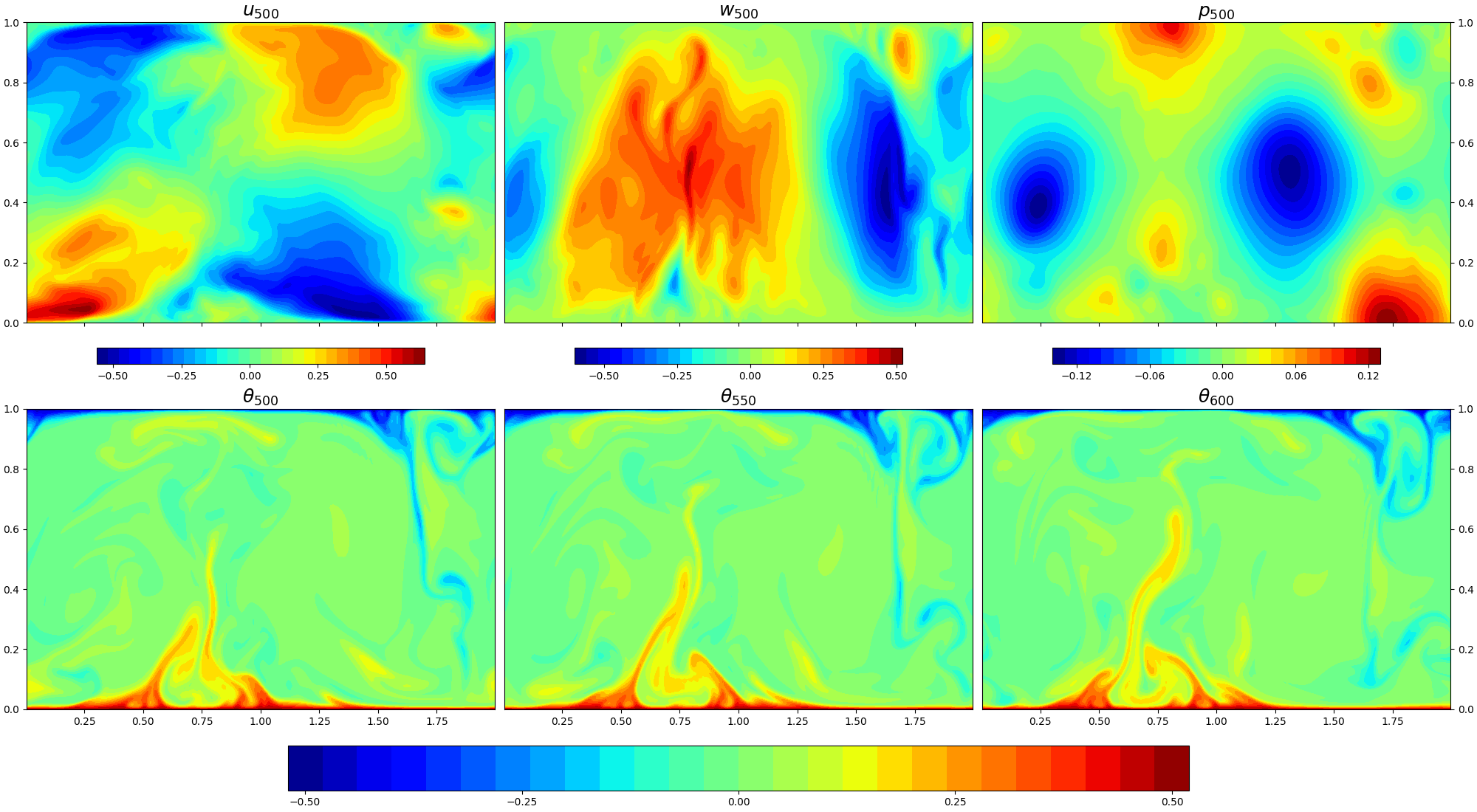}
    \caption{Sample observation $\bm{Y}_{t}$ of Rayleigh-B\'enard convection \eqref{eq:rbc}. The two velocity components, $u_{t}$ and $w_{t}$, and pressure $p_{t}$ are on the top row, while the bottom row shows the temporal evolution of temperature $\theta_{t}$ to highlight the convection.}
    \label{fig:rbc_sample}
\end{figure}

The two-dimensional RBC is governed by the Navier-Stokes equations under the Boussinesq approximation, which encode mass conservation, momentum, and energy conservation for a thermally stratified fluid \citep{navier1822memoire, boussinesq1903, rbc}. The governing (nondimensional) equations for RBC are
\begin{subequations} \label{eq:rbc}
\begin{gather}
    \frac{\partial u_{t}}{\partial x} + \frac{\partial w_{t}}{\partial z} = 0, \label{eq:mc} \\
    \frac{\partial u_{t}}{\partial t} + u_{t} \frac{\partial u_{t}}{\partial x} + w_{t} \frac{\partial u_{t}}{\partial z} = - \frac{\partial p_{t}}{\partial x} + \sqrt{\frac{\text{Pr}}{\text{Ra}}} \left( \frac{\partial^{2} u_{t}}{\partial x^{2}} + \frac{\partial^{2} u_{t}}{\partial z^{2}} \right), \label{eq:momentum_u} \\
    \frac{\partial w_{t}}{\partial t} + u_{t} \frac{\partial w_{t}}{\partial x} + w_{t} \frac{\partial w_{t}}{\partial z} = - \frac{\partial p_{t}}{\partial z} + \sqrt{\frac{\text{Pr}}{\text{Ra}}} \left( \frac{\partial^{2} w_{t}}{\partial x^{2}} + \frac{\partial^{2} w_{t}}{\partial z^{2}} \right) + \theta_{t}, \label{eq:momentum_w} \\
    \frac{\partial \theta_{t}}{\partial t} + u_{t} \frac{\partial \theta_{t}}{\partial x} + w_{t} \frac{\partial \theta_{t}}{\partial z} = \frac{1}{\sqrt{\text{PrRa}}} \left( \frac{\partial^{2} \theta_{t}}{\partial x^{2}} + \frac{\partial^{2} \theta_{t}}{\partial z^{2}} \right), \label{eq:temp}
\end{gather}
\end{subequations}
where \eqref{eq:mc} is mass conservation, \eqref{eq:momentum_u}-\eqref{eq:momentum_w} is momentum conservation, \eqref{eq:temp} is energy conservation, and Pr and Ra are the Prandtl and Rayleigh numbers respectively \citep{prandtl1904, rbc}. The Pr and Ra numbers are ratios of physical constants and a result of nondimensionalization. Specifically  the Pr number is the ratio of momentum diffusivity and thermal diffusivity, and for this simulation is set to $7.4 \times 10^{-1}$. The Rayleigh number, which for this data is $2.54 \times 10^{8}$, is a measure of the strength of the buoyancy and the tendency for turbulent flow. The RBC flow for these choices of Pr and Ra resembles the behavior of air. The constant temperature, before nondimensionalization, at the bottom and top plate are $\theta^{*}_{\text{bot}} = 299^{\circ}$K (25.85$^{\circ}$C) and $\theta^{*}_{\text{top}} = 280^{\circ}$K (6.85$^{\circ}$C), respectively. The turnover rate, i.e., the time required for a fluid particle to complete one full convection cycle, can be estimated from the physical constants associated with this data, and for this flow is $\tau_{c} = 5.4\text{s}$ (or 54 observations). One example observation is in Figure \ref{fig:rbc_sample}, where we show $u_{t_{1}},w_{t_{1}}$ and $p_{t_{1}}$ (the top row) and $\theta_{t_{1}}, \theta_{t_{2}}$, and $\theta_{t_{3}}$ (bottom row), with $t_{1}=500, t_{2}=550$, and $t_{3}=600$, to provide visual evidence for convection.   

If we denote the \textit{dimensional} (spatial and temporal) coordinates by $\left(x^{*}, z^{*}, t^{*} \right)$ and the RBC observation at $t^{*}$ by $\bm{Y}^{*}_{t^{*}} = \left(u_{t^{*}}^{*}, w_{t^{*}}^{*}, p_{t^{*}}^{*}, \theta_{t^{*}}^{*} \right)$, then the governing equations are
\begin{subequations} \label{eq:rbc_dimensional_starred} 
\begin{gather}
    \frac{\partial u_{t^{*}}^{*}}{\partial x^{*}} + \frac{\partial w_{t^{*}}^{*}}{\partial z^{*}} = 0, \label{eq:mc_dim_star} \\
    \frac{\partial u_{t^{*}}^{*}}{\partial t^{*}} + u_{t^{*}}^{*} \frac{\partial u_{t^{*}}^{*}}{\partial x^{*}} + w_{t^{*}}^{*} \frac{\partial u_{t^{*}}^{*}}{\partial z^{*}} = - \frac{1}{\rho_{0}} \frac{\partial p_{t^{*}}^{*}}{\partial x^{*}} + \nu \left( \frac{\partial^{2} u_{t^{*}}^{*}}{\partial x^{*2}} + \frac{\partial^{2} u_{t^{*}}^{*}}{\partial z^{*2}} \right), \label{eq:momentum_u_dim_star} \\
    \frac{\partial w_{t^{*}}^{*}}{\partial t^{*}} + u_{t^{*}}^{*} \frac{\partial w_{t^{*}}^{*}}{\partial x^{*}} + w_{t^{*}}^{*} \frac{\partial w_{t^{*}}^{*}}{\partial z^{*}} = - \frac{1}{\rho_{0}} \frac{\partial p_{t^{*}}^{*}}{\partial z^{*}} + \nu \left( \frac{\partial^{2} w_{t^{*}}^{*}}{\partial x^{*2}} + \frac{\partial^{2} w_{t^{*}}^{*}}{\partial z^{*2}} \right) -g\left( 1 - \alpha \left( \theta_{t^{*}}^{*} - \theta_{\text{bot}}^{*} \right) \right), \label{eq:momentum_w_dim_star} \\
    \frac{\partial \theta_{t^{*}}^{*}}{\partial t^{*}} + u_{t^{*}}^{*} \frac{\partial \theta_{t^{*}}^{*}}{\partial x^{*}} + w_{t^{*}}^{*} \frac{\partial \theta_{t^{*}}^{*}}{\partial z^{*}} = \kappa \left( \frac{\partial^{2} \theta_{t^{*}}^{*}}{\partial x^{*2}} + \frac{\partial^{2} \theta_{t^{*}}^{*}}{\partial z^{*2}} \right), \label{eq:temp_dim_star}
\end{gather}
\end{subequations}
where \eqref{eq:mc_dim_star}-\eqref{eq:temp_dim_star} represent mass conservation, momentum, and energy conservation, respectively. The physical parameters are the kinematic viscosity $\nu= 10^{-5}~m^{2}/s$, gravity $g=5.45 \times 10^{-1}~m/s^{2}$, the thermal expansion coefficient $\alpha = 3.33 \times 10^{-3}~\mathrm{K}^{-1}$, and thermal diffusivity $\kappa = 1.39 \times 10^{-5}~m^{2}/s$. As specified in the main text, the constant temperature at the bottom and top plate are $\theta^{*}_{\text{bot}} = 299^{\circ}$K (25.85$^{\circ}$C) and $\theta^{*}_{\text{top}} = 280^{\circ}$K (6.85$^{\circ}$C), respectively, and $\rho_{0}=1~\text{kg}/m^{3}$ is the density of the fluid at $\theta^{*}_{\text{bot}}$.

Nondimensionalization is a crucial step in the analysis of RBC, as it simplifies the equations and highlights the governing physical mechanisms. By rescaling variables using characteristic length, time, velocity, and temperature scales, the equations are rewritten in a form that reduces the number of physical parameters and reveals the relative importance of competing forces in the system. To nondimensionalize the equations \eqref{eq:rbc_dimensional_starred}, we introduce characteristic scales:
\begin{itemize}
    \item Length (distance between plates):  $H$,
    \item Temperature (temperature difference between plates): $\Delta \theta^{*} = \theta^{*}_{\text{bot}}- \theta^{*}_{\text{top}}$ ,
    \item Velocity: $U = \sqrt{\alpha g H \Delta \theta^{*}}$,
    \item Time: $\tau = \frac{H}{U}$,
    \item Pressure: $P = \rho_{0} U^{2}$,
\end{itemize}
and define nondimensional variables as
\begin{gather*}
    x = \frac{x^{*}}{H}, \qquad z = \frac{z^{*}}{H}, \qquad t = \frac{t^{*}}{\tau}, \\
    u_{t} = \frac{u_{t}^{*}}{U}, \qquad w_{t} = \frac{w_{t}^{*}}{U}, \qquad
    \theta_{t} = \frac{\theta_{t}^{*} - \theta_{\text{top}}^{*}}{\Delta \theta^{*}}, \qquad p_{t} = \frac{p_{t}^{*}}{P}.
\end{gather*}
The nondimensional equations can be obtained by  substituting each scaled term in \eqref{eq:rbc_dimensional_starred}. 

The nondimensionalization for mass conservation \eqref{eq:mc_dim_star} is simply
\begin{subequations}
\begin{gather}
    \frac{U}{H} \left(\frac{\partial u_{t}}{\partial x} + \frac{\partial w_{t}}{\partial z} \right) = 0 \nonumber \\ \quad \frac{\partial u_{t}}{\partial x} + \frac{\partial w_{t}}{\partial z} = 0. \label{eq:mc_ndim}
\end{gather}
In the case of the momentum equation for horizontal velocity \eqref{eq:momentum_u_dim_star}, 
\begin{align}
    \frac{U^{2}}{H} \left( \frac{\partial u_{t}}{\partial t} + u_{t}\frac{\partial u_{t}}{\partial x} + w_{t} \frac{\partial u_{t}}{\partial z} \right) &= - \frac{\rho_{0} U^{2} }{\rho_{0}} \frac{1}{H} \frac{\partial p_{t}}{\partial x}  +  \frac{\nu U}{H^{2}} \left( \frac{\partial^{2} u_{t}}{\partial x^{2}} + \frac{\partial^{2} u_{t}}{\partial z^{2}} \right) \nonumber \\
    \frac{\partial u_{t}}{\partial t} + u_{t} \frac{\partial u_{t}}{\partial x} + w_{t} \frac{\partial u_{t}}{\partial z} &= - \frac{\partial p_{t}}{\partial x}  +  \frac{\nu}{UH} \left( \frac{\partial^{2} u_{t}}{\partial x^{2}} + \frac{\partial^{2} u_{t}}{\partial z^{2}} \right) \nonumber \\
    \frac{\partial u_{t}}{\partial t} + u_{t} \frac{\partial u_{t}}{\partial x} + w_{t} \frac{\partial u_{t}}{\partial z} &= - \frac{\partial p_{t}}{\partial x}  +  \sqrt{\frac{\nu}{\kappa}} \sqrt{\frac{\nu \kappa}{\alpha g H^{3} \Delta \theta^{*}}} \left( \frac{\partial^{2} u_{t}}{\partial x^{2}} + \frac{\partial^{2} u_{t}}{\partial z^{2}} \right) \nonumber \\
    \frac{\partial u_{t}}{\partial t} + u_{t} \frac{\partial u_{t}}{\partial x} + w_{t} \frac{\partial u_{t}}{\partial z} &= - \frac{\partial p_{t}}{\partial x}  +  \sqrt{\frac{\text{Pr}}{\text{Ra}}} \left( \frac{\partial^{2} u_{t}}{\partial x^{2}} + \frac{\partial^{2} u_{t}}{\partial z^{2}} \right), \label{eq:momentum_u_ndim}
\end{align}
and similarly for vertical velocity \eqref{eq:momentum_w_dim_star},
\begin{align}
    \frac{U^{2}}{H} \left( \frac{\partial w_{t}}{\partial t} + u_{t} \frac{\partial w_{t}}{\partial x} + w_{t} \frac{\partial w_{t}}{\partial z} \right)
    &= - \frac{\rho_{0} U^{2}}{\rho_{0}} \frac{1}{H} \frac{\partial p_{t}}{\partial z}
    + \frac{\nu U}{H^{2}} \left( \frac{\partial^{2} w_{t}}{\partial x^{2}} + \frac{\partial^{2} w_{t}}{\partial z^{2}} \right)
    - g \alpha \Delta \theta^{*} \, \theta_{t} \nonumber \\
    \frac{\partial w_{t}}{\partial t} + u_{t} \frac{\partial w_{t}}{\partial x} + w_{t} \frac{\partial w_{t}}{\partial z}
    &= - \frac{\partial p_{t}}{\partial z}
    + \frac{\nu}{UH} \left( \frac{\partial^{2} w_{t}}{\partial x^{2}} + \frac{\partial^{2} w_{t}}{\partial z^{2}} \right)
    - \frac{\alpha g H \Delta \theta^{*}}{U^{2}} \theta_{t} \nonumber \\
    \frac{\partial w_{t}}{\partial t} + u_{t} \frac{\partial w_{t}}{\partial x} + w_{t} \frac{\partial w_{t}}{\partial z}
    &= - \frac{\partial p_{t}}{\partial z}
    + \sqrt{\frac{\text{Pr}}{\text{Ra}}} \left( \frac{\partial^{2} w_{t}}{\partial x^{2}} + \frac{\partial^{2} w_{t}}{\partial z^{2}} \right)
    - \theta_{t} \label{eq:momentum_w_ndim}
\end{align}
Lastly, for energy conservation \eqref{eq:temp_dim_star},
\begin{align}
    \frac{\partial \theta_{t}}{\partial t}
    + u_{t} \frac{\partial \theta_{t}}{\partial x}
    + w_{t} \frac{\partial \theta_{t}}{\partial z}
    &= \frac{\kappa}{UH} \left( \frac{\partial^{2} \theta_{t}}{\partial x^{2}} + \frac{\partial^{2} \theta_{t}}{\partial z^{2}} \right) \nonumber \\
    \frac{\partial \theta_{t}}{\partial t}
    + u_{t} \frac{\partial \theta_{t}}{\partial x}
    + w_{t} \frac{\partial \theta_{t}}{\partial z} &= \sqrt{\frac{1}{\text{Ra} \text{Pr}}} \left( \frac{\partial^{2} \theta_{t}}{\partial x^{2}} + \frac{\partial^{2} \theta_{t}}{\partial z^{2}} \right). \label{eq:temp_ndim}
\end{align}
\end{subequations}
So the nondimensional governing equations for RBC are \eqref{eq:mc_ndim}-\eqref{eq:temp_ndim}. The physical constant $\text{Pr} = \nu / \kappa$ is the Prandtl number and $\text{Ra} = \frac{\alpha g H^{3} \Delta \theta^{*}}{\nu \kappa}$ is the Rayleigh number and they are the two characteristic constants of RBC \citep{prandtl1904, rbc}. 

\section{Methods}

\begin{figure}[!ht]
    \centering
    \includegraphics[width=\linewidth]{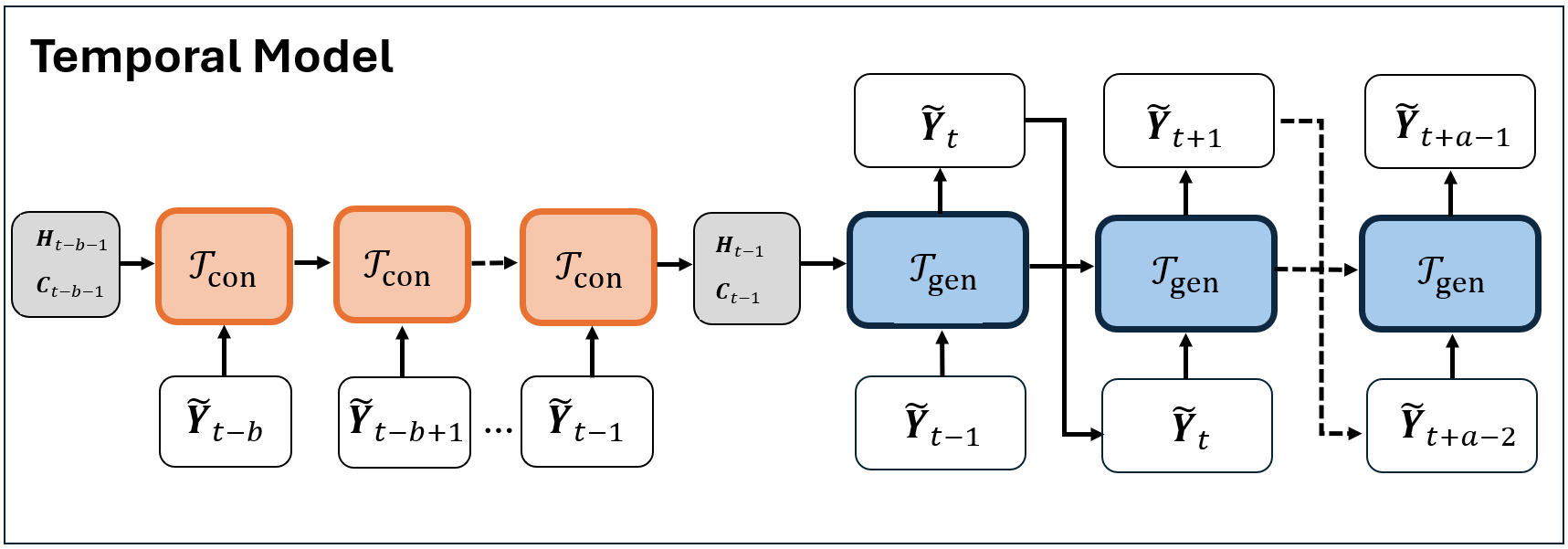}
    \caption{Illustration of the temporal model, composed of sub-models $\mathcal{T}_{\text{con}}$ and $\mathcal{T}_{\text{gen}}$. The input sequence $\tilde{\mathcal{Y}}_{t;b}^{-}$ is processed by $\mathcal{T}_{\text{con}}$, storing the short- and long-term dependencies, over the entirety of $\tilde{\mathcal{Y}}_{t;b}^{-}$,  in $\bm{H}_{t-1}$ and $\bm{C}_{t-1}$, respectively. The sequence generator $\mathcal{T}_{\text{gen}}$ is initialized with $\bm{H}_{t-1}$ and $\bm{C}_{t-1}$ and generates the output sequence (in reduced space).}
    \label{fig:s2_temporal_model}
\end{figure}

Figure \ref{fig:s2_temporal_model} shows an illustration of the temporal model, highlighting how the context-builder $\mathcal{T}_{\text{con}}$ and sequence-generator $\mathcal{T}_{\text{gen}}$ interact.

\section{Inference}

Physics-informed neural networks are known to suffer from ``gradient pathologies" driven by the different scales and variabilities of dimensional velocity, pressure, and temperature \citep{wang_understanding_2020}. For example, temperature is measured in degrees Kelvin (in our case ranging from $280^{\circ}$K to $290^{\circ}$K) and exhibits strong gradients, while pressure is smoother and on the order of magnitude of $10^{-4}$. We address the different scales by nondimensionalizing the data, as we detail in Section \ref{sec:governing_pde}. However, nondimensionalization alone does not eliminate differences in spatial and temporal variability across the variables. To mitigate this, we employ an adaptive loss weighting strategy during training, wherein the relative contributions of each variable to the total loss are dynamically adjusted. By adapting the weights over time, the model balances learning across variables with differing smoothness and signal-to-noise characteristics, thereby preventing any one variable from dominating the optimization process.

Since we train the CAE and RNN separately, we tailor the adaptive loss strategy to each model. Section \ref{sec:hyperpars} outlines the chosen model hyperparameters. We detail the dynamic loss balancing strategy in the training of the CAE, in Section \ref{sec:cae_training_supp}, and of the temporal model, in Section \ref{sec:rnn_training_supp}. 

\subsection{Hyperparameters} \label{sec:hyperpars}

In the CAE, we assume that the spatial encoder and decoder each have $L=4$ layers, and the number of kernels, which are all $r_{x}^{\ell}=r_{z}^{\ell}=3$, are $K_{1}=K_{2}=K_{3}=K_{4}=64$, so the reduced space representation for any observation $\bm{Y}_{t} \in \mathbb{R}^{n_{x} \times n_{z} \times d}$ is  $\tilde{\bm{Y}}_{t} \in \mathbb{R}^{16 \times 16 \times 64}$, $n_{x}=n_{z}=256$, resulting in a dimension reduction of 93.75\%. The activation function $g$ is the leaky rectified linear unit, a modified version of the rectified linear unit that allows some negative values, in every layer except for the last, where $g$ is the hyperbolic tangent function function \citep{goodfellow2016}. In the RNN, the context builder, $\mathcal{T}_{\text{con}}$, and sequence generator, $\mathcal{T}_{\text{gen}}$, each have $\tilde{K}=128$ kernels, where each kernel is $\tilde{r}_{x}=\tilde{r}_{z}=1$, i.e., the temporal evolution is modeled ``pixel-wise'' in the reduced space. For the lengths of the input and output sequences, $b$ and $a$, respectively, we leveraged the turnover time parameter $\tau_{c}$, defined as the time required for a fluid particle to complete one full convection cycle, which for this flow is $\tau_{c} = 5.4s$ (or 54 observations). We selected $b=80$ and $a=60$ to ensure that both the input and output windows span at least one turnover time. This provides the model with enough temporal context to capture the dominant dynamical structures and evolve them meaningfully into the future. So the temporal model (in reduced space) takes an input sequence of length $b=80$ time steps, $\tilde{\mathcal{Y}}^{-}_{t;b}$, and outputs the next $a=60$ (reduced) observations, $\tilde{\mathcal{Y}}^{+}_{t;a}$.

\subsection{Convolutional Autoencoder Inference} \label{sec:cae_training_supp}

We find $\widehat{\bm{W}}^{\text{space}}$ by minimizing the mean squared error with respect to the data,
\begin{gather}
    \widehat{\bm{W}}^{\text{space}} = \argmin_{\bm{W}^{\text{space}}} \text{MSE}_{\text{space}}\left(\bm{W}^{\text{space}}\right) = \argmin_{\bm{W}^{\text{space}}} \left\{ \sum_{i} \lambda^{\text{space}}_{i}\mathcal{L}^{\text{space}}_{i} \left(\bm{W}^{\text{space}}\right) \right\},
\end{gather}
where $\bm{\lambda}^{\text{space}}= \left\{\lambda^{\text{space}}_{i} : i=u, w, p, \theta \right\}$, are weights balancing the four terms' respective contributions. Denote by $\bm{Y}\left(\bm{W}^{\text{space}}\right) = \left(u\left(\bm{W}^{\text{space}}\right), w\left(\bm{W}^{\text{space}}\right), p\left(\bm{W}^{\text{space}}\right), \theta\left(\bm{W}^{\text{space}}\right) \right)$ the data reconstruction conditional on $\bm{W}^{\text{space}}=\left(\bm{W}_{\text{en}}, \bm{W}_{\text{de}} \right)$, where
\begin{gather*}
    \bm{Y}\left(\bm{W}^{\text{space}}\right) = \text{CAE}\left(\bm{Y} \mid \bm{W}^{\text{space}}\right) = \mathcal{SD}\left( \mathcal{SE} \left( \bm{Y} \mid \bm{W}_{\text{en}} \right) \mid \bm{W}_{\text{de}} \right).
\end{gather*}
The loss balancing strategy we implement is a form of dynamic weight averaging, where we compute the ratio of each loss term over consecutive epochs and apply a softmax function to determine the weights \citep{zhang2018end}. Intuitively speaking, this approach ensures that, at any optimization step, the loss terms that are decreasing faster are weighted less while the ones that are decreasing more slowly get weighted more. The approach is detailed in Algorithm \ref{alg:cae_loss_balance_training}.

\begin{algorithm}[hbt!]
\setstretch{1.25}
\caption{CAE Training with dynamically balanced losses} 
\SetKwInOut{Parameter}{Parameters}

\label{alg:cae_loss_balance_training}

\Parameter{$\bm{W}^{\text{space}}$}

\SetKwInOut{Parameter}{Hyperparameters }
\Parameter{$\bm{\lambda}^{\text{space}}$ } 

$e \gets 1$ \tcp*[r]{Epoch counter}

$\bm{W}^{\text{space}}_{e} \gets \text{random initialization}$ \tcp*[r]{Parameters at epoch $e$}

$\bm{\lambda}^{\text{space}}_{e} \gets (1/4,1/4,1/4,1/4)$ \tcp*[r]{Loss weights at epoch $e$}

$\mathcal{L}^{\text{space}}_{e} \gets \left(\mathcal{L}^{\text{space}}_{e,u}, \mathcal{L}^{\text{space}}_{e,w}, \mathcal{L}^{\text{space}}_{e,p}, \mathcal{L}^{\text{space}}_{e,\theta} \right)$ \tcp*[r]{Losses at epoch $e$}

\While{$e \leq E$ \tcp*[r]{$E$ is the number of epochs}}{
    $\bm{Y}(\bm{W}^{\text{space}}_{e}) = \text{CAE} \left( \bm{Y} \mid \bm{W}^{\text{space}}_{e} \right)$ \tcp*[r]{Forward pass}
    $\mathcal{L}^{\text{space}}_{e} \gets \mathcal{L}^{\text{space}} \left( \bm{W}^{\text{space}}_{e} \right)$ \tcp*[r]{Compute losses}
    \If{$e \geq 2$}{$\bm{\lambda}^{\text{space}}_{e+1} \gets \text{softmax} \left( \frac{\mathcal{L}^{\text{space}}_{e}}{\mathcal{L}^{\text{space}}_{e-1}} \right)$ ;}
    $\text{MSE}_{\text{space}} \left( \bm{W}^{\text{space}}_{e} \right) = \sum_{i} \lambda^{\text{space}}_{e,i} \mathcal{L}^{\text{space}}_{e,i} \left(\bm{W}^{\text{space}}_{e} \right) $ \tcp*[r]{Balance losses}
    $\bm{W}^{\text{space}}_{e+1} \leftarrow \text{Backpropagation} \left( \text{MSE}_{\text{space}} \left( \bm{W}^{\text{space}}_{e} \right)\right)$ ; \\
    $e \gets e + 1$ ;
}

$\widehat{\bm{W}}^{\text{space}} \gets \bm{W}^{\text{space}}_{E}$ \tcp*[r]{Estimated parametrs}
\end{algorithm}

\subsection{Recurrent Neural Network Physics-Informed Inference} \label{sec:rnn_training_supp}

We estimate the temporal parameters $\widehat{\bm{W}}^{\text{time}}$ \textit{conditional on} $\widehat{\bm{W}}^{\text{space}}$, the estimated CAE parameters. We find $\widehat{\bm{W}}^{\text{time}}$ by minimizing a penalized target function $\mathcal{L}\left(\bm{W}^{\text{time}} \right)$:
\begin{align}
    \widehat{\bm{W}}^{\text{time}} &= \argmin_{\bm{W}^{\text{time}}} \mathcal{L}\left(\bm{W}^{\text{time}} \right) \nonumber \\
    &= \argmin_{\bm{W}^{\text{time}}} \left\{ \sum_{i} \lambda^{\text{Data}}_{i}\mathcal{L}^{\text{Data}}_{i}\left( \bm{W}^{\text{time}} \right) + \sum_{j} \lambda_{j}^{\text{PDE}} \mathcal{L}^{\text{PDE}}_{j}(\bm{W}^{\text{time}}) \right\}, \label{eq:rnn_loss}
\end{align}
where $i \in \{u, w, p, \theta\}$ and $j \in \{\text{Mass},~ \text{Mom-}u,~ \text{Mom-}w,~ \text{Energy}\}$ and denote 
\begin{gather*}
    \bm{\lambda}^{\text{Data}} = \left(\lambda^{\text{Data}}_{u}, \lambda^{\text{Data}}_{w}, \lambda^{\text{Data}}_{p}, \lambda^{\text{Data}}_{\theta} \right), \quad \bm{\lambda}^{\text{PDE}} = \left(\lambda^{\text{PDE}}_{\text{Mass}}, \lambda^{\text{PDE}}_{\text{Mom-}u}, \lambda^{\text{PDE}}_{\text{Mom-}w}, \lambda^{\text{PDE}}_{\text{Energy}} \right).
\end{gather*}
The term $\mathcal{L}^{\text{Data}}_{i}$ represents the loss terms with respect to the observed data and are similar to the ones in Section \ref{sec:cae_training_supp}, while the terms $\mathcal{L}^{\text{PDE}}_{j}$ represent the loss terms with respect to the (nondimensional) governing PDEs. Further, denote 
\begin{gather*}
    \mathcal{L}^{\text{Data}}(\bm{W}^{\text{time}}) = \sum_{i} \lambda^{\text{Data}}_{i}\mathcal{L}^{\text{Data}}_{i}\left( \bm{W}^{\text{time}} \right), \quad \mathcal{L}^{\text{PDE}}(\bm{W}^{\text{time}}) = \sum_{j} \lambda_{j}^{\text{PDE}} \mathcal{L}^{\text{PDE}}_{j}(\bm{W}^{\text{time}}),
\end{gather*}
and the reduced space input and output sequences in the training set $\tilde{\mathcal{D}}_{\text{train}}^{\text{RNN}}$, as defined in Section 4.1 of the manuscript, by 
\begin{gather*}
    \left(\tilde{\mathcal{Y}}^{-}, \tilde{\mathcal{Y}}^{+} \right) = \left\{\left( \tilde{\mathcal{Y}}^{-}_{t;b}, \tilde{\mathcal{Y}}^{+}_{t;a} \right) : t = b, b+o, b+2o, \ldots, n-a \right\}.
\end{gather*}
Similarly, denote the predicted sequences (in the reduced space) conditional on $\bm{W}^{\text{time}}$ by $\tilde{\mathcal{Y}}^{+}\left(\bm{W}^{\text{time}} \right) = \mathcal{T}\left(\tilde{\mathcal{Y}}^{-} \mid \bm{W}^{\text{time}}  \right)$ and the predicted sequences in the original space, needed to compute the eight loss terms in \eqref{eq:rnn_loss}, by $\mathcal{Y}^{+}\left(\bm{W}^{\text{time}} \right) = \mathcal{SD}\left( \tilde{\mathcal{Y}}^{+}\left(\bm{W}^{\text{time}} \right) \mid \widehat{\bm{W}}_{\text{de}} \right)$.

The derivatives in $\mathcal{L}^{\text{PDE}}(\bm{W}^{\text{time}})$ are approximated using a central difference method. To compute the physics loss, we apply finite difference operators to the predicted fields, using central differences in the interior of the domain, and forward or backward differences near the boundaries to handle edge conditions. If we denote the $r$-th element of the predicted sequence starting at $t$ at spatial location $(x_{i},z_{j})$, $x_{i} \in G_{x}$ and $z_{j} \in G_{z}$, where $G_{x}$ and $G_{z}$ are as defined in Section \ref{sec:governing_pde}, by 
\begin{gather*}
    \widehat{\bm{Y}}_{t;r}(x_{i},z_{j}) = \left(\widehat{u}_{t;r}(x_{i},z_{j}), \widehat{w}_{t;r}(x_{i},z_{j}), \widehat{p}_{t;r}(x_{i},z_{j}), \widehat{\theta}_{t;r}(x_{i},z_{j}) \right),
\end{gather*}
we approximate mass conservation as
\begin{align*}
    \mathcal{L}^{\text{PDE}}_{\text{Mass}} \left(\bm{W}^{\text{time}}\right) &= \frac{1}{\tilde{n} \cdot a \cdot n_{x} \cdot n_{z}} \sum_{t,r,i,j} \left(\frac{\partial \widehat{u}_{t;r}(x_{i},z_{j})}{\partial x} + \frac{\partial \widehat{w}_{t;r}(x_{i},z_{j})}{\partial z} \right)^{2} \\
    &\approx \frac{1}{\tilde{n} \cdot a \cdot n_{x} \cdot n_{z}} \sum_{t,r,i,j} \left( \widehat{R}^{\text{PDE}}_{\text{Mass}}(t,r,i,j)  \right)^2, \\
    \widehat{R}^{\text{PDE}}_{\text{Mass}}(t,r,i,j) &= \frac{u_{t;r}(x_{i+1},z_j) - \widehat{u}_{t;r}(x_{i-1},z_j)}{x_{i+1}-x_{i-1}} + \frac{w_{t;r}(x_i,z_{j+1}) - \widehat{w}_{t;r}(x_i,z_{j-1})}{z_{j+1}-z_{j-1}},
\end{align*}
and $ \mathcal{L}^{\text{PDE}}_{\text{Mom-}u} \left(\bm{W}^{\text{time}}\right)$, $ \mathcal{L}^{\text{PDE}}_{\text{Mom-}w} \left(\bm{W}^{\text{time}}\right)$, and $ \mathcal{L}^{\text{PDE}}_{\text{Energy}} \left(\bm{W}^{\text{time}}\right)$ are similar.

The loss balancing strategy we implement is a mixture of the dynamic weight average framework outlined in Algorithm \ref{alg:cae_loss_balance_training} and a gradient-based method where the loss with respect to the data and the loss with respect to the physics are scaled by the norm of their respective gradients \citep{wang2023expert}. This addition is necessary since, especially in the early stages of training, the gradients with respect to $\mathcal{L}^{\text{PDE}}(\bm{W}^{\text{time}})$ are orders of magnitude larger than those with respect to $\mathcal{L}^{\text{Data}}(\bm{W}^{\text{time}})$, mostly because of squared terms and derivatives in $\mathcal{L}^{\text{PDE}}(\bm{W}^{\text{time}})$. As such, in absence of additional (possibly gradient-based) balancing, the parameter optimization would steer in the direction of trivial PDE solutions and virtually ignore the observation. The approach is summarized in Algorithm \ref{alg:rnn_loss_balance_training}

\begin{algorithm}[hbt!]
\setstretch{1.25}
\caption{RNN Training with dynamically balanced losses} 
\SetKwInOut{Parameter}{Parameters}

\label{alg:rnn_loss_balance_training}

\Parameter{$\bm{W}^{\text{time}}$}

\SetKwInOut{Parameter}{Hyperparameters}
\Parameter{$\bm{\lambda}^{\text{time}}$ } 

$e \gets 1$ \tcp*[r]{Epoch counter}

$\bm{W}^{\text{time}}_{e} \gets \text{random initialization}$ \tcp*[r]{Parameters at epoch $e$}

$\bm{\lambda}^{\text{time}}_{e} \gets \left( \bm{\lambda}^{\text{Data}}_{e}, \bm{\lambda}^{\text{PDE}}_{e} \right)$ \tcp*[r]{Loss weights at epoch $e$}

$\bm{\lambda}^{\text{time}}_{e}\gets (1/8,1/8,1/8,1/8,1/8,1/8,1/8,1/8)$ ; \\
$\mathcal{L}^{\text{Data}}_{e} \gets \left(\mathcal{L}^{\text{Data}}_{e,u}, \mathcal{L}^{\text{Data}}_{e,w}, \mathcal{L}^{\text{Data}}_{e,p}, \mathcal{L}^{\text{Data}}_{e,\theta} \right)$ \tcp*[r]{Data losses at epoch $e$} 

$\mathcal{L}^{\text{PDE}}_{e} \gets \left(\mathcal{L}^{\text{PDE}}_{e,\text{Mass}}, \mathcal{L}^{\text{PDE}}_{e,\text{Mom-}u}, \mathcal{L}^{\text{PDE}}_{e,\text{Mom-}w}, \mathcal{L}^{\text{PDE}}_{e,\text{Energy}} \right)$ \tcp*[r]{Physics losses at epoch $e$}

\While{$e \leq E$ \tcp*[r]{$E$ is the number of epochs}}{
    $\tilde{\mathcal{Y}}^{+}\left(\bm{W}^{\text{time}} \right) = \mathcal{T}\left(\tilde{\mathcal{Y}}^{-} \mid \bm{W}^{\text{time}}  \right)$ \tcp*[r]{Forward pass of the temporal model}
    $\mathcal{Y}^{+}\left(\bm{W}^{\text{time}} \right) = \mathcal{SD}\left( \tilde{\mathcal{Y}}^{+}\left(\bm{W}^{\text{time}} \right) \mid \widehat{\bm{W}}_{\text{de}} \right)$ \tcp*[r]{Reconstruction}
    $\left(\mathcal{L}^{\text{Data}}_{e} , \mathcal{L}^{\text{PDE}}_{e}\right)\gets \left(\mathcal{L}^{\text{Data}}(\bm{W}^{\text{time}}),~ \mathcal{L}^{\text{PDE}}(\bm{W}^{\text{time}}) \right)$ \tcp*[r]{Compute losses}
    $\mathcal{L}^{\text{time}}_{e} \gets \left(\mathcal{L}^{\text{Data}}_{e} , \mathcal{L}^{\text{PDE}}_{e}\right)$ ; \\
    \If{$e \geq 2$}{ \texttt{// Two-step balancing} \\
    $\left( \bm{\lambda}^{\text{Data}}_{e+1}, \bm{\lambda}^{\text{PDE}}_{e+1} \right) = \bm{\lambda}^{\text{time}}_{e+1} \gets \text{softmax} \left( \frac{\mathcal{L}^{\text{time}}_{e}}{\mathcal{L}^{\text{time}}_{e-1}} \right)$ \tcp*[r]{DWA}
    $\bm{\lambda}^{\text{Data}}_{e+1} \gets  \bm{\lambda}^{\text{Data}}_{e+1} \cdot \left(\lVert \nabla \mathcal{L}^{\text{Data}}(\bm{W}^{\text{time}}) \rVert_{2} \right)^{-1}$  \tcp*[r]{Gradient-based for Data}
    $\bm{\lambda}^{\text{PDE}}_{e+1} \gets  \bm{\lambda}^{\text{PDE}}_{e+1} \cdot \left(\lVert \nabla \mathcal{L}^{\text{PDE}}(\bm{W}^{\text{time}}) \rVert_{2} \right)^{-1}$ \tcp*[r]{Gradient-based for PDE}}
    
    $\mathcal{L}\left(\bm{W}^{\text{time}} \right) = \sum_{i} \lambda^{\text{Data}}_{i}\mathcal{L}^{\text{Data}}_{i}\left( \bm{W}^{\text{time}} \right) + \sum_{j} \lambda_{j}^{\text{PDE}} \mathcal{L}^{\text{PDE}}_{j}(\bm{W}^{\text{time}})$ ; \\
    $\bm{W}^{\text{time}}_{e+1} \leftarrow \text{Backpropagation} \left( \mathcal{L}\left(\bm{W}^{\text{time}} \right)\right)$ ; \\
    $e \gets e + 1$ ;
}

$\widehat{\bm{W}}^{\text{time}} \gets \bm{W}^{\text{time}}_{E}$ \tcp*[r]{Estimated parametrs}
\end{algorithm}

Finally, Algorithm \ref{alg:forecast_loop} outlines the forecasting framework for generating spatiotemporal predictions over an arbitrary horizon using PI-CRNN. The method first applies a spatial encoder to reduce the dimensionality of the input sequence and initializes temporal context states using a conditioning network. These states are then iteratively advanced by a generative temporal model, producing reduced-space predictions that are decoded back to the original spatial resolution at each step and assembled into a long-range forecast.

\begin{algorithm}[hbt!]
\setstretch{1.25}
\caption{Forecasting sequences of arbitrary length}
\label{alg:forecast_loop}

\SetKwInOut{Input}{Input}
\SetKwInOut{Parameter}{Parameters}
\SetKwInOut{Output}{Output}

\Input{$\mathcal{Y}_{t;b}^{-} = \left(\bm{Y}_{t-b}, \bm{Y}_{t-b+1}, \ldots, \bm{Y}_{t-1}\right)$}
\Parameter{$\widehat{\bm{W}}_{\text{en}}, \widehat{\bm{W}}_{\text{con}}, \widehat{\bm{W}}_{\text{gen}}, \widehat{\bm{W}}_{\text{de}}$}
\Output{$\widehat{\mathcal{Y}}^{+}_{t;R}$}

$\widehat{\tilde{\mathcal{Y}}}_{t;b}^{-}
\gets \mathcal{SE}\!\left(
\mathcal{Y}_{t;b}^{-} \mid \widehat{\bm{W}}_{\text{en}}
\right)$ \tcp*[r]{Spatial data reduction}

$\left(\bm{H}_{t-1}, \bm{C}_{t-1}\right)
\gets \mathcal{T}_{\text{con}}\!\left(
\widehat{\tilde{\mathcal{Y}}}_{t;b}^{-} \mid \widehat{\bm{W}}_{\text{con}}
\right)$ \tcp*[r]{Context matrices}

$t \gets 0$ \tcp*[r]{Forecast step counter}

\While{$t < R$}{
    $\bm{\widehat{\tilde{Y}}}_{t}, \bm{H}_{t}, \bm{C}_{t}
    \gets \mathcal{T}_{\text{gen}}\!\left(
    \bm{\widehat{\tilde{Y}}}_{t-1},
    \bm{H}_{t-1},
    \bm{C}_{t-1}
    \mid \widehat{\bm{W}}_{\text{gen}}
    \right)$ \tcp*[r]{Reduced-space prediction}

    $\widehat{\bm{Y}}_{t}
    \gets \mathcal{SD}\!\left(
    \bm{\widehat{\tilde{Y}}}_{t}
    \mid \widehat{\bm{W}}_{\text{de}}
    \right)$ \tcp*[r]{Decompression to original scale}

    $t \gets t + 1$ ;
}

$\widehat{\mathcal{Y}}^{+}_{t;R}
\gets \left(
\widehat{\bm{Y}}_{t},
\ldots,
\widehat{\bm{Y}}_{t+R-1}
\right)$ \tcp*[r]{Long-range forecast}

\end{algorithm}

\section{Uncertainty Quantification} \label{sec:uq}

The high sensitivity of RBC with respect to the initial condition makes predicting its \textit{specific} dynamics exceptionally challenging. Infinitesimally small perturbations, e.g., in thermal fluctuations or boundary layer instabilities, create diverging dynamics, thereby limiting the reliability of uncertainty quantification (UQ) over long forecasts \citep{LEUTBECHER20083515, RevModPhys.81.503, lorenz1963}. For this reason, we limit UQ to the first two turnover times of the forecast (corresponding to $T_{\text{uq}}=108$ time steps), beyond which the rapid divergence of trajectories renders probabilistic forecasts impractical.

We quantify the uncertainty using a bootstrap-based calibration method, which relies on conformal prediction intervals \citep{conformal, bonas, bonas_quantile}. Conformal prediction intervals are constructed by estimating residual uncertainty using quantiles, thereby not relying on the Gaussianity assumption. Since the model was trained up to $n$, we generate an ensemble of $E=50$ forecasts over two turnover times ($T_{\text{uq}}$ total time steps) given input sequences $\mathcal{Y}^{-}_{t;b}$, $n \leq t \leq T-T_{\text{uq}}$. So, the $T_{\text{uq}}$ predicted observations are all within the testing set. Denote the $j$-th element of the long-range forecast starting at $e$ by $\widehat{\bm{Y}}_{t_{e}+j}$, $e=1,\ldots,E$. The $E$ forecasts are then grouped according to their position within the estimated sequence, $\bm{f}_{1} = \left(\widehat{\bm{Y}}_{t_{1}+1}, \widehat{\bm{Y}}_{t_{2}+1}, \ldots, \widehat{\bm{Y}}_{t_{E}+1} \right)$, $\bm{f}_{2} = \left(\widehat{\bm{Y}}_{t_{1}+2}, \widehat{\bm{Y}}_{t_{2}+2}, \ldots, \widehat{\bm{Y}}_{t_{E}+2} \right)$, etc... So $\bm{f}_{1}$ contains all the one-step ahead forecasts, $\bm{f}_{2}$ contains all the two-step ahead forecasts, and so on. Then, we compute the absolute value of the residuals for each $t$-step ahead, $\bm{r}_{t} = \left(\bm{R}_{t_{1}+t}, \bm{R}_{t_{1}+t}, \ldots, \bm{R}_{t_{E}+t} \right)$, where $\bm{R}_{t_{e}+t} = \left| \bm{Y}_{t_{e}+t} - \widehat{\bm{Y}}_{t_{e}+t} \right|,~ \bm{R}_{t_{e}+t} = \left(\bm{R}_{u,t_{e}+t}, \bm{R}_{w,t_{e}+t}, \bm{R}_{p,t_{e}+t}, \bm{R}_{\theta, t_{e}+t} \right),~t=1,\ldots,T_{\text{uq}}, ~ e=1, \ldots, E$. The samples $\bm{r}_{t}$ are used to estimate the $\alpha$-quantile $\bm{q}_{j,t}$, where $j=u,w,p,\theta$, formally,
\begin{gather*}
    \bm{\hat{q}}_{u,t} = \inf \left\{ \bm{r}_{u,t}: \frac{1}{E} \sum_{e=1}^{E} \mathbb{I} \left(\bm{R}_{u,t_{e}+t} \leq \bm{r}_{u,t} \right) \geq 1- \alpha \right\},~ \bm{\hat{q}}_{u,t} \in \mathbb{R}^{n_{x} \times n_{z}},
\end{gather*}
and similarly for $\bm{\hat{q}}_{w,t}$, $\bm{\hat{q}}_{p,t}$, and $\bm{\hat{q}}_{\theta, t}$, $\hat{\bm{q}}_{t} = \left(\bm{\hat{q}}_{u,t}, \bm{\hat{q}}_{w,t}, \bm{\hat{q}}_{p,t}, \bm{\hat{q}}_{\theta, t} \right)$. This conformal prediction framework thus enables the construction of valid uncertainty intervals that naturally expand over time, reflecting increasing forecast uncertainty in RNN predictions. By grouping the $E$ forecasts based on their lead time, the method allows residuals to be computed separately for each time step in the forecast horizon, capturing the model’s historical error behavior at that specific lead time. Lastly, we perform an additional calibration step, for each variable, by multiplying the interval by a factor $\bm{\zeta}=\left(\zeta_{u}, \zeta_{w}, \zeta_{p}, \zeta_{\theta} \right)$, found via cross validation \citep{bonas_quantile}. So, $(1-\alpha)$\% prediction intervals can be obtained by construction as $P\left( \widehat{\bm{Y}}_{t} - \bm{\zeta}\bm{\hat{q}}_{t} \leq \bm{Y}_{t} \leq \widehat{\bm{Y}}_{t} + \bm{\zeta}\bm{\hat{q}}_{t} \right) \approx (1-\alpha)\%$.

Table \ref{tab:uq} presents the results for the quantification of uncertainty using the conformal methodology. Specifically, we evaluate the empirical coverage of prediction intervals (PIs) at three confidence levels-- 80\%, 90\%, and 95\% --for the first two turnovers of the ensemble of $E$ forecasts. The results demonstrate that the observed coverage rates closely align with their nominal levels, providing strong empirical support for the validity of the conformal approach detailed in Section 5 as applied to our PI-CRNN model. The close agreement between the empirical and nominal coverage probabilities (80\%, 90\%, and 95\%), all contained in the IQR, suggests that the model reliably captures the underlying uncertainty distribution.

\renewcommand{\arraystretch}{1.25}
\begin{table}[!ht]
    \centering
    \begin{tabular}{| l |c c c c | c|}
        \hline
        & \multicolumn{5}{c|}{Coverage (IQR)} \\ 
        P.I. & $u$ & $w$ & $p$ & $\theta$ & Total \\ \hline \hline
        80\% & $83.6~(1.4)$ & $82.4~(1.3)$ & $84.6~(4.5)$ & $81.2~(2.0)$ & $82.9~(2.3)$ \\ \hline
        90\% & $90.8~(0.9)$ & $90.0~(0.9)$ & $92.2~(2.2)$ & $89.3~(1.4)$ & $90.1~(1.3)$ \\ \hline
        95\% & $94.4~(0.9)$ & $93.8~(0.8)$ & $95.2~(1.5)$ & $93.2~(1.1)$ & $94.2~(1.1)$ \\ \hline
    \end{tabular}
    \caption{Uncertainty quantification of our PI-CRNN in terms of the 80\%, 90\%, and 95\% prediction interval across all spatial locations of the forecast.}
    \label{tab:uq}
\end{table}

\section{Results} \label{sec:results_supp}

\subsection{Spatial}

\subsubsection{Metrics of Assessment}

The spatial metrics assess to what extent the CAE leads to loss of information. As such, the metrics we employ are the mean squared error and the SSIM between the reconstruction and ground truth over the testing set. The SSIM was originally conceived to quantify the similarity between images by comparing visual fidelity with respect to the original data. The SSIM is effectively a pseudo correlation coefficient as it ranges from -1 to 1, with 1 implying a perfect reconstruction \citep{ssim}. 

We compute $\text{MSE}_{\text{space}}$ in equation (9) of the main text with $\lambda^{\text{space}}_{i}=1/4$, $i \in \{u,w,p,\theta\}$, so that the model is evaluated using an unweighted $\text{MSE}_{\text{space}}$ across the four RBC variables. This choice reflects the assumption that, for assessment purposes, all four variables are equally important. Using a weighted $\text{MSE}_{\text{space}}$ for evaluation could obscure poor reconstruction performance in some variables by compensating with better performance in others. For example, if the CAE reconstructs temperature very accurately but performs poorly on pressure, a heavily weighted loss on temperature could mask the pressure error in the overall metric. As a result, the model might appear to perform well even though it fails to accurately reconstruct all physical fields. An unweighted MSE avoids this issue by treating discrepancies in each variable equally. The SSIM is instead defined as
\begin{gather*}
    \text{SSIM} = \frac{1}{4} \left[\text{SSIM}(u) + \text{SSIM}(w) + \text{SSIM}(p) + \text{SSIM}(\theta) \right], \\
    \text{SSIM}(u) = \frac{1}{T-n} \sum_{t \in S_{\text{test}}} \frac{\left( 2\mu_{\hat{u}_{t}} \mu_{u_{t}} +C_{1} \right) \left( 2\sigma_{\hat{u}_{t} u_{t}} + C_{2} \right) }{\left( \mu^{2}_{\hat{u}_{t}} + \mu^{2}_{u_{t}} +C_{1} \right) \left( \sigma^{2}_{\hat{u}_{t}} + \sigma^{2}_{u_{t}} + C_{2} \right)}, \\
    S_{\text{test}} = \{n+1, \ldots, T \},
\end{gather*}
where $\mu_{\hat{u}_{t}}$ and $\sigma^{2}_{\hat{u}_{t}}$ are the mean and variance (over the spatial domain) of $u_{t}\left( \widehat{\bm{W}}^{\text{space}} \right)$, where  $\bm{Y}_{t}\left( \widehat{\bm{W}}^{\text{space}} \right) = \left( u_{t}\left( \widehat{\bm{W}}^{\text{space}} \right), w_{t}\left( \widehat{\bm{W}}^{\text{space}} \right), p_{t}\left( \widehat{\bm{W}}^{\text{space}} \right), \theta_{t}\left( \widehat{\bm{W}}^{\text{space}} \right) \right)$ is the reconstructed observation at $t$. The term $\sigma_{\hat{u}_{t} u_{t}}$ is the covariance between $u_{t}\left( \widehat{\bm{W}}^{\text{space}} \right)$ and $u_{t}$ and $C_{1}=C_{2}=10^{-5}$ are meant to prevent division by zero. $\text{SSIM}(w)$, $\text{SSIM}(p)$, and $\text{SSIM}(\theta)$ are similar to $\text{SSIM}(u)$.

\subsubsection{Gaussian Diagnostics}

\begin{figure}[!ht]
    \centering
    \includegraphics[width=\linewidth]{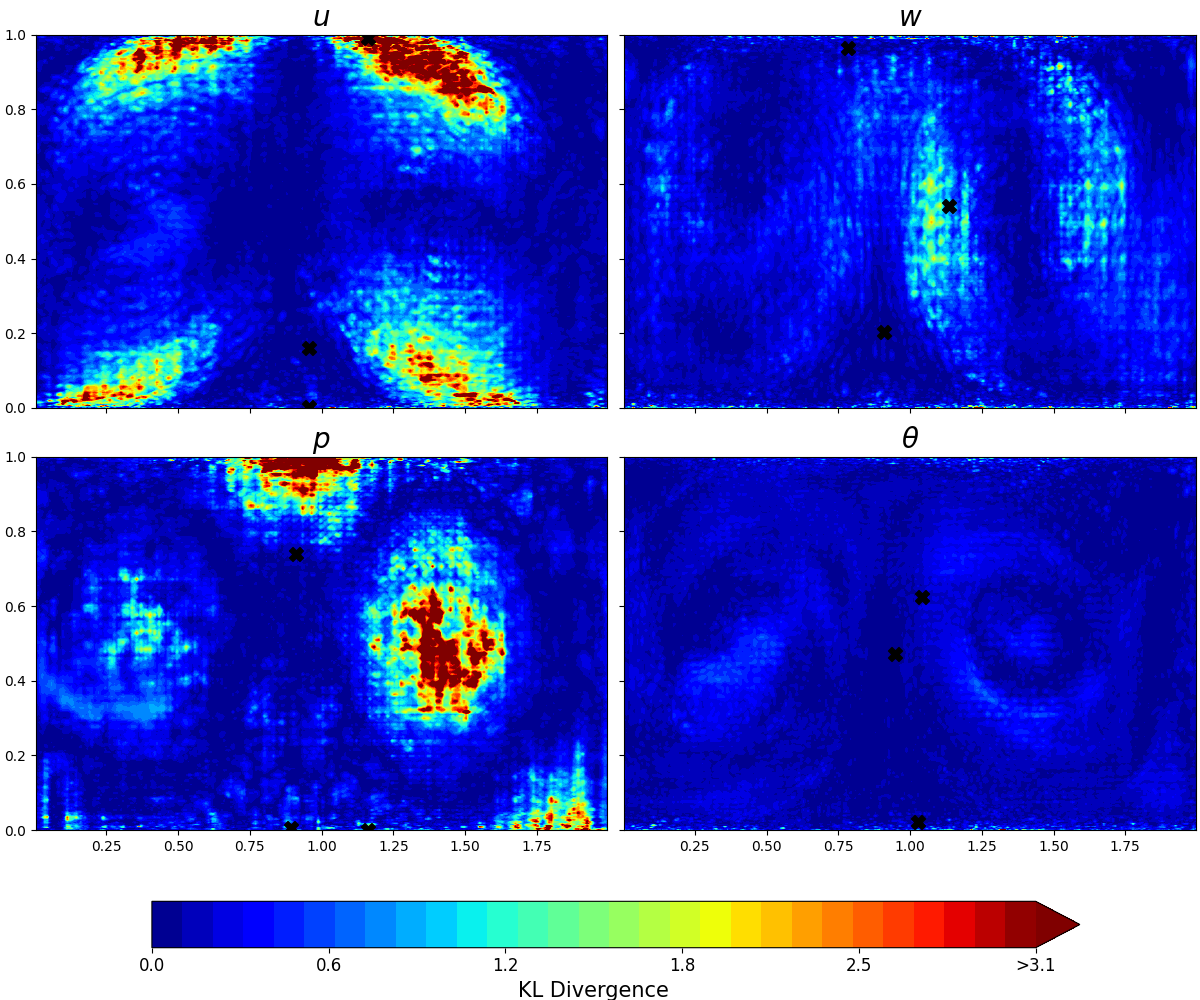}
    \caption{Assessing spatial normality: KL divergence between normalized residuals and a standard normal distribution over the test set, $t=n+1,\ldots,T$. The ``$\times$'' in each plot represent spatial locations for which we show Q-Q plots in Figure \ref{fig:s4_qq_cae}.}
    \label{fig:s3_kl_map}
\end{figure}

\begin{figure}[!ht]
    \centering
    \includegraphics[width=\linewidth]{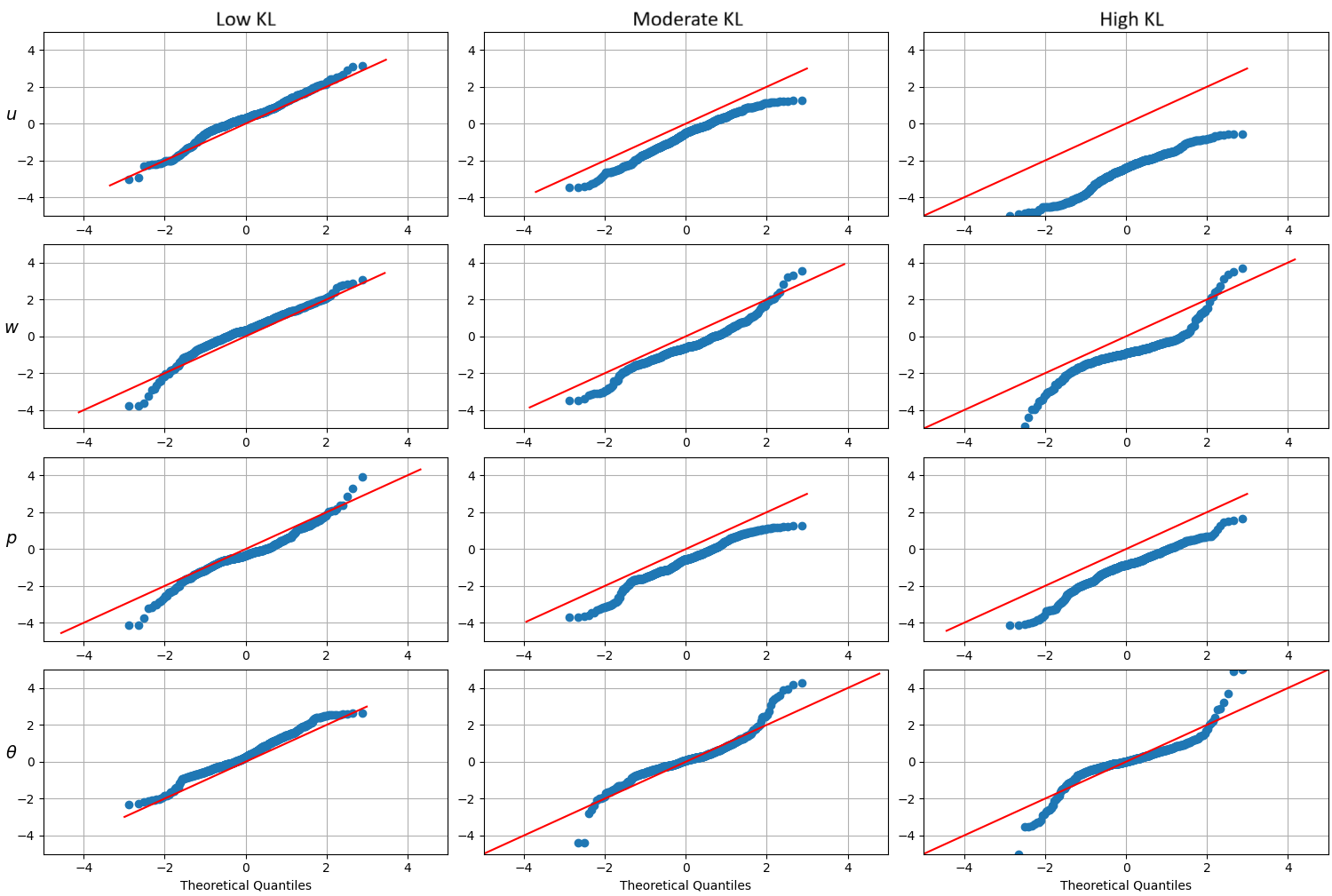}
    \caption{Assessing spatial normality: Q-Q plots for spatial locations corresponding to low, moderate, and high KL divergence between standardized residuals and a standard normal distribution. The three KL divergences were taken from each tercile and are marked by a ``$\times$'' symbol in Figure \ref{fig:s3_kl_map}.}
    \label{fig:s4_qq_cae}
\end{figure}

We verify the distributional assumptions in (1a), i.e., of $\bm{\eta}_{t}$, using the residuals obtained from the spatial compression over the test set,
\begin{gather*}
    \bm{e}_{t} = \bm{Y}_{t} - \mathcal{SD}\left(\mathcal{SE}\left(\bm{Y}_{t} \mid \widehat{\bm{W}}_{\text{en}}\right) \mid \widehat{\bm{W}}_{\text{de}} \right),~ t=n+1, \ldots, T,
\end{gather*}
where $\bm{e}_{t} \in \mathbb{R}^{n_{x} \times n_{z} \times d}$, $n_{x}=n_{z}=256$ and $d=4$. To best interpret the results, we compute the Kullback-Leibler (KL) divergence with respect to a normal distribution for every spatial location and every RBC variable, shown in Figure \ref{fig:s3_kl_map}. Figure \ref{fig:s4_qq_cae} presents Q-Q plots at representative grid points spanning low, intermediate, and high KL divergence regions for all four variables. While residuals do exhibit some degree of spatial dependence driven by the plumes of hot and cold air as seen in Figure \ref{fig:s3_kl_map}, the Q-Q plots in Figure \ref{fig:s4_qq_cae} show that even in spatial locations where the KL divergence is largest the residuals are in practice not that different from a normal distribution. This suggests that deviations from normality are mostly mild and primarily spatially organized rather than indicative of systematic non-Gaussian behavior.

\subsubsection{Other Spatial Methods} \label{sec:spatial_supp}

The three spatial methods used to compare performance with the CAE are PCA, ICA, FRK, and POD. Here we expand on each one to provide further context.

\subsubsection*{Principal Component Analysis}

PCA is a linear dimensionality reduction technique that transforms data into a new coordinate system. The goal of PCA is to identify the directions, or \textit{principal components}, that capture the maximum variance in the data. Let $\bm{X} \in \mathbb{R}^{n \times m}$ be a dataset consisting of $n$ observations and $m$ variables. The objective is to find a set of orthogonal vectors $\bm{v}_{1}, \bm{v}_{2}, \dots, \bm{v}_{k}$ that maximize the variance of the projected data. Mathematically, PCA seeks to solve the eigenvalue problem for the covariance matrix $\bm{\Sigma} = \frac{1}{n} \bm{X}^{\top} \bm{X}$, such that $\bm{\Sigma} \bm{v}_{i} = \lambda_{i} \mathbf{v}_{i}$, $i = 1, 2, \ldots, k$, where $\lambda_{i}$ are the eigenvalues representing the variance explained by each principal component $\bm{v}_{i}$. The components are ordered by their corresponding eigenvalues, with the first principal component explaining the maximum variance in the data \citep{jolliffe2002principal}.

In the context of our spatial analysis, PCA is employed to examine the spatiotemporal dynamics of temperature and velocity fields. Let $\bm{X} \in \mathbb{R}^{n \times n_{x} \times n_{z} \times {d}}$ represent the training dataset, where $n$ is the number of training time steps, $n_{x}$ and $n_{z}$ are the spatial dimensions, and $d$ represents the four RBC variables. To apply PCA, we first flatten the data into a matrix of dimensions $n \times (n_x n_z d)$. The resulting principal components, $\bm{v}_{1}, \bm{v}_{2}, \dots, \bm{v}_{k}$, ordered by their eigenvalues, reveal the dominant modes of variability across space and time.

\subsubsection*{Independent Component Analysis}

ICA is a statistical technique used to separate a multivariate signal into additive, independent components. Unlike PCA, which focuses on maximizing variance, ICA focuses on maximizing statistical independence between the components. Given a set of observed data $\bm{X} \in \mathbb{R}^{n \times m}$, which is assumed to be a linear mixture of $k$ independent sources $\bm{s}_1, \dots, \bm{s}_k$, ICA aims to estimate a demixing matrix $A$ such that $\bm{X} = \bm{A} \mathbf{s}$, where $\bm{s}$ represents the independent source signals and $\bm{A}$ is the unknown mixing matrix. ICA seeks to recover the independent components $\hat{\bm{s}}_1, \hat{\bm{s}}_2, \dots, \hat{\bm{s}}_k$ by maximizing their statistical independence. Mathematically, this independence is quantified by minimizing a cost function based on higher-order moments of the components \citep{comon1994independent}.

In the context of RBC, similarly to PCA, we apply ICA to the flattened data matrix $\bm{X} \in \mathbb{R}^{n \times (n_x n_z d)}$, where $n$ is the number of training time steps, $n_x$ and $n_z$ are the spatial dimensions, and $d$ represents the four RBC variables. This transformation allows ICA to separate the observed data into independent components that capture distinct physical processes occurring within the convection system.

\subsubsection*{Fixed Rank Kriging}

FKR is a method used in geostatistics to make spatial predictions when dealing with large datasets that would otherwise be computationally expensive due to the size of the covariance matrix \citep{Cressie2008}. The core idea is to approximate the full covariance matrix, typically dense and of size $n \times n$ (where $n$ is the number of spatial points), by decomposing it into a low-rank approximation. In standard kriging, the covariance matrix $\bm{C}$ of size $n \times n$ is inverted to obtain the weights for predicting at new locations. This inversion has a computational complexity of $\mathcal{O}(n^3)$, which is prohibitive for large datasets. In FRK, the covariance matrix $\bm{C}$ is approximated as $\bm{C} \approx \bm{L}\bm{L}^{\top}$, where $\bm{L}$ is a low-rank matrix with rank $r \ll n$. This allows the inversion to be performed in reduced space, reducing the computational cost significantly, specifically to $\mathcal{O}(nr^2)$, where $r$ is the fixed rank. The rank $r$ is typically chosen based on a balance between the desired computational efficiency and the quality of the approximation. The fixed rank assumption ensures that the number of parameters required to model the covariance structure is manageable while still capturing spatial correlations.

\subsubsection*{Proper Orthogonal Decomposition}

Proper Orthogonal Decomposition (POD) is a dimensionality reduction technique closely related to PCA and widely used in fluid dynamics. The main objective of POD is to extract a set of orthogonal basis functions, or \textit{modes}, that optimally represent the dynamics of a system in terms of energy content. Given a dataset $\bm{X} \in \mathbb{R}^{n \times m}$ with $n$ snapshots of dimensions $m$, POD seeks a low-dimensional representation by solving an eigenvalue problem for the temporal correlation matrix $\bm{C} = \frac{1}{n} \bm{X} \bm{X}^{\top}$. The resulting eigenvectors correspond to temporal coefficients, while the spatial structures (POD modes) are obtained by projecting the data onto these coefficients. Mathematically, the POD basis functions are defined as $\bm{\phi}_{i} = \frac{1}{\sqrt{\lambda_i}} \bm{X}^{\top} \bm{v}_{i}$, where $\lambda_i$ are the eigenvalues representing the energy captured by each mode and $\bm{v}_{i}$ are the eigenvectors of $\bm{C}$. The modes are ordered by their associated eigenvalues, with the first mode capturing the maximum energy in the system \citep{POD}.

In the context of our spatial analysis, POD is employed to identify the most energetic modes of the RBC fields. Let $\bm{X} \in \mathbb{R}^{n \times n_{x} \times n_{z} \times d}$ denote the training dataset, where $n$ is the number of time steps, $n_{x}$ and $n_{z}$ are the spatial dimensions, and $d$ represents the four RBC variables. The data are first reshaped into a matrix of size $n \times (n_x n_z d)$, and the POD decomposition yields spatial modes that capture the dominant energy-containing structures. These modes provide a compact representation of the flow dynamics, reducing the complexity of the system while retaining its essential spatiotemporal behavior.

\subsection{Temporal}

\subsubsection{Metrics of Assessment}

The first metric of assessment is the deviation with respect to the governing PDE \eqref{eq:rbc},
\begin{gather*}
    \text{MSE}_{\text{RBC}} = \frac{1}{4} \left(\mathcal{L}^{\text{PDE}}_{\text{Mass}} + \mathcal{L}^{\text{PDE}}_{\text{Mom-}u} + \mathcal{L}^{\text{PDE}}_{\text{Mom-}w} + \mathcal{L}^{\text{PDE}}_{\text{Energy}} \right),
\end{gather*}
where $\mathcal{L}^{\text{PDE}}_{(\cdot)}$ is as defined in equation (10) of the main manuscript. The reasoning behind using an unweighted $\text{MSE}_{\text{RBC}}$ is similar to that behind using an unweighted $\text{MSE}_{\text{space}}$. That is, a weighted $\text{MSE}_{\text{RBC}}$ for evaluation could obscure poor reconstruction performance in some physical equations by compensating with better performance in others. Since a forecast with a low $\text{MSE}_{\text{RBC}}$ does \textit{not} guarantee that it is Rayleigh-B\'enard convection (e.g., trivial solutions), we assess the effectiveness of our PI-CRNN model by also considering some key physical properties of Rayleigh-B\'enard convection that should be at least loosely preserved. 

The three physical metrics used to assess the forecast's statistical adherence to RBC are the Nusselt number, the probability density functions of the velocity component and temperature, and the dissipation. Below we expand on each one to provide further context on the physical meaning of each.

\subsubsection*{Nusselt number}

One of the most fundamental quantities used to characterize RBC is the Nu number. This dimensionless quantity assesses the efficiency of heat transfer due to convective motion relative to the purely conductive case. In the absence of fluid motion, all heat transfer occurs via conduction, and $\text{Nu}=1$. As gravity-driven convective plumes emerge and intensify, they transport additional heat leading to $\text{Nu} > 1$ \citep{verma2018physics}. For this data Nu $\approx31$, which, given the choices of Ra and Pr numbers, describes the behavior of air turbulence between $6^{\circ}$C and $25^{\circ}$C.  Physically, $\text{Nu}$ reflects how much more effectively the system transports heat due to fluid motion compared to conduction alone. In the context of RBC, where the fluid is heated from below and cooled from above, convective plumes carry warm fluid upwards and cold fluid downwards, enhancing the net vertical heat flux. As such, $\text{Nu}$ provides a global indicator of the intensity and structure of convective dynamics, making it a commonly used assessment metric \citep{paul2011transition, pandey2021robust, blass2021flow}.

Formally, it is defined as
\begin{gather*}
    \text{Nu} = 1 + \sqrt{\text{PrRa}} \left\{ \frac{1}{T \cdot n_{x}} \sum_{t,z} w_{t}(x,0.5)\left[ \theta_{t}(x,0.5) - \overline{\theta} \right]  \right\},  \\
    \overline{\theta} = \frac{1}{T \cdot n_{x}} \sum_{t,x} \theta_{t}(x,0.5),
\end{gather*}
and we follow established conventions by estimating $\text{Nu}$ at $z=0.5$, i.e., the mid-point of the vertical domain \citep{emran2012conditional}. Thus, the metric we employ is the mean squared error between the true Nusselt number Nu and the forecast's Nusselt number $\widehat{\text{Nu}}$. We quantify the uncertainty in $\widehat{\text{Nu}}$ by generating multiple forecasts (i.e., with different input sequences chosen over the testing set).

\begin{figure}[!ht]
    \centering
    \includegraphics[width=\linewidth]{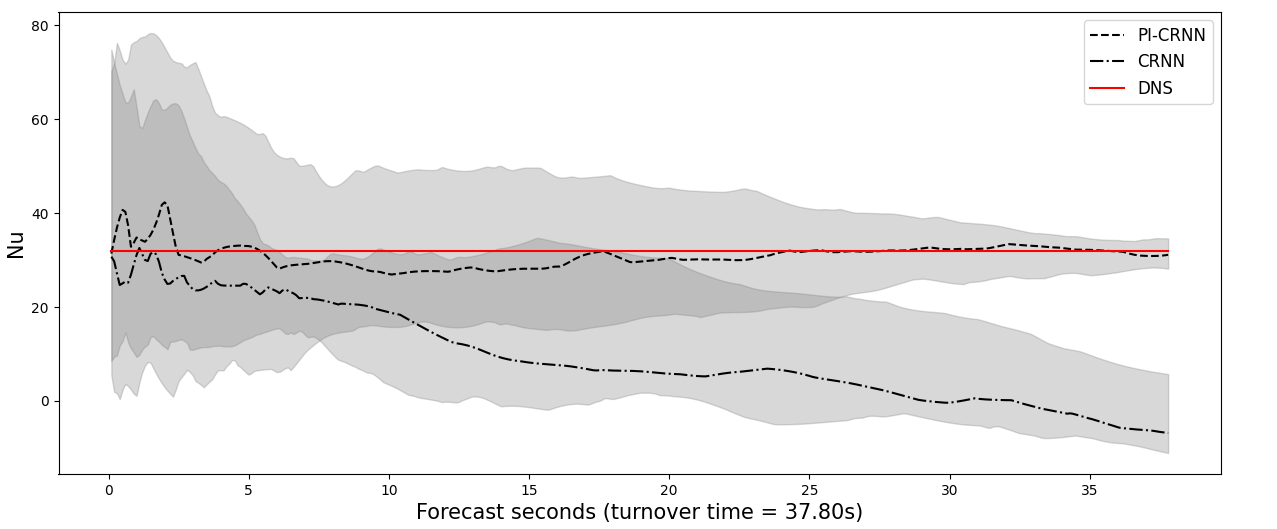}
    \caption{Nusselt number comparison between PI-CRNN and CRNN. The horizontal line is the true Nu estimated over the entire dataset. The dotted lines, identifiable in the legend, show the cumulative Nu of the forecasts and the data \textit{up to that point in time}. The bands represent the IQR. The mean squared error for PI-CRNN and CRNN is 22.11 and 2,544.28, respectively.}
    \label{fig:S5_Nusselt}
\end{figure}

Figure \ref{fig:S5_Nusselt} shows a comparison between the cumulative Nusselt number (in time) estimated from the PI-CRNN and CRNN forecasts, along with the overall Nusselt number, estimated from averaging over the entire time series of DNS observations. The bands around each represent the IQR with respect to the ensemble of forecasts as defined in Section 5.3 of the main text. The PI-CRNN forecast's Nu demonstrates good agreement with that of the DNS, whereas CRNN diverges to an unphysical value, since Nu is strictly positive by definition. Nonetheless, the rate at which the IQR on PI-CRNN shrinks is considerably slower than that of DNS. This is primarily driven by small-scale deviations from a perfectly physical forecast in the early stages of the forecast, which progressively propagate to larger scales. Indeed, the internal variability of PI-CRNN, which is tied to the initial input sequence, highlights RBC's significant sensitivity to infinitesimal changes in the initial conditions.

\subsubsection*{Probability density function}

PDFs are another widely used metric in the analysis of RBC to characterize the statistical properties of turbulent flow and temperature fields. PDFs in fact provide valuable insights into the nature of heat and momentum transport across different regions of the domain. For example, the PDF of the vertical velocity $w_{t}$ under turbulent regimes is wide and non-Gaussian (suggesting frequent mixing between hot and cold plumes), as opposed to laminar (non-turbulent) regimes when the PDF tends to be Gaussian \citep{emran2012conditional, data_driven_RBC}. Moreover, comparing PDFs between true and predicted fields is a powerful diagnostic for evaluating whether the forecast faithfully captures the underlying physical variability of RBC, including extreme temperature events \citep{blass2021flow}.

Since a closed-form expression for these PDFs does not exist, we implement kernel density estimation with a Gaussian kernel. The PDF is formally estimated as
\begin{gather*}
    \hat{f}(u) = \frac{1}{Rn_{x} \sqrt{2 \pi}} \sum_{t=1}^{R} \sum_{x \in G_{x}} \ \exp \left( - \frac{\left[u-\widehat{u}_{n+t}(x,0.5) \right]^{2}}{2h_{u}^{2}} \right), \\
    h_{u}=\hat{\sigma}_{u}(Rn_{x})^{-1/5}, \qquad
    \hat{\sigma}_{u} = \sqrt{\text{Var}(\widehat{u}_{n+t}(x,0.5))},
\end{gather*}
where $h_{u}$ is chosen according to Scott's rule and $G_{x}$ is the original grid defined in Section \ref{sec:governing_pde} and $R = 378$ \citep{scott_rule}. The procedure for $\hat{f}(w)$ and $\hat{f}(\theta)$ is similar. We denote the true PDFs by $f(u), f(w)$, and $f(\theta)$. The PDF for the pressure is generally not relevant in incompressible flows (such as this RBC) so we do not consider it \citep{siggia1994}. 

We assess the discrepancy in the forecast's PDFs using the KL divergence, which measures the ``distance" between two distributions \citep{kullback1951information}. Formally, we compute
\begin{gather*}
    \text{KL}_{u} = \int_{-\infty}^{\infty} f(u) \log\left(\frac{f(u)}{\hat{f}(u)} \right),
\end{gather*}
and $\text{KL}_{w}$ and $\text{KL}_{\theta}$ are similar. So the metric of assessment is the average KL divergence across the three variables of interest,
\begin{gather*}
    \text{KL} = \frac{1}{3} \left( \text{KL}_{u} + \text{KL}_{w} + \text{KL}_{\theta}\right). \label{eq:metric_kl}
\end{gather*}

\begin{figure}[!ht]
    \centering
    \includegraphics[width=\linewidth]{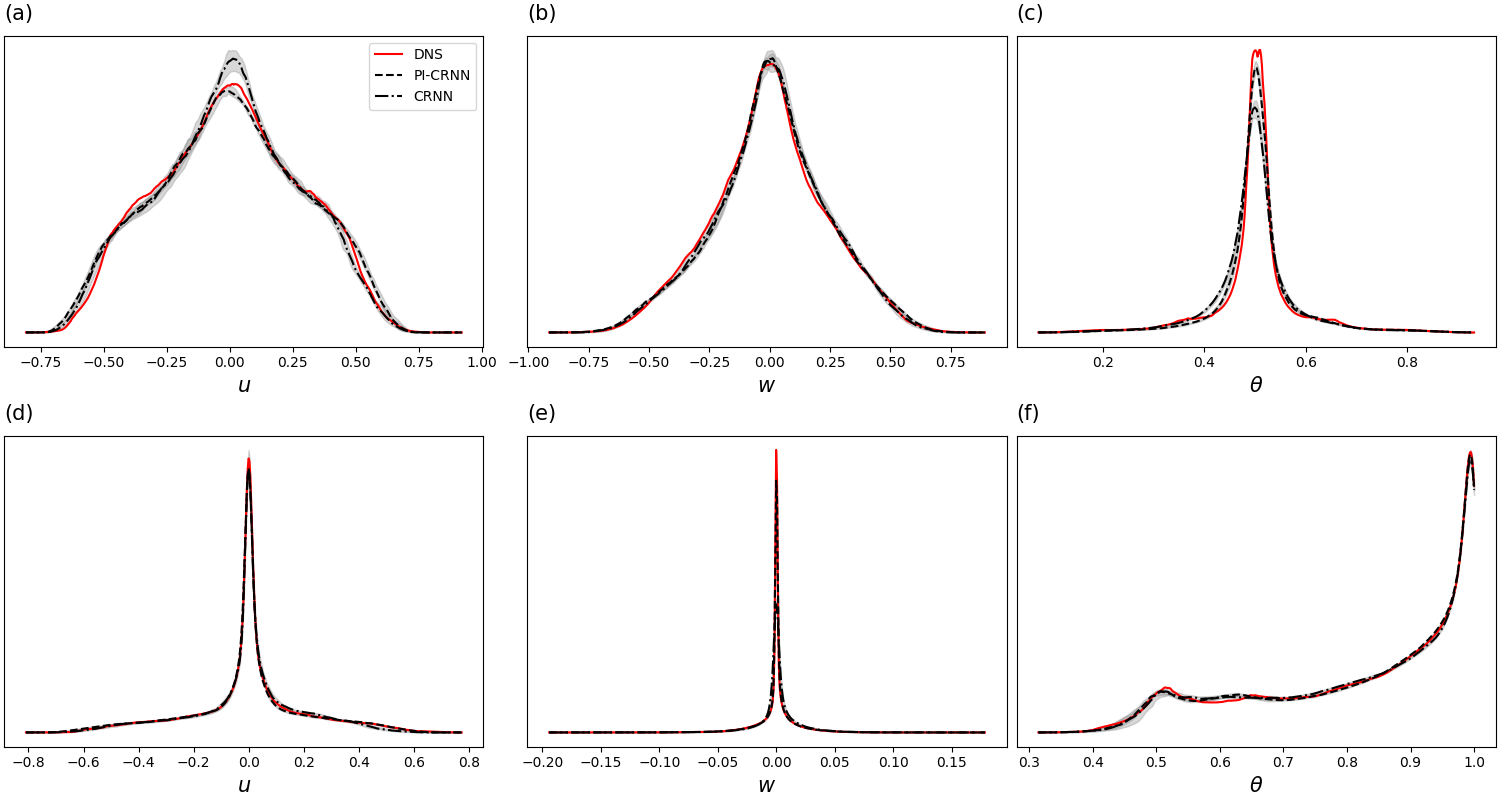}
    \caption{PDF comparison between PI-CRNN, CRNN, and and DNS in the bulk, in panels (a)-(c), and the bottom boundary layer, in panels (d)-(f). In both cases, the bands represent the IQR. For PI-CRNN, the KLs for $u_{t}$, $w_{t}$, and $\theta_{t}$ are 2.39, 1.33, and 11.45, respectively. For CRNN, the KLs are 1.94, 2.28, and 26.25, respectively.}
    \label{fig:S6_pdf}
\end{figure}

Figure \ref{fig:S6_pdf} shows the comparison of PDFs for the velocity components and temperature in the bulk and the boundary layer, defined as $\frac{1}{2\text{Nu}} < z < 1-\frac{1}{2\text{Nu}}$ and $z < \frac{1}{2\text{Nu}}$, respectively \citep{GROSSMANN_LOHSE_2000}. Specifically, panels (a)-(c) show the PDF comparison for $\widehat{u}_{t}$, $\widehat{w}_{t}$, and $\widehat{\theta}_{t}$ in the bulk, while panels (d)-(f) show the same for the boundary layer. The boundary layer PDFs show good agreement with ground truth, with the largest discrepancy being exhibited by the temperature, in panel (f). Panels (a-c) show that PI-CRNN is able to retain most of the flow characteristics in the bulk. Specifically, the PDF of $\widehat{u}_{t}$ is centered around 0 and the PDF of $\widehat{\theta}_{t}$ is centered around 0.5 (recall, nondimensional temperature is constant and equal to 0 and 1 at the top and bottom of the spatial domain, respectively). The PDF of the wall-normal velocity, $\widehat{w}_{t}$, is (expectedly) non-Gaussian, which is a result of mixing between hot and cold plumes in the middle region of the spatial domain. As such, the PDF estimated by PI-CRNN captures distributional characteristics well, but cannot optimally reproduce the true PDE of $\widehat{w}_{t}$.

\subsubsection*{Scalar dissipation}

The dissipation of thermal fluctuations reflects how turbulence and diffusion jointly act to smooth out temperature differences within the fluid. In RBC at high Rayleigh numbers, such as the data used for this work, plumes transport heat nonlinearly through the bulk (away from the plates). Over time, thermal diffusion smooths out sharp temperature contrasts, whose effect is magnified at smaller spatial scales where fluctuations are strongest \citep{emran2008fine}. This metric complements global statistics like Nu and PDFs by focusing not just on the net transport of heat or its distribution, but also on how the fine structure of the temperature field evolves \citep{emran2008fine, schumacher2007sub}. As such, the dissipation is a much stricter benchmark, since it depends on accurately capturing small-scale gradients, and it presents a more difficult challenge for any emulator to capture.

For a given $t$, the dissipation is formally defined as
\begin{gather*}
    \varepsilon^{\theta}_{t}(x,z) = \kappa \left[\left( \frac{\partial \theta^{*}_{t}(x,z)}{\partial x} \right)^{2} + \left( \frac{\partial \theta^{*}_{t}(x,z)}{\partial z} \right)^{2} \right],
\end{gather*}
where, recall, $\theta^{*}_{t}(x,z)$ denotes \textit{dimensional} temperature and $\kappa$ is the (known \textit{a priori}) thermal diffusivity parameter. The turbulence exhibited in Rayleigh-B\'enard convection with $10^{7} < \text{Ra} < 10^{9}$ (as is the case in this data), is \textit{fully developed and chaotic}. For reasons similar to those for Nusselt number, we focus our analysis of this metric in the center of the spatial domain, where turbulent mixing is prevalent \citep{emran2012conditional}.

\begin{figure}[!ht]
    \centering
    \includegraphics[width=\linewidth]{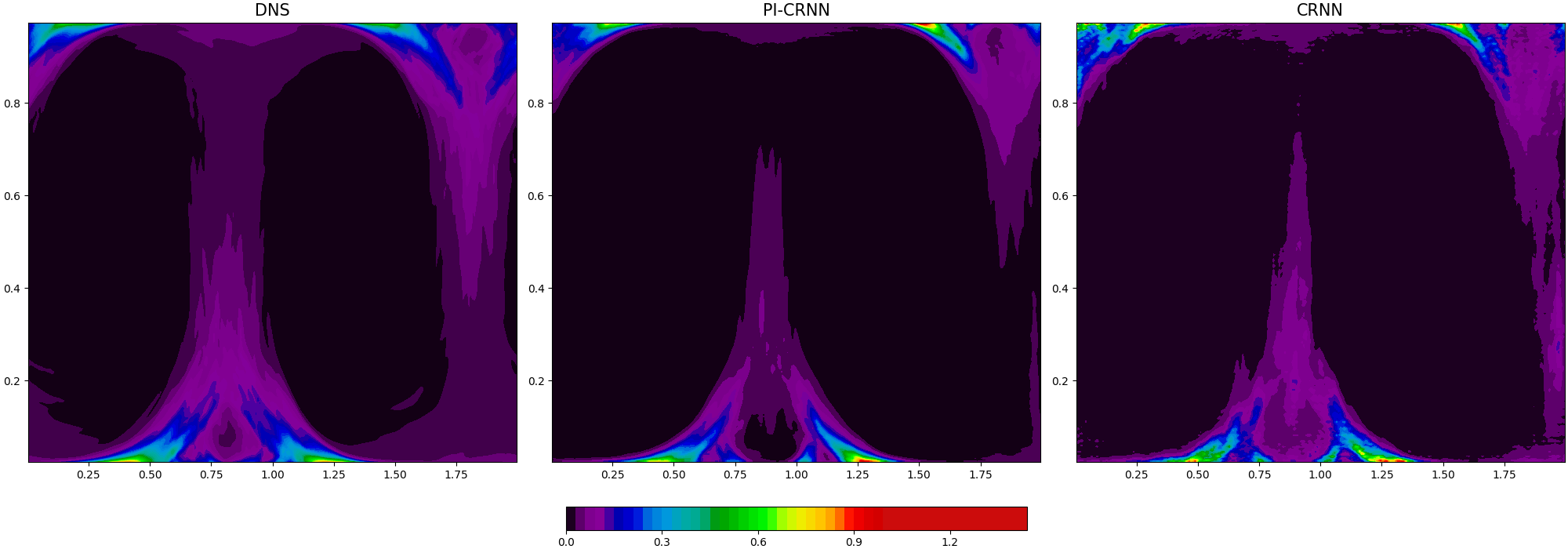}
    \caption{Dissipation rate of thermal fluctuations comparison between DNS, PI-CRNN, and CRNN, all averaged over time.  The mean squared error for PI-CRNN and CRNN is 3.55 $\times 10^{-3}$ and 4.89 $\times 10^{-3}$, respectively.}
    \label{fig:S7_dissipation}
\end{figure}

Figure \ref{fig:S7_dissipation} shows a comparison of the dissipation rate of thermal fluctuations among DNS, PI-CRNN, and CRNN. Since it is a much stricter metric, for the aforementioned reasons, PI-CRNN suboptimally captures the spatial variability of the dissipation, as the difference between the data and the forecast is evident, particularly in regions of high-gradient activity where the model tends to underestimate the intensity of dissipation structures. This discrepancy suggests that while the PI-CRNN is capable of learning large-scale dynamics such as Nu and the PDFs, it struggles to accurately reproduce small-scale, highly intermittent features that contribute significantly to dissipation. The dissipation figure for CRNN is considerably fragmented, showing thermal fluctuations in regions of the spatial domain where they should not be. This is almost exclusively driven by the marked deviation of CRNN from physical reality, as seen in the Nusselt number results of Figure \ref{fig:S5_Nusselt}.

\subsubsection{Forecast reliability}

In highly turbulent systems such as RBC, evaluating a surrogate model purely on a time-step basis can be misleading. Because RBC exhibits strong sensitivity to initial conditions, trajectories starting from nearly identical states can diverge rapidly, so the system' states at individual time steps becomes increasingly unpredictable over longer horizons \citep{lorenz1963, siggia1994}. Consequently, metrics based on individual time steps fail to capture the surrogate’s ability to reproduce the underlying statistical and structural properties of the flow. Instead, it is more appropriate to assess the model using physically meaningful timescales, such as the turnover time, which reflects the natural evolution of coherent structures in RBC. Below we reformulate the notion of turnover time in a statistical sense and formalize the reliability window of our model.

The turnover time $\tau_{c}=H / U$, where $H$ and $U$ are as defined in Section \ref{sec:governing_pde}, is a characteristic timescale that estimates how long it takes for a fluid parcel to travel the height of the spatial domain \citep{siggia1994}. Statistically, $\tau_{c}$ represents the amount of time over which the flow at a fixed point in space remains significantly correlated with itself. So observations that are at least $\tau_{c}$ seconds apart may be approximately considered independent.

We evaluate how many turnover times the model can successfully generate while maintaining physically consistent dynamics. Successfully forecasting multiple turnover times demonstrates that the model can effectively `reproduce itself' well beyond the training output window (the model was trained to predict $a=60$ observations, approximately corresponding to one turnover time). Indeed, beyond the output window of length $a$ the flow's structures are no longer \textit{physically} constrained by the input sequence (of $b=80$ observations), yet the model continues to produce physically consistent behavior. Using turnover time as a benchmark thus allows us to emphasize the model’s ability to capture the essential statistical and structural features of RBC, rather than focusing on exact pointwise accuracy, which is particularly appropriate given the system’s inherent chaotic dynamics.

To assess how far into the future the model yields physically reliable predictions, we propose an ensemble-based validation analysis using mass conservation as a criterion. In RBC the spatially averaged velocities should remain zero at all times, reflecting the incompressibility of the flow and periodicity of the domain. Monitoring the global means provides a natural diagnostic: if the forecasts remain consistently near zero, the dynamics are preserved, whereas a systematic drift signals that the model is effectively `leaking mass' and the forecast has become \textit{unphysical}. This criterion offers a physically interpretable way to determine the time horizon over which the forecasts remain trustworthy.

\begin{figure}[!ht]
    \centering
    \includegraphics[width=\linewidth]{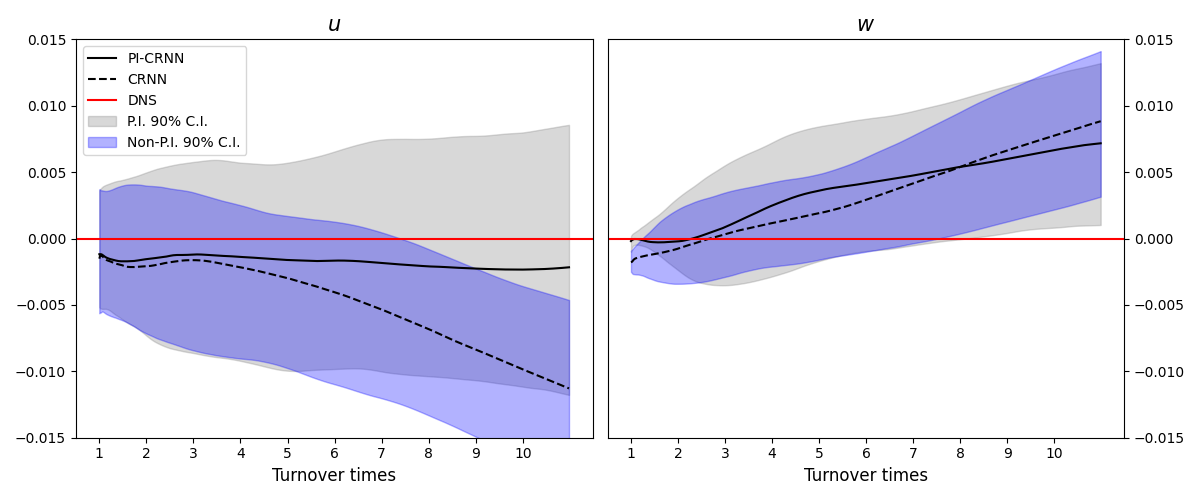}
    \caption{Cumulative spatial means of the horizontal ($u$) and vertical ($w$) velocity components over ensemble forecasts. The shaded regions indicate the 90\% empirical confidence intervals across ensemble members. Once either confidence interval no longer contains zero, the drift is considered statistically significant (i.e., violation of mass conservation) and the forecast is no longer considered reliable.}
    \label{fig:S8_reliability}
\end{figure}

Specifically, we selected multiple input sequences (all of length $b$) over the validation set and generated an ensemble of $E$ forecasts, each one of $\tau_c=10$ turnover times ($R=10\tau_{c} / \Delta_{t}=540$ time steps). For each forecast trajectory $e$, $e=1,\ldots,E$, we computed the cumulative spatial means of the two velocity components. We chose this approach to assess whether the model correctly captures the long-term statistical properties of the system. Denote by $\hat{u}_{e,r}$ and $\hat{w}_{e,r}$ the spatial average of the $e$-th ensemble member at the $r$-th forecast step, $r=1,\ldots,R$, for the horizontal and vertical velocity components, respectively. We obtain
\begin{subequations} \label{eq:uw_means}
\begin{gather}
    \overline{\bm{u}} = \left(\bm{\overline{u}}_{1}, \ldots, \bm{\overline{u}}_{R} \right), \qquad \overline{\bm{w}} = \left(\bm{\overline{w}}_{1}, \ldots, \bm{\overline{w}}_{R} \right),\\
    \overline{\bm{u}}_{t} = \left(\overline{u}_{1,r}, \ldots, \overline{u}_{E,r} \right)^{\top}, \qquad \overline{\bm{w}}_{r} = \left(\overline{w}_{1,r}, \ldots, \overline{w}_{E,r} \right)^{\top}, \\
    \overline{u}_{e,r} = \frac{1}{r} \sum_{i=1}^{r} \hat{u}_{e,i}, \qquad \overline{w}_{e,r} = \frac{1}{r} \sum_{i=1}^{r} \hat{w}_{e,i}.
\end{gather}
\end{subequations}
From the matrices $\bm{\overline{u}}, \bm{\overline{w}} \in \mathbb{R}^{E \times R}$, we estimate the column-wise 5-th and 95-th percentiles, i.e., the 90\% empirical confidence intervals for $\overline{u}_{e,r}$ and $\overline{u}_{e,r}$ at each forecast step $r$. We regard the forecast as reliable for as many turnover times as both intervals simultaneously contain zero throughout the entire turnover window, leading us to conclude that the forecast is physically reliable for \textit{seven turnover times}. Figure \ref{fig:S8_reliability} shows visual evidence for our conclusion and a comparison with the non-physics-informed version (CRNN) of our model.

\subsubsection*{Forecast Fidelity with Respect to DNS}

\begin{figure}[!ht]
    \centering
    \includegraphics[width=\linewidth]{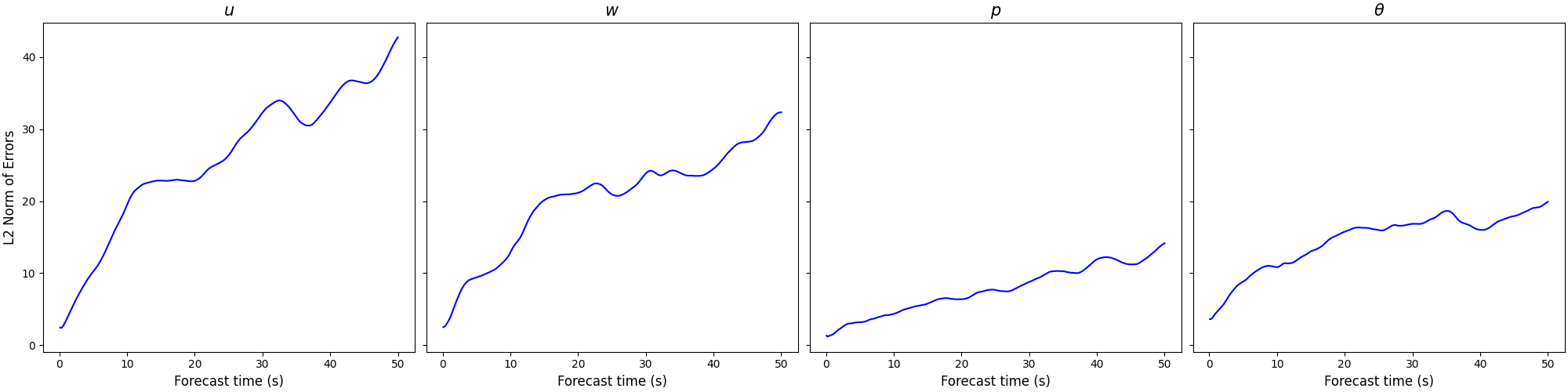}
    \caption{Time evolution of the L2 error between the forecasts and DNS. Each panel shows the error for each variable as a function of time, quantifying how deviations accumulate as the forecast horizon extends. This metric provides a direct assessment of trajectory-level accuracy, complementing the physical metrics of Figures S5-S7.}
    \label{fig:S9_dns_error}
\end{figure}

Figure \ref{fig:S9_dns_error} quantifies the forecast fidelity of PI-CRNN with respect to the DNS using a spatial L2 norm for each variable \citep{siam_survey}. In the case of $u$, formally,
\begin{gather*}
    \mathrm{L2}(u) = \sqrt{\frac{1}{N_x N_z} \sum_{i=1}^{N_x} \sum_{j=1}^{N_z} \left( u_{i,j}(t) - \hat{u}_{i,j}(t) \right)^2 },
\end{gather*}
and similarly for $w$, $p$, and $\theta$. As expected, the errors grow over time, reflecting the increasing difficulty of accurately predicting the full flow state at longer horizons. However, even as pointwise deviations accumulate, progressively amplifying small-scale errors, the forecast may follow alternative trajectories that are physically plausible, still capturing the essential spatiotemporal structures of RBC, as shown in Figures S5-S7. This perspective complements the turnover-time-based analysis in the main text, providing a direct, quantitative measure of trajectory-level accuracy while recognizing the inherent non-uniqueness of physically consistent solutions in nonlinear flows.

\subsubsection{Gaussian Diagnostics}

\begin{figure}[!ht]
    \centering
    \includegraphics[width=\linewidth]{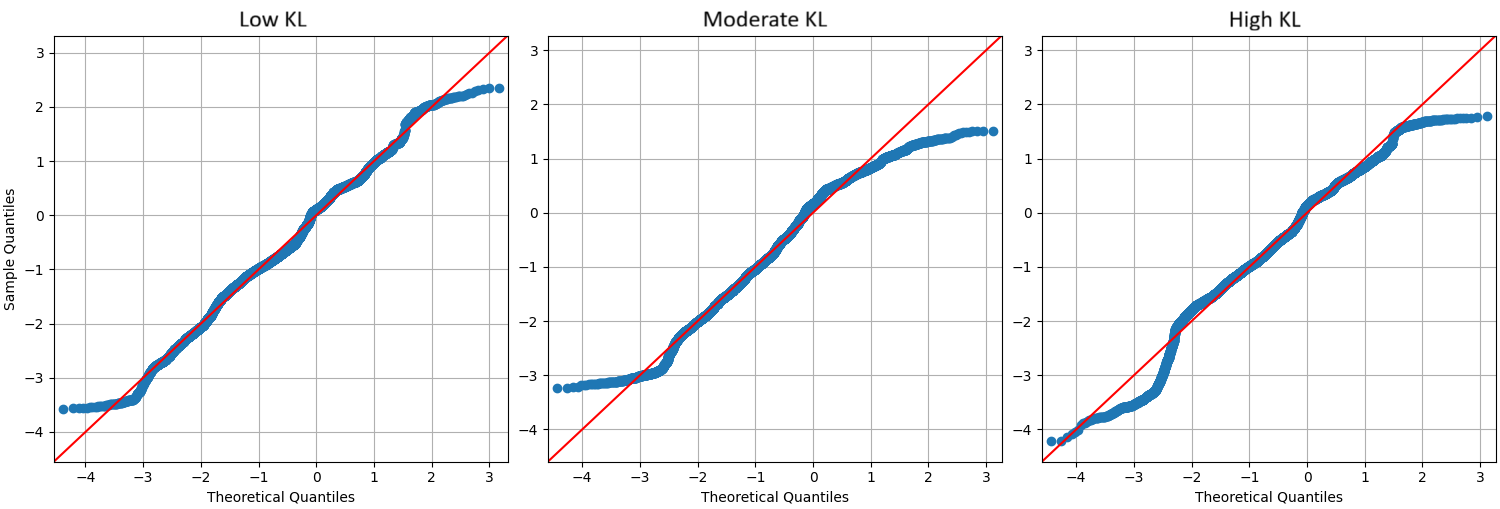}
    \caption{Q-Q plots for reduced-space entries corresponding to low, moderate, and high KL divergence between standardized residuals and a standard normal distribution. The three KL divergences were taken from each tercile. The residuals exhibit approximately Gaussian behavior, though not always centered around zero.}
    \label{fig:s10_qq_rnn}
\end{figure}

We verify the distributional assumptions in (1b), i.e., of $\tilde{\bm{\epsilon}}_{t}$, using the reduce-space residuals over the test set,
\begin{align*}
    \tilde{\bm{e}}_{t} &= \tilde{\bm{Y}}_{t} - \widehat{\tilde{\bm{Y}}}_{t} \\ 
    &= \mathcal{SE}\left(\bm{Y}_{t} \mid \widehat{\bm{W}}_{\text{en}}\right) - \mathcal{T}\left( \mathcal{SE}\left( \mathcal{Y}^{-}_{t;b}  \mid \widehat{\bm{W}}_{\text{en}}\right) \mid \widehat{\bm{W}}^{\text{time}} \right),~ t=n+b, \ldots, T-a,
\end{align*}
where $\mathcal{Y}^{-}_{t;b} = \left(\bm{Y}_{t-b}, \bm{Y}_{t-b+1}, \ldots, \bm{Y}_{t-1} \right)$, $\tilde{\bm{e}}_{t} \in \mathbb{R}^{\tilde{n}_{x} \times \tilde{n}_{z} \times \tilde{d}}$, $\tilde{n}_{x}=\tilde{n}_{z}=16$ and $\tilde{d}=64$, $a=60$, and $b=80$. We only show representative Q-Q plots in Figure \ref{fig:s10_qq_rnn}, as the projection onto a lower-dimensional space removes many of the visual references present in the full high-dimensional field. In fact, we construct Q-Q plots for each reduced grid cell $(i,j)$, $i = 1,\ldots,\tilde{n}_{x}$ and $j=1,\ldots,\tilde{n}_{z}$, by aggregating residuals across the feature channels. In other words, we examine whether the model errors are Gaussian in the reduced space while accounting for multiple factors. The Q-Q plots in Figure \ref{fig:s10_qq_rnn}, also spanning low, intermediate, and high KL divergence, provide visual evidence that the Gaussian assumption on $\tilde{\bm{\epsilon}}_{t}$ is approximately correct. The median KL divergence (taken across the $\tilde{n}_{x} \times \tilde{n}_{z}$ reduced-space grid) is 0.24 with an IQR of 0.02.

\subsubsection{Other Temporal Methods} \label{sec:temporal_supp}

The temporal methods used to compare performance with PI-CRNN are ARIMA, PI-ESN, CRNN. Here we expand on each one to provide further context.

\subsubsection*{ARIMA}

ARIMA is a time series model that captures autocorrelations within a univariate sequence and is often used because of its simplicity \citep{brockwell2016introduction}. It combines three components: autoregression (AR), differencing (I) to ensure stationarity, and a moving average (MA) of past errors. An ARIMA $(p,d,q)$ model forecasts a value based on its past values (lags up to order $p$), differenced values (to degree $d$), and past forecast errors (lags up to order $q$). We implemented ARIMA for every reduced space spatial location for each of the reduced space feature maps. 

\subsubsection*{Echo State Network}

An ESN is a type of RNN designed to efficiently process temporal data while mitigating the difficulties associated with training traditional RNNs \citep{jaeger2001}. The ESN consists of a reservoir $\bm{W}$ and a state matrix $\bm{H}_{t} \in \mathbb{R}^{N}$, where $N$ is the number of neurons. The state is updated based on the previous state and a nonlinear transformation of the input signal $\tilde{\bm{Y}}_{t-1}$, which is mapped through an input weight matrix $\mathbf{W}_{\text{in}}$ as follows:
\begin{gather*}
    \bm{H}_{t} = g\left( \bm{W}_{\text{in}} \tilde{\bm{Y}}_{t-1} + \bm{W} \bm{H}_{t-1} \right).
\end{gather*}
Crucially, each entry of both $\bm{W}_{\text{in}}$ and $\bm{W}$ are realizations from a spike-and-slab prior and \textit{kept fixed during training}, providing significant computational relief compared to other RNNs \citep{mitchell1988bayesian}. The network's output $\widehat{\tilde{\bm{Y}}}_{t} \in \mathbb{R}^{m}$ is computed via
\begin{gather*}
   \widehat{\tilde{\bm{Y}}}_{t} = \mathbf{W}_{\text{out}} \, [\bm{H}_{t}; \tilde{\bm{Y}}_{t-1}] + \mathbf{b},
\end{gather*}
where $[\bm{H}_{t}; \tilde{\bm{Y}}_{t-1}]$ denotes the concatenation of the state and the current input, $\mathbf{W}_{\text{out}}$ is a trainable weight matrix, and $\mathbf{b}$ is a bias vector.

\subsubsection*{Visual Comparison with Second Best Model (CRNN)}

\begin{figure}[!ht]
    \centering
    \textbf{PI-CRNN}
    \vspace{5pt}
    \includegraphics[width=\linewidth]{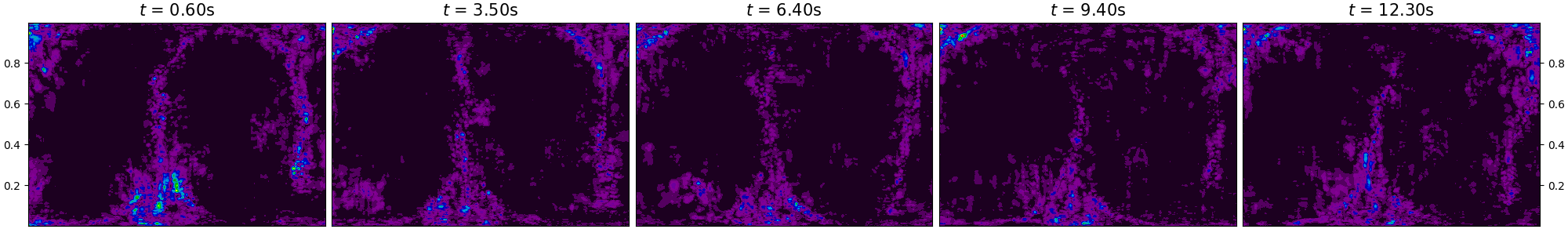} \\
    \textbf{CRNN}
    \includegraphics[width=\linewidth]{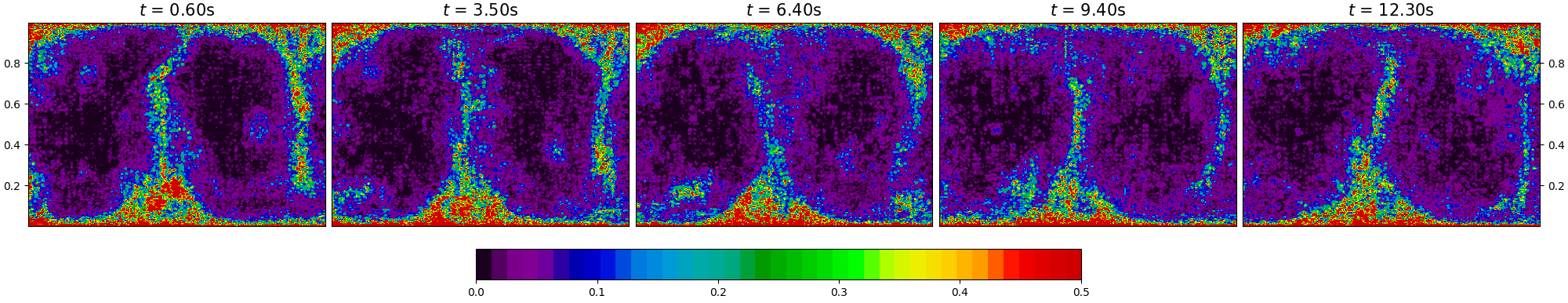} 
    \caption{Results comparison between PI-CRNN and the second best model, CRNN, in terms of the discrepancy with respect to the governing PDE, MSE$_{\text{RBC}}$.}
    \label{fig:s11_pde_comparison}
\end{figure}

Figure \ref{fig:s11_pde_comparison} provide additional visual evidence underscoring the superior performance of PI-CRNN compared to the second-best method, CRNN, i.e., the purely data-driven counterpart to our physics-informed approach. Specifically, Figure \ref{fig:s11_pde_comparison} shows the fidelity of the predictions with respect to the underlying physics, containing the corresponding errors based on the MSE$_{\text{RBC}}$ metric. The advantage of PI-CRNN is especially pronounced in this figure, which reveals that CRNN, lacking any physics in the inferential process, deviates considerably from the governing PDE. This difference persists consistently across the entire forecast, highlighting the critical role of physics-based regularization in ensuring long-term physical plausibility.

\subsubsection*{Simplified Single-Step Temporal Model}

We show the results for a model with a convolutional autoencoder whose dynamics is assumed to be autoregressive and that predicts only a single future timestep at each iteration. There is only one ConvLSTM temporal component requiring (every time) the entire input sequence $\tilde{\mathcal{Y}}^{-}_{t;b}$ to predict $\tilde{\bm{Y}}_{t}$. The temporal model in PI-CRNN instead processes the input sequence, via the context-builder, \textit{once} and the sequence-generator predicts an arbitrary number of future time steps. The approach may be summarized as
\begin{gather*}
    \overbrace{\mathcal{Y}^{-}_{t;b} \xrightarrow{\shortstack{\text{Spatial} \\ \text{Encoder}}} \tilde{\mathcal{Y}}^{-}_{t;b}  \xrightarrow{\shortstack{\text{Temporal} \\ \text{Model}}} \tilde{\bm{Y}}_{t} \xrightarrow{\shortstack{\text{Spatial} \\ \text{Decoder}}} \bm{Y}_{t}}^{ \text{Autoregressive Spatiotemporal Model}}.
\end{gather*}

\begin{figure}[!ht]
    \centering
    \includegraphics[width=\linewidth]{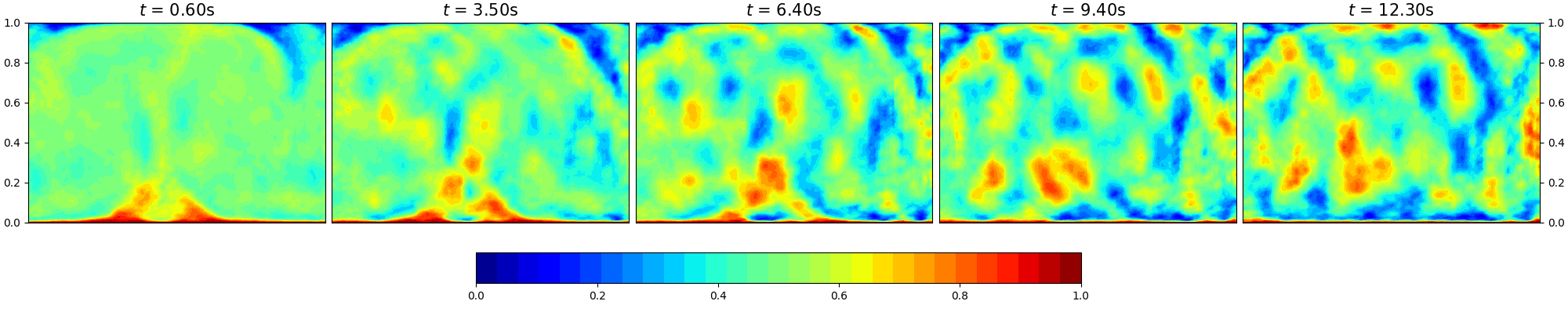}
    \includegraphics[width=\linewidth]{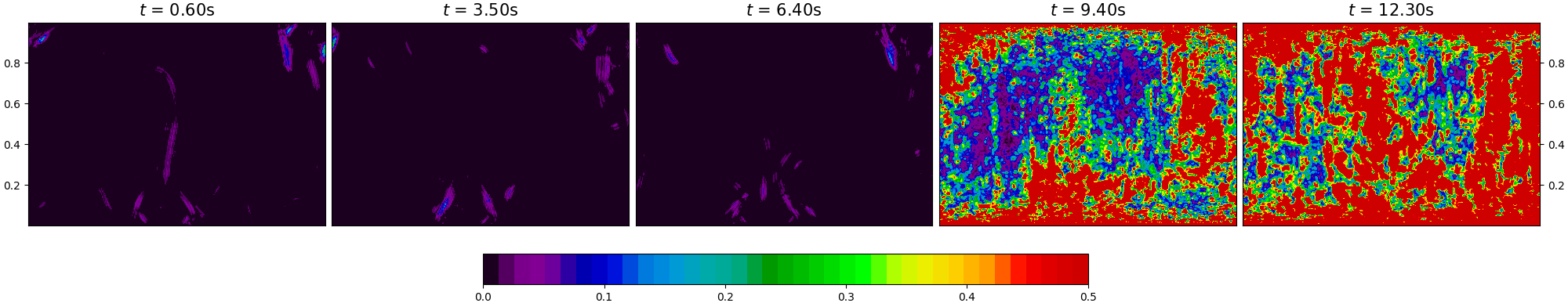}
    \caption{Temperature forecast (top row) and PDE squared residuals (bottom row) of RBC using an autoregressive model that perform iterative forecasting. The model recursively feeds its own predictions as input during inference, resulting in accumulated errors and degraded physical performance over time, particularly in capturing long-term turbulent dynamics.}
    \label{fig:s12_snapshots_bad}
\end{figure}

Figure \ref{fig:s12_snapshots_bad} presents the temperature forecast generated by the autoregressive model, using the same time steps as in Figure 3 of the manuscript. The model employs a single temporal component that predicts only the next time step, effectively neglecting long-term dynamics. During inference, the model requires an input sequence, so predictions are performed recursively. This iterative strategy has modeling and practical limitations, especially for turbulent convection. Specifically, it suboptimally captures the long-term temporal structure of RBC, and the repeated use of the model’s own predictions leads to rapid error accumulation and nonphysical predictions. The discrepancy with respect to the PDE, i.e., $\text{MSE}_{\text{RBC}}$ as in the first column of Table 2 of the manuscript, is $32.05$, three orders of magnitude larger than that of PI-CRNN. 

\subsubsection{Three Dimensional Case}

To further demonstrate the utility of the approach in more computationally intensive settings, we have performed additional experiments on a three-dimensional variant of RBC. Importantly, the 3D simulations, still obtained via DNS, were performed \textit{at the same Rayleigh number} as the 2D, i.e., the strength of the buoyancy forcing is the same. These 3D experiments require \textit{significantly} greater computational effort and more closely resemble real-world turbulent convection studies. Surrogate models for DNS of 3D RBC at high Ra are rare, as most previous studies have focused on significantly lower regimes, typically around $\text{Ra} < 10^{6}$, where the flow is better described as unsteady laminar flow rather than fully turbulent. To our knowledge, the closest related work is the recent study by \citet{Fromme2025}, which employed an equivariant autoencoder architecture for 3D RBC at moderate Rayleigh numbers.

Below we highlight the differences of the three dimensional model with respect to the two dimensional case and we present the results.

\subsubsection*{Methods}

In the 3D case the convolution operation in the spatial encoder, context builder, sequence generator, and spatial decoder is performed using $3 \times 3$ kernels. To avoid redundancy, here we only give the formal definitions of the three dimensional convolution and deconvolution.

The three-dimensional spatial encoder consists of $\ell=1, \ldots, L$ convolutional layers with $K^{\ell}$ kernels in the $\ell$-th layer, where each kernel is a three-dimensional matrix $r^{\ell}_{x} \times r^{\ell}_{y} \times r^{\ell}_{z}$. The kernels each perform a local convolution of the input every $s=2$ elements (\textit{strides}) and are summed across the kernels from the $(\ell-1)$-th layer. As such, the $\ell$-th convolutional layer with $K^{\ell}$ kernels and stride $s$ maps yields $K^{\ell}$ \textit{feature maps}, each one of dimensions $n_{x}^{\ell} \times n_{y}^{\ell} \times n_{z}^{\ell}$, i.e., a four-dimensional matrix $n_{x}^{\ell} \times n_{y}^{\ell} \times n_{z}^{\ell} \times K^{\ell}$, where $n_{x}^{\ell} = \lfloor n_{x}^{\ell-1} / s \rfloor$, $n_{y}^{\ell} = \lfloor n_{y}^{\ell-1} / s \rfloor$, and $n_{z}^{\ell} = \lfloor n_{z}^{\ell-1} / s \rfloor$. In the $\ell$-th layer, denote by $Z^{\ell}_{i,j,k,f}$ the $(i,j,k)$ entry of the $f$-th feature map, by $w^{\ell}_{a,b,c,\tilde{f},f}$ the $(a,b,c)$ entry of the $f$-th kernel for the $\tilde{f}$-th channel of the previous layer, $\tilde{f} \in \{1,\ldots,K^{\ell-1} \}$, and by $w^{\ell}_{0,f}$ the bias term. The three-dimensional spatial encoder performs the following:
\begin{gather}
    Z^{\ell}_{i,j,k,f} = g \left( \sum_{\tilde{f}=1}^{K^{\ell-1}} \sum_{a=1}^{r_{x}^{\ell}} \sum_{b=1}^{r_{y}^{\ell}} \sum_{b=1}^{r_{z}^{\ell}} Z^{\ell-1}_{si+a,sj+b,sk+c;\tilde{f}} \cdot w^{\ell}_{a,b;\tilde{f},f} + w^{\ell}_{0,f} \right). \label{eq:convolution_3d}
\end{gather}
If we denote the weights in the $\ell$-th layer of the three-dimensional spatial encoder by 
\begin{align*}
    \bm{W}^{\ell} &= \Big\{ \left(w^{\ell}_{a,b,c,\tilde{f},f}, w^{\ell}_{0,r} \right) : a=1, \ldots, r_{x}^{\ell}, ~b=1, \ldots, r_{y}^{\ell},~c=1, \ldots, r_{z}^{\ell}, \\
    &\qquad ~\tilde{f}=1,\ldots, K^{\ell-1},~f=1,\ldots,K^{\ell} \Big\},
\end{align*}
and $\bm{Z}^{\ell} = \{Z^{\ell}_{i,j,k,f} : i = 1, \ldots, n^{\ell}_{x},~ j = 1, \ldots, n^{\ell}_{y},~ k = 1, \ldots, n^{\ell}_{z},~f=1,\ldots,K^{\ell} \}$, then, the spatial encoder $\mathcal{SE}(\cdot \mid \bm{W}_{\text{en}})$ can be written in matrix form
\begin{gather}
    \bm{Z}^{\ell} = g\left( \bm{Z}^{\ell-1} \ast \bm{W}^{\ell} \right),~ \ell=1,\ldots,L, \label{mod:spatial_encoder_3d}
\end{gather}
where $\ast$ denotes the strided ($s$) convolution operation \eqref{eq:convolution_3d}. Additionally, note that $\bm{Z}^{0}=\bm{Y}$, $\bm{Z}^{L}=\bm{\tilde{Y}}$, with $\left(n_{x}^{L}, n_{y}^{L},n_{z}^{L}, K^{L} \right) = \left(\Tilde{n}_{x}, \Tilde{n}_{y}, \Tilde{n}_{z}, \tilde{d} \right)$. So for $\ell=1$ we have $K^{0} = d$, the five RBC variables in the three-dimensional case, and for $\ell=L$ we have $K^{L} = \tilde{d}$, i.e., the number of feature maps in $\bm{\tilde{Y}}$ (the reduced space representation of $\bm{Y}$). We denote the parameters in the spatial encoder by $\bm{W}_{\text{en}} = \left\{\bm{W}^{\ell} : \ell=1,\ldots,L\right\}$ and the number of parameters to estimate  is
\begin{gather}
    \left| \bm{W}_{\text{en}} \right| = \sum_{\ell=1}^{L} \left(r_{x}^{\ell} \cdot r_{y}^{\ell} \cdot r_{z}^{\ell} \cdot K^{\ell-1} + 1 \right) \cdot K^{\ell}. \label{eq:num_pars_conv_3d}
\end{gather}

The three-dimensional spatial decoder consists of $L+1$ deconvolutional layers, which are similar to convolutional layers but \textit{increase} the number of dimensions of the feature maps. In the $\ell$-th deconvolutional layer, denote the number of kernels by  $\dot{K}^{\ell}$ and the size of each kernel by $\dot{r}_{x}^{\ell} \times \dot{r}_{y}^{\ell} \times \dot{r}_{z}^{\ell}$. In the first $L$ layers of the decoder, the number of kernels in the $\ell$-th layer is the same as the number of kernels in the $(L-\ell-1)$-th layer of the encoder, so the spatial encoder and spatial decoder ``mirror'' each other, where $( \dot{n}_{x}^{\ell}, \dot{n}_{y}^{\ell}, \dot{n}_{z}^{\ell}, \dot{K}^{\ell} ) = \left(n_{x}^{L-\ell-1}, n_{y}^{L-\ell-1}, n_{z}^{L-\ell-1}, K^{L-\ell-1} \right)$. Since the $L$-th layer of the decoder yields $K^{1}$ feature maps (i.e., the number of feature maps in the first layer of the encoder), the spatial decoder includes one final (output) layer with $\dot{K}^{L+1}=d$ kernels, so that the output matches the dimensions of the data, $n_{x} \times n_{y} \times n_{z} \times d$.  The deconvolution operation is similar to \eqref{eq:convolution_3d} except it increases the dimensions of the feature maps. Indeed, the $\ell$-th deconvolutional layer with $\dot{K}^{\ell}$ kernels and stride equal to $s$ yields a feature map of dimensions  $\dot{n}_{x}^{\ell} \times \dot{n}_{y}^{\ell} \times \dot{n}_{z}^{\ell} \times \dot{K}^{\ell}$, where $\dot{n}_{x}^{\ell} = s\dot{n}_{x}^{\ell-1}$, $\dot{n}_{y}^{\ell} = s\dot{n}_{y}^{\ell-1}$, and $\dot{n}_{z}^{\ell} = s\dot{n}_{z}^{\ell-1}$. Similar to the 2D case, the input to the spatial decoder is the $L$-th layer's output in the spatial encoder, $\bm{Z}^{L} = \bm{\tilde{Y}}$. In the $\ell$-th layer, if denote by $\dot{Z}^{\ell}_{i,j,k,f}$ the $(i,j,k)$ entry of the $f$-th feature map, by $\dot{w}^{\ell}_{a,b,c,\tilde{f},f}$ the $(a,b,c)$ entry of the $f$-th kernel for the $\tilde{f}$-th channel of the previous layer, $\tilde{f} \in \{1,\ldots,\dot{K}^{\ell-1} \}$, and by $\dot{w}^{\ell}_{0,f}$ the bias term. The spatial decoder performs the following:
\begin{gather*}
    \dot{Z}^{\ell}_{i,j,k,f} = g \left( \sum_{\tilde{f}=1}^{\dot{K}^{\ell-1}} \sum_{a=1}^{\dot{r}_{x}^{\ell}} \sum_{b=1}^{\dot{r}_{y}^{\ell}} \sum_{c=1}^{\dot{r}_{z}^{\ell}} \dot{Z}^{\ell-1}_{\lfloor \frac{i}{s} \rfloor +a,\lfloor \frac{j}{s} \rfloor+b, \lfloor \frac{k}{s} \rfloor+c;\tilde{f}} \cdot \dot{w}^{\ell}_{a,b,c,\tilde{f},f} + \dot{w}^{\ell}_{0,f} \right).
\end{gather*}
If we denote the weights in the $\ell$-th layer of the spatial decoder by 
\begin{align*}
    \dot{\bm{W}}^{\ell} &= \Big\{ \left(\dot{w}^{\ell}_{a,b,c,\tilde{f},f}, \dot{w}^{\ell}_{0,f} \right) : a=1, \ldots, \dot{r}_{x}^{\ell},~b=1, \ldots, \dot{r}_{y}^{\ell},~c=1, \ldots, \dot{r}_{z}^{\ell} \\
    &\qquad \tilde{f}=1,\ldots, K^{\ell-1},~f=1,\ldots,\dot{K}^{\ell} \Big\}
\end{align*}
and $\dot{\bm{Z}}^{\ell} = \{\dot{Z}^{\ell}_{i,j,k,f} : i = 1, \ldots, \dot{n}^{\ell}_{x},~ j = 1, \ldots, \dot{n}^{\ell}_{y},~k = 1, \ldots, \dot{n}^{\ell}_{z},~f=1,\ldots,\dot{K}^{\ell} \}$, then, the spatial decoder can be written in matrix form
\begin{subequations} 
\begin{gather*}
    \dot{\bm{Z}}^{\ell} = g\left( \dot{\bm{Z}}^{\ell-1} \dot{\ast} \dot{\bm{W}}^{\ell} \right),~ \ell=1,\ldots,L, \\
    \widehat{\bm{Y}} = g \left( \dot{\bm{Z}}^{L} \dot{\ast} \dot{\bm{W}}^{L+1} \right),
\end{gather*}
\end{subequations}
where $\dot{\ast}$ denotes the strided ($s$) deconvolution operation. Additionally, $\dot{\bm{Z}}^{0}=\bm{\tilde{Y}}$, $\dot{\bm{Z}}^{L}=\bm{Z}^{1}$, with $(\dot{n}_{x}^{L}, \dot{n}_{y}^{L}, \dot{n}_{z}^{L}, \dot{K}^{L}) = \left(n_{x}, n_{y}, n_{z}, K^{1}\right)$. We denote the parameters of the spatial decoder by $\bm{W}_{\text{de}} = \left\{\dot{\bm{W}}^{\ell}: \ell=1,\ldots,L,L+1 \right\}$ and the number or parameters in $\bm{W}_{\text{de}}$ can be derived along the same lines of \eqref{eq:num_pars_conv_3d}. 

\subsubsection*{Results}

The three dimensional analysis, more so than the two dimensional case, was constrained by the \textit{large} amount of memory required. The spatial compression we achieved is of a factor of 5 (by comparison we were able to compress the two dimensional RBC by a factor of 16). The lengths of the input and output sequences, $b$ and $a$, respectively, were set to $b = 10$ and $a = 5$, reflecting a compromise between temporal coverage and computational feasibility. In 2D, we were able to use input and output sequence lengths of $b = 80$ and $a = 60$, respectively; such long sequences were impossible to handle in 3D due to memory constraints and the high Rayleigh number of $\text{Ra} = 2.54 \times 10^8$. Training the three-dimensional surrogate also required \textit{significantly longer} time than the 2D, owing to both the increased data dimensionality and the higher memory footprint of the convolutional and recurrent layers.

Despite these challenges, the physics-informed model largely captures the essential large-scale convective structures. However, it suboptimally reproduced the small-scale features that are critical for long-term forecasting, where strong turbulence and mixing dominate. This represents an initial step toward 3D surrogate modeling under such extreme conditions, highlighting both the potential and the current limitations of physics-informed learning in this regime. For comparison purposes, we also trained a non-physics-informed version of the model. The results are summarized in Table \ref{tab:3d_pi_vs_nonpi}.

\begin{table}[!ht]
\centering
\begin{tabular}{lccc}
\hline
Model & Data Loss ($10^{-4}$) & Physics Loss ($10^{-2}$) & Runtime \\
\hline
\textbf{PI-CRNN} &  \textbf{4.06} &   \textbf{5.11} &  3.07s (0.6s per step) \\
CRNN &  5.15 &  52.13 &  3.05s (0.6s per step) \\
\hline
\end{tabular}
\caption{Comparison between 3D PI-CRNN and 3D CRNN (the non-physics informed version of our approach). The physics-informed model outperforms the physics-agnostic model both in terms of predicting the data \textit{and} maintaining key physical structures of 3D RBC. The difference in runtime is negligible.}
\label{tab:3d_pi_vs_nonpi}
\end{table}

\begin{figure}[H]
    \centering
    \includegraphics[height=12cm, width=\linewidth]{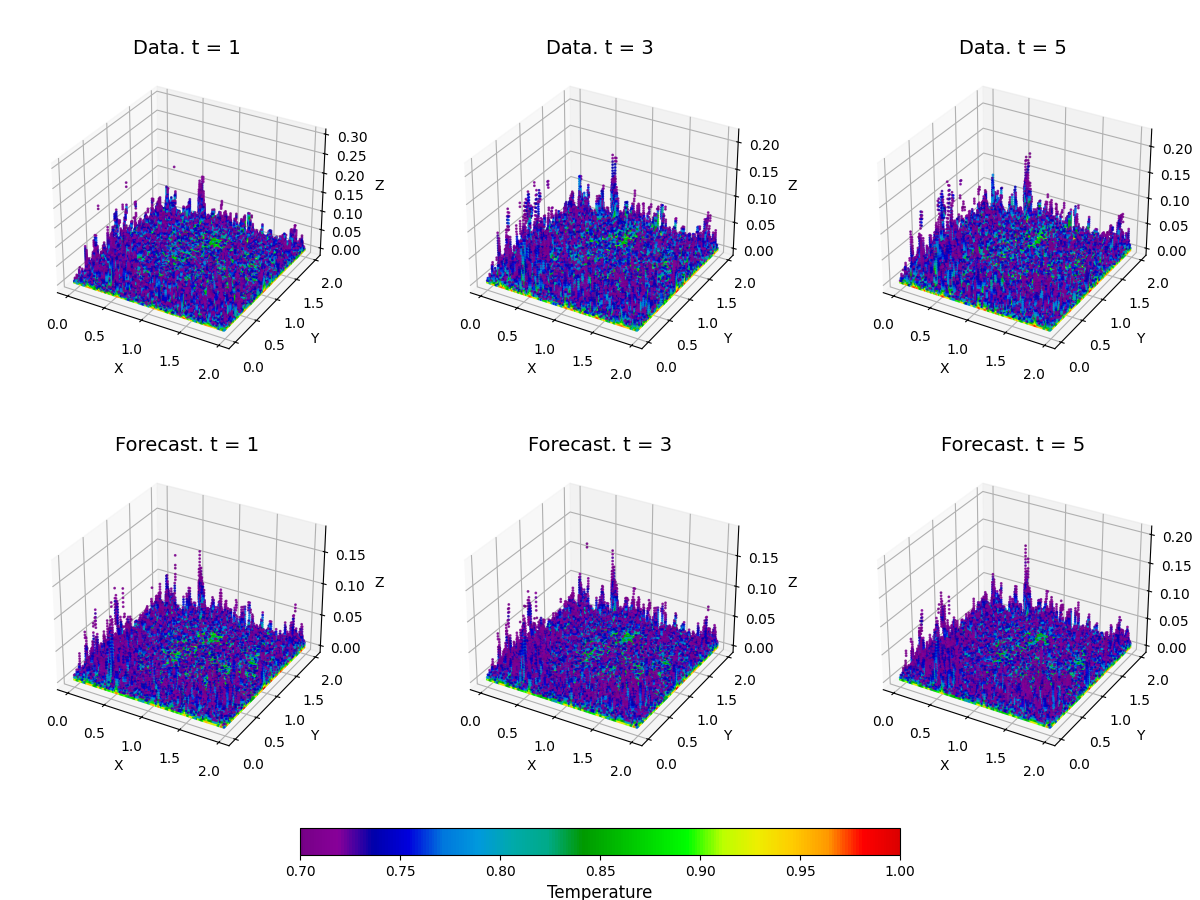}
    \caption{Snapshots of the 3D temperature field comparing the PI-CRNN forecast (bottom row) with the corresponding observations (top row) at three time steps.}
    \label{fig:s13_3d}
\end{figure}

Figure \ref{fig:s13_3d} shows three snapshots of the PI-CRNN forecast alongside the corresponding observations, highlighting the model’s ability to capture the key flow features over time. Despite the inherent difficulties posed by 3D turbulent convection, the physics-informed approach demonstrates improved stability and accuracy, which constitutes a promising strategy for future work.

\end{document}